\documentclass[journal]{IEEEtran}

\usepackage{amsmath,amssymb,amsmath,amsfonts}
\usepackage{bm}
\usepackage{bbm}
\usepackage{graphicx}
\usepackage{cite}
\usepackage{epstopdf}
\usepackage{balance}
\usepackage{gensymb}
\usepackage{color}
\usepackage{lipsum}
\usepackage{float}
\usepackage{epstopdf}
\usepackage[mathcal]{eucal}
\usepackage{subfiles}
\usepackage[T1]{fontenc}
\usepackage{siunitx}
\usepackage[hyphens]{url}
\usepackage[breaklinks=true]{hyperref}
\usepackage{tabulary}
\usepackage{epsfig,graphics,graphicx,subfig,amssymb,amstext,amsmath,algorithm,algorithmic,wrapfig,multirow}
\usepackage{array}
\usepackage{adjustbox}
\usepackage[dvipsnames]{xcolor}
\usepackage{cite}
\usepackage{times}
\usepackage{placeins}
\usepackage{textcomp}
\usepackage[font=small]{caption}
\usepackage[breaklinks=true]{hyperref}
\usepackage{flushend}
\usepackage{color, colortbl}
\usepackage{tabularx}
\usepackage{bm}
\usepackage{enumerate}
\usepackage{tikz}
\usepackage{pgfplots}
\usepackage{epstopdf}
\usetikzlibrary{shapes,arrows}
\usepackage[utf8]{inputenc}
\usepackage{mathtools}
\usepackage[utf8]{inputenc}
\usepackage{xcolor}
\usepackage{tikz}
\usetikzlibrary{chains,arrows,calc,positioning}
\usepackage{amssymb}
\usepackage{pifont}
\usepackage{soul}
\usepackage{breqn} 
\usepackage{booktabs} 
\usepackage{multirow}
\usepackage{enumitem}
\usepackage{tabulary}
\usepackage[normalem]{ulem}
\usepackage{dblfloatfix}
\usepackage{makecell}
\usepackage{lipsum}
\usepackage{amsthm}
\usepackage{comment}

\newtheoremstyle{mystyle}
  {}
  {}
  {\itshape}
  {}
  {\bfseries}
  {.}
  { }
  {}
\theoremstyle{mystyle}
\newtheorem{takeaway}{Takeaway}

\def\ISDs   {{\mathsf{ISD}_{\sf s}}}
\def\ISDd   {{\mathsf{ISD}_{\sf d}}}

\newcommand{\cmark}{\ding{51}}%
\newcommand{\xmark}{ }%

\newlength \figwidth
\pgfplotsset{}
\definecolor{bittersweet}{rgb}{1.0, 0.44, 0.37}
\definecolor{glaucous}{rgb}{0.38, 0.51, 0.71}
\definecolor{gainsboro}{rgb}{0.86, 0.86, 0.86}
\definecolor{babyblueeyes}{rgb}{0.63, 0.79, 0.95}
\definecolor{silver}{rgb}{0.75, 0.75, 0.75}
\definecolor{neoncarrot}{rgb}{1.0, 0.64, 0.26}
\definecolor{Gray}{gray}{0.9}
\definecolor{LightCyan}{rgb}{0.88,1,1}
\definecolor{BackgroundLightBlue}{rgb}{0.97,0.97,1}
\definecolor{BackgroundGray}{gray}{0.98}

\newcommand{\red}[1]{{\textcolor[rgb]{1,0,0}{#1}}}
\newcommand{\green}[1]{{\textcolor[rgb]{0,0.5,0}{#1}}}
\newcommand{\brown}[1]{{\textcolor[rgb]{0.65,0.16,0.16}{#1}}}

\newcommand{\gio}[1]{\noindent \green{ {{$\blacktriangleright$ 
   {\textsf{[Gio]: #1}} $\blacktriangleleft$}}}}

\newcommand{\adrian}[1]{\noindent \red{ {{$\blacktriangleright$ 
   {\textsf{[Adrian]: #1}} $\blacktriangleleft$}}}}
\newcommand{\angel}[1]{\noindent \brown{ {{$\blacktriangleright$ 
   {\textsf{[Angel]: #1}} $\blacktriangleleft$}}}}

\usepackage[acronyms,nonumberlist,nopostdot,nomain,nogroupskip]{glossaries}
\newacronym{quic}{QUIC}{Quick UDP Internet Connections}
\newacronym{3gpp}{3GPP}{3rd Generation Partnership Project}
\newacronym{adc}{ADC}{Analog to Digital Converter}
\newacronym{5g}{5G}{5th generation}
\newacronym{aimd}{AIMD}{Additive Increase Multiplicative Decrease}
\newacronym{am}{AM}{Acknowledged Mode}
\newacronym{amc}{AMC}{Adaptive Modulation and Coding}
\newacronym{aqm}{AQM}{Active Queue Management}
\newacronym{awgn}{AGWN}{Additive White Gaussian Noise}
\newacronym{afd}{AFD}{Austin Fire Department}
\newacronym{balia}{BALIA}{Balanced Link Adaptation}
\newacronym{bdp}{BDP}{Bandwidth-Delay Product}
\newacronym{bf}{BF}{Beamforming}
\newacronym{cc}{CC}{Congestion Control}
\newacronym{cdf}{CDF}{Cumulative Distribution Function}
\newacronym{cn}{CN}{Core Network}
\newacronym{cqi}{CQI}{Channel Quality Information}
\newacronym{cp}{CP}{Control Plane}
\newacronym{csirs}{CSI-RS}{Channel State Information - Reference Signal}
\newacronym{dc}{DC}{Dual Connectivity}
\newacronym{dce}{DCE}{Direct Code Execution}
\newacronym{dci}{DCI}{Downlink Control Information}
\newacronym{dl}{DL}{Downlink}
\newacronym{dmr}{DMR}{Deadline Miss Ratio}
\newacronym{dmrs}{DMRS}{DeModulation Reference Signal}
\newacronym{e2e}{E2E}{End-to-End}
\newacronym{ecn}{ECN}{Explicit Congestion Notification}
\newacronym{edf}{EDF}{Earliest Deadline First}
\newacronym{enb}{eNB}{evolved Node Base}
\newacronym{epc}{EPC}{Evolved Packet Core}
\newacronym{es}{ES}{Edge Server}
\newacronym{fdma}{FDMA}{Frequency Division Multiple Access}
\newacronym{fdd}{FDD}{Frequency Division Duplexing}
\newacronym[firstplural=Radio Access Technologies (RATs)]{rat}{RAT}{Radio Access Technology}
\newacronym{fs}{FS}{Fast Switching}
\newacronym{ftp}{FTP}{File Transfer Protocol}
\newacronym{gnb}{gNB}{Next Generation Node Base}
\newacronym{harq}{HARQ}{Hybrid Automatic Repeat reQuest}
\newacronym{hetnet}{HetNet}{Heterogeneous Network}
\newacronym{hh}{HH}{Hard Handover}
\newacronym{hol}{HOL}{Head-of-Line}
\newacronym{ia}{IA}{Initial Access}
\newacronym{imt}{IMT}{International Mobile Telecommunication}
\newacronym{iot}{IoT}{Internet of Things}
\newacronym{los}{LOS}{Line of Sight}
\newacronym{lte}{LTE}{Long Term Evolution}
\newacronym{m2m}{M2M}{Machine to Machine}
\newacronym{mac}{MAC}{Medium Access Control}
\newacronym{mc}{MC}{Multi-Connectivity}
\newacronym{mcs}{MCS}{Modulation and Coding Scheme}
\newacronym{mec}{MEC}{Mobile Edge Cloud}
\newacronym{mi}{MI}{Mutual Information}
\newacronym{mimo}{MIMO}{Multiple Input, Multiple Output}
\newacronym{mmwave}{mmWave}{millimeter wave}
\newacronym{mr}{MR}{Maximum Rate}
\newacronym{mss}{MSS}{Maximum Segment Size}
\newacronym{mtd}{MTD}{Machine-Type Device}
\newacronym{mtu}{MTU}{Maximum Transmission Unit}
\newacronym{nfv}{NFV}{Network Function Virtualization}
\newacronym{nlos}{NLOS}{Non Line of Sight}
\newacronym{nr}{NR}{New Radio}
\newacronym{ofdm}{OFDM}{Orthogonal Frequency Division Multiplexing}
\newacronym{pdcch}{PDCCH}{Physical Downlonk Control Channel}
\newacronym{pdcp}{PDCP}{Packet Data Convergence Protocol}
\newacronym{pdsch}{PDSCH}{Physical Downlink Shared Channel}
\newacronym{pdu}{PDU}{Packet Data Unit}
\newacronym{pf}{PF}{Proportional Fair}
\newacronym{pgw}{PGW}{Packet Gateway}
\newacronym{phy}{PHY}{Physical}
\newacronym{pbch}{PBCH}{Physical Broadcast Channel}
\newacronym[plural=\gls{mme}s,firstplural=Mobility Management Entities (MMEs)]{mme}{MME}{Mobility Management Entity}
\newacronym{prb}{PRB}{Physical Resource Block}
\newacronym{pss}{PSS}{Primary Synchronization Signal}
\newacronym{pucch}{PUCCH}{Physical Uplink Control Channel}
\newacronym{pusch}{PUSCH}{Physical Uplink Shared Channel}
\newacronym{rach}{RACH}{Random Access Channel}
\newacronym{ran}{RAN}{Radio Access Network}
\newacronym{red}{RED}{Robotics Emergency Deployment}
\newacronym{rf}{RF}{Radio Frequency}
\newacronym{rlc}{RLC}{Radio Link Control}
\newacronym{rlf}{RLF}{Radio Link Failure}
\newacronym{rrc}{RRC}{Radio Resource Control}
\newacronym{rrm}{RRM}{Radio Resource Management}
\newacronym{rr}{RR}{Round Robin}
\newacronym{rs}{RS}{Remote Server}
\newacronym{rsrp}{RSRP}{Reference Signal Received Power}
\newacronym{rss}{RSS}{Received Signal Strength}
\newacronym{rtt}{RTT}{Round Trip Time}
\newacronym{rw}{RW}{Receive Window}
\newacronym{rx}{RX}{Receiver}
\newacronym{sa}{SA}{standalone}
\newacronym{sack}{SACK}{Selective Acknowledgment}
\newacronym{sap}{SAP}{Service Access Point}
\newacronym{sch}{SCH}{Secondary Cell Handover}
\newacronym{scoot}{SCOOT}{Split Cycle Offset Optimization Technique}
\newacronym{sdma}{SDMA}{Spatial Division Multiple Access}
\newacronym{sinr}{SINR}{Signal to Interference plus Noise Ratio}
\newacronym{sm}{SM}{Saturation Mode}
\newacronym{snr}{SNR}{Signal to Noise Ratio}
\newacronym{son}{SON}{Self-Organizing Network}
\newacronym{ss}{SS}{Synchronization Signal}
\newacronym{srs}{SRS}{Sounding Reference Signal}
\newacronym{sss}{SSS}{Secondary Synchronization Signal}
\newacronym{tb}{TB}{Transport Block}
\newacronym{tcp}{TCP}{Transmission Control Protocol}
\newacronym{tdd}{TDD}{Time Division Duplexing}
\newacronym{tdma}{TDMA}{Time Division Multiple Access}
\newacronym{tfl}{TfL}{Transport for London}
\newacronym{tm}{TM}{Transparent Mode}
\newacronym{trp}{TRP}{Transmitter Receiver Pair}
\newacronym{tti}{TTI}{Transmission Time Interval}
\newacronym{ttt}{TTT}{Time-to-Trigger}
\newacronym{tx}{TX}{Transmitter}
\newacronym{ue}{UE}{User Equipment}
\newacronym{ul}{UL}{Uplink}
\newacronym{uml}{UML}{Unified Modeling Language}
\newacronym{um}{UM}{Unacknowledged Mode}
\newacronym{utc}{UTC}{Urban Traffic Control}
\newacronym{vm}{VM}{Virtual Machine}
\newacronym{rsrq}{RSRQ}{Reference Signal Received Quality}
\newacronym{rssi}{RSSI}{Received Signal Strength Indicator}
\newacronym{crs}{CRS}{Cell Reference Signal}
\newacronym{comp}{CoMP}{Coordinated Multi-Point}
\newacronym{cran}{C-RAN}{Cloud \acrlong{ran}}
\newacronym{ca}{CA}{Carrier Aggregation}
\newacronym{cco}{CC}{Carrier Component}
\newacronym{nsa}{NSA}{Non Stand Alone}
\newacronym{embb}{eMBB}{Enhanced Mobility Broadband}
\newacronym{bsr}{BSR}{Buffer Status Report}
\newacronym{srb}{SRB}{Service Radio Bearer}
\newacronym{scm}{SCM}{Spatial Channel Model}
\newacronym{sctp}{SCTP}{Stream Control Transmission Protocol}
\newacronym{mptcp}{MPTCP}{Multi-path TCP}
\newacronym{ietf}{IETF}{Internet Engineering Task Force}
\newacronym{os}{OS}{Operating System}
\newacronym{tls}{TLS}{Transport Layer Security}
\newacronym{rfc}{RFC}{Request for Comments}
\newacronym{http}{HTTP}{HyperText Transfer Protocol}
\newacronym{nat}{NAT}{Network Address Translation}
\newacronym{api}{API}{Application Programming Interface}
\newacronym{rto}{RTO}{Retransmission Timeout}
\newacronym{psc}{PSC}{Public Safety Communication}
\newacronym{rpgm}{RPGM}{Reference Point Group Mobility}
\newacronym{ic}{IC}{Incident Command}
\newacronym{rsu}{RSU}{Road Side Unit}
\newacronym{uav}{UAV}{unmanned aerial vehicle}
\newacronym{usv}{USV}{Unmanned Surface Vehicle}
\newacronym{uas}{UAS}{Unmanned Aerial System}
\newacronym{iab}{IAB}{Integrated Access and Backhaul}
\newacronym{qoe}{QoE}{Quality of Experience}
\newacronym{ssim}{SSIM}{Structural Similarity Index}
\newacronym{psnr}{PSNR}{Peak Signal to Noise Ratio}
\newacronym{bs}{BS}{Base Station}
\newacronym{mu}{MU}{Multiple User}
\newacronym{ag}{AG}{Air-to-Ground}
\newacronym{af}{AF}{Array Factor}
\newacronym{ula}{ULA}{Uniform Linear Array}
\newacronym{upa}{UPA}{Uniform Planar Array}
\newacronym{lcs}{LCS}{Local Coordinate System}
\newacronym{psd}{PSD}{Power Spectral Density}
\newacronym{vq}{VQ}{vector quantization}
\newacronym{a2g}{A2G}{air-to-ground}
\newacronym{em}{EM}{electromagnetic}
\newacronym{vae}{VAE}{variational autoencoder}

\makeatletter

 \let\oldforeign@language\foreign@language
 \DeclareRobustCommand{\foreign@language}[1]{%
   \lowercase{\oldforeign@language{#1}}}

\def\nb0{{\mathbf{0}}}
\def\nb1{{\mathbf{1}}}

\makeatother

\IEEEoverridecommandlockouts

\begin{document}
\title{What Will the Future of\\UAV Cellular Communications Be?\\A Flight from 5G to 6G}


\author{
Giovanni~Geraci,~\IEEEmembership{Senior~Member,~IEEE,} Adrian~Garcia-Rodriguez,~\IEEEmembership{Member,~IEEE,} 
M.~Mahdi~Azari,~\IEEEmembership{Member,~IEEE,}
Angel~Lozano,~\IEEEmembership{Fellow,~IEEE,} 
Marco~Mezzavilla,~\IEEEmembership{Senior~Member,~IEEE,}
Symeon~Chatzinotas,~\IEEEmembership{Senior~Member,~IEEE,}
Yun~Chen, \\
Sundeep~Rangan,~\IEEEmembership{Fellow,~IEEE,}
and Marco~Di~Renzo,~\IEEEmembership{Fellow,~IEEE} 
\thanks{G.~Geraci and A.~Lozano are with University Pompeu Fabra, 08018 Barcelona, Spain (\{giovanni.geraci, angel.lozano\}@upf.edu).}
\thanks{A.~Garcia-Rodriguez is with the Mathematical and Algorithmic Sciences Lab, Huawei Technologies, 92100 Boulogne-Billancourt, France (a.garciarodriguez.2013@ieee.org).}
\thanks{M.~M.~Azari and S.~Chatzinotas are with the Interdisciplinary Centre for Security,
Reliability and Trust (SnT), University of Luxembourg, L-1855 Luxembourg (\{mohammadmahdi.azari, symeon.chatzinotas\}@uni.lu).}
\thanks{M.~Mezzavilla and S.~Rangan are with NYU WIRELESS, Tandon School of Engineering, New York University, Brooklyn, NY 11201, USA (\{mezzavilla, srangan\}@nyu.edu).}
\thanks{Y.~Chen is with North Carolina State University, Raleigh, NC 27606, USA (yunchen@utexas.edu).}
\thanks{M. Di Renzo is with Universit\'e Paris-Saclay, CNRS, CentraleSup\'elec, Laboratoire des Signaux et Syst\`emes, 3 Rue Joliot-Curie, 91192 Gif-sur-Yvette, France (marco.di-renzo@universite-paris-saclay.fr).}
}

\maketitle

\begin{abstract}
\emph{What will the future of UAV cellular communications be?} In this tutorial article, we address such a compelling yet difficult question by embarking on a journey from 5G to 6G and sharing a large number of realistic case studies supported by original results. We start by overviewing the status quo on UAV communications from an industrial standpoint, providing fresh updates from the 3GPP and detailing new 5G NR features in support of aerial devices. We then show the potential and the limitations of such features.
In particular, we demonstrate how sub-6\,GHz massive MIMO can successfully tackle cell selection and interference challenges, we showcase encouraging mmWave coverage evaluations in both urban and suburban/rural settings, and we examine the peculiarities of direct device-to-device communications in the sky. 
Moving on, we sneak a peek at next-generation UAV communications, listing some of the use cases envisioned for the 2030s.
We identify the most promising 6G enablers for UAV communication, those expected to take the performance and reliability to the next level.
For each of these disruptive new paradigms (non-terrestrial networks, cell-free architectures, artificial intelligence, reconfigurable intelligent surfaces, and THz communications), we gauge the prospective benefits for UAVs and discuss the main technological hurdles that stand in the way. 
All along, we distil our numerous findings into essential takeaways, and we identify key open problems worthy of further study.
\end{abstract}

\section{Introduction}
\label{sec:introduction}


What would our beloved grandmothers have thought if they had received a bunch of roses delivered by a UAV?\footnote{Unmanned aerial vehicle, also commonly referred to as drone.} Their initial disbelief and skepticism would have soon made room for sheer gratitude, acknowledging the fond gesture. Likewise, even the most tech-reluctant of us will eventually come to terms with UAVs. Barely seen in action movies until a decade ago, their progressive
blending into our daily lives will enhance safety and greatly impact labor and leisure activities alike.


\subsection{Motivation and Background}

Most stakeholders regard reliable connectivity as a must-have for the UAV ecosystem to thrive. Exciting and socially impactful use cases, including automated crowd, weather, and traffic monitoring and live virtual reality experiences, mandate handling UAV-originated high-resolution video. Venture capitalists, willing to invest in advanced aerial mobility, and also governments, are betting on the wireless industry to address the unmet needs of critical UAV command and control (C2) links \cite{WSJ21}. A breakthrough in reliability could persuade legislators to ease regulations on civilian pilotless flights, giving the green light to new vertical markets. 

As a result, cellular communications involving UAVs have witnessed a surge of interest, following two philosophies epitomized as \emph{what can UAVs do for networks} and \emph{what can networks do for UAVs}, respectively \cite{GerGarLin2019}. The former, pioneered by academia, advocates enhancing cellular networks with UAV-mounted base stations (BSs), wirelessly backhauled to the ground. For disaster assistance, border surveillance, or hotspot events, UAVs carrying a radio access node could be promptly dispatched, cheaply maintained, and easily manoeuvred.
The latter, at first predominant in the standardization fora, triggered a parallel stream of research that aims at supporting UAV end-devices through cellular networks \cite{3GPP36777}. 

Whether featuring UAVs as data beneficiaries or suppliers, the fly-and-connect dream faces showstoppers. As ground cellular BSs are typically downtilted, UAVs are only reached by their upper antenna sidelobes and experience sharp signal fluctuations. Flying above buildings, UAVs receive/create line-of-sight (LoS) interfering signals from/to a plurality of cells, respectively hindering correct decoding of latency-sensitive C2 and overwhelming the weaker signals of ground users (GUEs). The pronounced signal strength fluctuations experienced by UAVs, combined with their high speeds and rapid veering, create unique mobility management challenges, causing frequent and sometimes unnecessary handovers. Technical solutions to these problems, among others, have been introduced in 4G Long-term Evolution (LTE) to deal with a handful of connected UAVs, but as they proliferate, UAV cellular communications will struggle to 
`take off' by solely relying on a network built for the ground. Aware of these hurdles, and acknowledging the importance of aerial communications, the wireless research community has been rolling up its sleeves to drive a native and long-lasting support for UAVs in 5G New Radio (NR) and beyond \cite{ZenGuvZha2020,SaaBenMoz2020}.

\subsection{Contribution}


\begin{table*}[!t] 
\vspace{1mm}
\caption{Detailed content comparison between this article and relevant surveys and tutorials (covered: \cmark, partially covered: $\partial$).} 
\label{tab:otherTutorials}
\centering
\colorbox{BackgroundGray}{%
\begin{tabular}{ |l|c|c|c|c|c|c|c|c|c| } 
\toprule 
\rowcolor{BackgroundLightBlue}
 \hspace{65mm} \shortstack{\textbf{Reference $\rightarrow$}}  & \cite{wu20205g} &  \cite{mozaffari2019tutorial} &  \cite{fotouhi2019survey} & \cite{zeng2019accessing} & \cite{li2018uav} &  \cite{mishra2020survey} &  \cite{vinogradov2019tutorial} &  \cite{shakhatreh2019unmanned} & \cite{lahmeri2020machine}  \\
 \rowcolor{BackgroundLightBlue}
 \shortstack{\textbf{Content  $\downarrow$}} &   &   &   &   &   &   &   & &   \\ \midrule 
 Sec.~\ref{sec:5G}: UAV Cellular Communications in 5G NR   &  &   &   &   &    &   &   &  & \\
 \quad \ref{subsec:3Dhighways}: 3D Aerial Highways and the Delivery of Tomorrow & & & & & & & & & \\
 \quad \ref{subsec:NRusecases}: NR Use cases and Requirements  & $\partial$ & \xmark  &  \xmark & $\partial$  & \xmark  & $\partial$  & \xmark  & \xmark  & \xmark \\
 \quad \ref{subsec:NRfeatures}: NR Features in Support of UAVs & $\partial$ & \xmark  &  \xmark & \xmark  & \xmark  & $\partial$  & \xmark   & \xmark & \xmark \\ \midrule 
 \rowcolor{BackgroundLightBlue}
Sec.~\ref{sec:massiveMIMO}: 5G NR Massive MIMO for Enhanced UAV Cellular Support    &  &   &   &   &   &   &   &  & \\
\rowcolor{BackgroundLightBlue}
  \quad \ref{subsec:massivMIMO1}: Tackling the Initial Access and Cell Selection Challenges & \xmark & \xmark  & \xmark  & \xmark  &  \xmark  & \xmark &  \xmark & \xmark & \xmark \\
 \rowcolor{BackgroundLightBlue}
  \quad \ref{subsec:massivMIMO2}: Tackling the Interference Challenge &  $\partial$ & \xmark  & \xmark  & \xmark  & \xmark  & $\partial$ & \xmark    &   \xmark & \xmark  \\ \midrule 
 Sec.~\ref{sec:mmWave}: 5G NR mmWave for UAV Capacity Boost    &  &   &   &   &   &   &   &  & \\
 \quad \ref{subsec:mmWaveUrban}: Urban mmWave Coverage & \xmark  & \xmark  & \xmark  & \xmark  &  \xmark   & \xmark   &  \xmark  & \xmark & \xmark \\
 \quad\ref{subsec:mmWaveRural}: Rural/suburban mmWave Coverage & \xmark & \xmark  & \xmark  &  \xmark & \xmark   & \xmark  &  \xmark  & \xmark & \xmark  \\ \midrule 
 \rowcolor{BackgroundLightBlue}
 Sec.~\ref{sec:U2U}: UAV-to-UAV Cellular Communications    &  &   &   &   &   &   &   &  & \\
\rowcolor{BackgroundLightBlue}
   \quad \ref{subsec:U2Uapp}: Applications and Distinct Features of Aerial D2D & & & & & & & & & \\
\rowcolor{BackgroundLightBlue}
   \quad \ref{subsec:U2Udesign}: U2U Communications: Key Design Parameters & & & & & & & & & \\
\rowcolor{BackgroundLightBlue}
   \quad \ref{subsec:U2Uperformance}: Performance Analysis of U2U Communications & & & & & & & & & \\
\rowcolor{BackgroundLightBlue}
   \quad \ref{subsec:U2UmmWave}: mmWave U2U Communications & & & & & & & & &  \\ \midrule 
 Sec.~\ref{sec:6G}: UAV Cellular Communications Beyond 5G  & \xmark & \xmark  & \xmark  &  \xmark & \xmark   & \xmark  &  \xmark  & \xmark & \xmark  \\ \midrule 
 \rowcolor{BackgroundLightBlue}
Sec.~\ref{sec:NTN}: Towards Resilient UAV Communications with NTN  &  &   &   &   &   &   &   & &  \\
\rowcolor{BackgroundLightBlue}
     \quad \ref{subsec:NTNUAVtosat}: UAV-to-Satellite Communications  & \xmark & $\partial$  & $\partial$ & $\partial$  &  \cmark  &  \xmark &  $\partial$  & $\partial$  & \xmark \\
 \rowcolor{BackgroundLightBlue}
     \quad \ref{subsec:NTNroadblocks}: Roadblocks to an Integrated Ground-Air-Space Network & & & & & & & & &  \\ \midrule 
 Sec.~\ref{sec:cellfree}: Cell-free Architectures for UAVs   & $\partial$ & \xmark  & \xmark   & \xmark  &  \xmark  &  \xmark &  \xmark  & \xmark  & \xmark   \\ \midrule 
 \rowcolor{BackgroundLightBlue}
Sec.~\ref{sec:AI}: AI to Model and Enhance UAV Communications   &  &   &   &   &   &   &   &  & \\
\rowcolor{BackgroundLightBlue}
  \quad \ref{subsec:AIchannel}: AI for Aerial Channel Modeling   & \xmark & \xmark  & \xmark   & \xmark  & \xmark   & \xmark  & \xmark   &  \xmark & $\partial$ \\
\rowcolor{BackgroundLightBlue}
  \quad \ref{subsec:AImobility}: AI for UAV Mobility Management & \xmark & \xmark  & \xmark  & $\partial$   & \xmark   & $\partial$  & \xmark   & \xmark & \xmark  \\ \midrule 
 Sec.~\ref{sec:RIS}: Assisting UAV Cellular Communications with RISs   & $\partial$ & \xmark  & \xmark   & \xmark  & \xmark   & \xmark  &  \xmark  & \xmark  & $\partial$  \\ \midrule 
 \rowcolor{BackgroundLightBlue}
Sec.~\ref{sec:THz}: UAV Communications at THz Frequencies   & \xmark &  \xmark & \xmark  & \xmark  & \xmark   & $\partial$  &  \xmark  &   \xmark  & \xmark  \\ 
\bottomrule 
\end{tabular}
}
\end{table*}

\begin{table}[t!]
    \vspace{1mm}
    \caption{List of acronyms used throughout this article.}
    \label{tab:acronyms}
    \centering
    \colorbox{BackgroundGray}{%
    \begin{tabular}{l|l}
    \toprule 
    \rowcolor{BackgroundLightBlue}
    \shortstack{\textbf{Acronym}} 
    & \shortstack{\textbf{Definition}} \\ \midrule 
3GPP & Third-generation partnership project\\
\rowcolor{BackgroundLightBlue}
AI & Artificial intelligence\\
BS & Base station\\
\rowcolor{BackgroundLightBlue}
    BVLoS & Beyond visual line of sight\\
C2 & Command and control\\
\rowcolor{BackgroundLightBlue}
    CDF & Cumulative distribution function\\
    CSI & Channel state information\\
\rowcolor{BackgroundLightBlue}
D2D & Device-to-device communications\\
    DL & Downlink\\
\rowcolor{BackgroundLightBlue}
DVB & Digital video broadcasting\\
    eVTOL & Electric vertical takeoff and landing vehicle\\
\rowcolor{BackgroundLightBlue}
    GEO & Geostationary equatorial orbit\\
GUE & Ground user equipment\\
\rowcolor{BackgroundLightBlue}
    HAP & High-altitude platform\\
HARQ & Hybrid automatic repeat request\\
\rowcolor{BackgroundLightBlue}
    HD & High definition\\
IoT & Internet of Things\\
\rowcolor{BackgroundLightBlue}
    ISD & Intersite distance\\
    LEO & Low Earth orbit\\
\rowcolor{BackgroundLightBlue}
LoS & Line of sight\\
    LTE & Long-term Evolution\\
\rowcolor{BackgroundLightBlue}
MF & Matched filter\\
    mMIMO & Massive MIMO\\
\rowcolor{BackgroundLightBlue}
MMSE & Minimum mean-square error\\
    mmWave & Millimeter wave\\
\rowcolor{BackgroundLightBlue}
    NLoS & Non line of sight\\
NN & Neural network\\
\rowcolor{BackgroundLightBlue}
NTN & Non-terrestrial network\\
    NR & New Radio\\
\rowcolor{BackgroundLightBlue}
PRB & Physical resource block\\
    RAN & Radio access network \\
\rowcolor{BackgroundLightBlue}
    RIS & Reconfigurable intelligent surface\\
    RSRP & Reference signal received power\\
\rowcolor{BackgroundLightBlue}
    SINR & Signal-to-interference-plus-noise ratio\\
SNR & Signal-to-noise ratio\\
\rowcolor{BackgroundLightBlue}
    SRS & Sounding reference signal\\
SSB & Synchronization signal block\\
\rowcolor{BackgroundLightBlue}
    SU & Single user\\
TDD & Time division duplexing\\
\rowcolor{BackgroundLightBlue}
    U2U & UAV to UAV\\
UAV & Unmanned aerial vehicle\\
\rowcolor{BackgroundLightBlue}
    UL & Uplink\\
    UMa & Urban Macro\\
\rowcolor{BackgroundLightBlue}
UMi & Urban Micro\\
    URA & Uniform rectangular array \\
\rowcolor{BackgroundLightBlue}
    UTM & UAV traffic management\\
UxNB & Radio access node on-board UAV\\
\rowcolor{BackgroundLightBlue}
VLoS & Visual line of sight\\
    \bottomrule 
    \end{tabular}
    }
\noindent
\end{table}

UAV cellular communications have been all the rage and the present one is
far from the first tutorial on the topic.
However, this one-of-a-kind article is arguably the first merging industrial and academic views to forecast, through supporting evidence, the future of UAV cellular communications. This is the most appropriate time to do so, given the significant UAV-related updates in the third-generation partnership project (3GPP), e.g., the recent definition of new UAV applications for NR Rel. 17 \cite{3GPP22.125}. Moreover, the time is ripe to quantify the benefits and limitations of key 5G NR features such as massive MIMO (mMIMO) and millimeter wave (mmWave), among others. Lastly, the societal role to be played by UAVs in the coming decades, as well as the corresponding 6G empowering technologies, have become clearer than ever.

Abraham Lincoln is often quoted as having said that the best way to predict the future is by creating it. Likewise in this article, it is by showcasing our own recent research findings that we navigate the broad readership from 5G to 6G UAV use cases, requirements, and enablers. Our novel results---abundantly supplied, carefully selected, and accessibly presented---have been obtained in compliance with the assumptions specified by the 3GPP in \cite{3GPP36777, 3GPP38901}. Through these results, we share numerous lessons learnt and provide essential guidelines, defining a complete picture of the upcoming UAV cellular landscape and, we hope, acting as a catalyst for much-needed new research.

Presented in Table~\ref{tab:otherTutorials} is a detailed content comparison between this article and other relevant surveys and tutorials \cite{wu20205g,mozaffari2019tutorial,fotouhi2019survey,zeng2019accessing,li2018uav,mishra2020survey,vinogradov2019tutorial,shakhatreh2019unmanned,lahmeri2020machine}. Shorter overview articles can be found in \cite{cao2018airborne,frew2008airborne,zeng2018cellular,oubbati2020softwarization,hentati2020comprehensive,AbdPowMar2020}, whereas \cite{boccadoro2020internet,motlagh2016low,HayYanMuz2016,gupta2015survey,bekmezci2013flying,shakeri2019design,song2020survey,KhaGuvMat2019,khuwaja2018survey,zhang2019cooperation,bor20195g,AbdMar2020,noor2020review,hassanalian2017classifications,naqvi2018drone,al2020uavs,tahir2019swarms,nawaz2020uav,kodheli2020satellite,liu2018space,sharma2020communication} comprise other surveys or tutorials with less relevance to the present article.

\subsection{Outline}

The rest of this article is organized as follows. Sec.\ II overviews the status quo on UAV communications in 5G NR, distilling fresh updates from the 3GPP and detailing new 5G NR features that come---directly or indirectly---in support of UAVs. Sec.\ III focuses on how mMIMO can successfully tackle the cell selection and interference challenges. In Sec.\ IV, we move to higher frequencies and evaluate the mmWave coverage in both urban and suburban/rural settings. Sec.\ V is devoted to device-to-device links in the sky, that is, direct UAV-to-UAV cellular communications. In Sec.\ VI, we take a first peek beyond 5G, reviewing the most likely use cases envisioned for the next decade. Sec.\ VII to Sec.\ XI deal with five promising 6G disruptive paradigms (non-terrestrial networks (NTNs), cell-free architectures, artificial intelligence (AI), reconfigurable intelligent surfaces (RISs), and THz communications), quantifying their potential benefits for UAVs, and discussing the main technological hurdles yet to be overcome. We wrap up the article in Sec.\ IX and, all along, we extract our broad vision into takeaways, also highlighting open research questions for the interested reader.

The acronyms employed throughout the text are listed and described in Table \ref{tab:acronyms}. 
\section{UAV Cellular Communications in 5G NR} 
\label{sec:5G}


\begin{table*}[t!]
    \vspace{1mm}
    \caption{Requirements for selected 5G UAV use cases \cite{3GPP22125,3GPP22829}. Less demanding use cases have been omitted for brevity.}
    \label{tab:5G_use_cases}
    \centering
    \colorbox{BackgroundGray}{%
    \begin{tabular}{c|l|l|c|c|c|c}
    \toprule 
    \rowcolor{BackgroundLightBlue}
    \shortstack{\textbf{Command and control}} 
    & \shortstack{\textbf{Message interval}} 
    & \shortstack{\textbf{E2E latency}} 
    & \shortstack{\textbf{UAV speed}} 
    & \shortstack{\textbf{Message size}} 
    & \shortstack{\textbf{Reliability}}
    & \shortstack{\textbf{Remarks}} \\ 
    \midrule 
    {\multirow{2}{*}{\makecell{Waypoints control\\(cruising altitude)}}} 
    & {\multirow{2}{*}{\SI{1}{s}}} & {\multirow{2}{*}{\SI{1}{s}}} & {\multirow{2}{*}{\SI{300}{km/h}}} & {DL: \SI{100}{B}} & {\multirow{2}{*}{99.9\%}} & {\multirow{2}{*}{\makecell{---}}} \\ \cmidrule{5-5}
    & {\multirow{2}{*}{}} & {\multirow{2}{*}{}} & {\multirow{2}{*}{}}  & {UL: \SI{84}{}--\SI{140}{B}} & {\multirow{2}{*}{}} & {\multirow{2}{*}{}} \\
    \midrule 
    { \multirow{2}{*}{\makecell{Direct steering control\\(takeoff, landing, danger)}}}
    &  {\multirow{2}{*}{\SI{40}{ms}}} & {\multirow{2}{*}{\SI{40}{ms}}} &  {\multirow{2}{*}{\SI{60}{km/h}}} & {DL: \SI{24}{B}} &  {\multirow{2}{*}{99.9\%}} &  {\multirow{2}{*}{\makecell{Video feedback\\required$^{\star}$}}} \\ \cmidrule{5-5}
    & {\multirow{2}{*}{}} &  {\multirow{2}{*}{}} & {\multirow{2}{*}{}}  & {UL: \SI{84}{}--\SI{140}{B}} & {\multirow{2}{*}{}} & {\multirow{2}{*}{}} \\
    \midrule 
    {\multirow{2}{*}{\makecell{Autonomous flight via\\traffic management system}}}
    & {\multirow{2}{*}{\SI{1}{s}}} & {\multirow{2}{*}{\SI{5}{s}}} & {\multirow{2}{*}{\SI{300}{km/h}}} & {DL: \SI{10}{kB}} & {\multirow{2}{*}{99.9\%}} & {\multirow{2}{*}{\makecell{Nominal mission. Unusual\\events require extra DL.}}} \\ \cmidrule{5-5}
    & {\multirow{2}{*}{}} & {\multirow{2}{*}{}} & {\multirow{2}{*}{}}  & {UL: \SI{1500}{B}} & {\multirow{2}{*}{}} & {\multirow{2}{*}{}} \\
    \bottomrule 
    \toprule 
    \rowcolor{BackgroundLightBlue}
    \shortstack{\textbf{Payload data}} 
    & \shortstack{\textbf{Bit rate}} 
    & \shortstack{\textbf{E2E latency}} 
    & \shortstack{\textbf{Altitude}} 
    & \shortstack{\textbf{Service region}} 
    & \shortstack{\textbf{Positioning}}
    & \shortstack{\textbf{Remarks}} \\ \midrule 
    {\multirow{2}{*}{\makecell{$^{\star}$Video feedback (UL) for\\direct steering control}}}
    & VLoS: \SI{2}{Mbps} & VLoS: \SI{1}{s} & {\multirow{2}{*}{---}} & {\multirow{2}{*}{All}} & {\multirow{2}{*}{---}} & VLoS reliability 99.9\% \\ \cmidrule{2-3}\cmidrule{7-7} 
    & BVLoS: \SI{4}{Mbps} & BVLoS: \SI{140}{ms} & {\multirow{2}{*}{}}  & {\multirow{2}{*}{}} & {\multirow{2}{*}{}} & BVLoS reliability 99.99\% \\
    \midrule 
    {\multirow{2}{*}{\makecell{8K video live\\broadcast}}}
    & UL: \SI{100}{Mbps} & UL: \SI{200}{ms} & {\multirow{2}{*}{<\SI{100}{m}}} & {\multirow{2}{*}{Urban, scenic}} & {\multirow{2}{*}{\SI{0.5}{m}}} & {\multirow{2}{*}{---}} \\ \cmidrule{2-3}
    & DL: \SI{600}{kbps} & DL: \SI{20}{ms} & {\multirow{2}{*}{}}  & {\multirow{2}{*}{}} & {\multirow{2}{*}{}} & {\multirow{2}{*}{}} \\
    \midrule 
    {\multirow{2}{*}{\makecell{Laser mapping/\\HD patrol}}}
    & UL: \SI{120}{Mbps} & UL: \SI{200}{ms} & {\multirow{2}{*}{\SI{30}{}--\SI{300}{m}}} & {\multirow{2}{*}{\makecell{Urban, rural,\\scenic}}} & {\multirow{2}{*}{\SI{0.5}{m}}} & {\multirow{2}{*}{\makecell{Average speed\\\SI{60}{km/h}}}} \\ \cmidrule{2-3} 
    & DL: \SI{300}{kbps} & DL: \SI{20}{ms} & {\multirow{2}{*}{}}  & {\multirow{2}{*}{}} & {\multirow{2}{*}{}} & {\multirow{2}{*}{}} \\
    \midrule 
    {\multirow{2}{*}{4$\times$4K AI surveillance}}
    & UL: \SI{120}{Mbps} & UL: \SI{20}{ms} & {\multirow{2}{*}{<\SI{200}{m}}} & {\multirow{2}{*}{Urban, rural}} & {\multirow{2}{*}{0.1~m}} & {\multirow{2}{*}{---}} \\ \cmidrule{2-3} 
    & DL: \SI{50}{Mbps} & DL: \SI{20}{ms} & {\multirow{2}{*}{}}  & {\multirow{2}{*}{}} & {\multirow{2}{*}{}} & {\multirow{2}{*}{}} \\
    \midrule 
    {\multirow{2}{*}{\makecell{Remote UAV controller\\through HD video}}}
    & UL: \SI{25}{Mbps} & UL: \SI{100}{ms} & {\multirow{2}{*}{<\SI{300}{m}}} & {\multirow{2}{*}{Urban, rural}} & {\multirow{2}{*}{\SI{0.5}{m}}} & {\multirow{2}{*}{\makecell{Maximum speed\\ \SI{160}{km/h}}}} \\ \cmidrule{2-3} 
    & DL: \SI{300}{kbps} & DL: \SI{20}{ms} & {\multirow{2}{*}{}}  & {\multirow{2}{*}{}} & {\multirow{2}{*}{}} & {\multirow{2}{*}{}} \\
    \bottomrule 
    \end{tabular}
    }
\noindent
\end{table*}

The current decade will see an unprecedented growth in the number of UAVs employed for public and industrial purposes and, likewise, in the amount of data they generate and transfer in real-time to and from ground control stations. Among other examples, this will be the case for high-resolution aerial mapping, for UAVs inspecting and monitoring tens of thousands of kilometers of pipelines or railway networks \cite{QuantumSystems}, and even for UAVs serving as aerial radio access nodes (i.e., BSs) and demanding high-speed wireless backhaul. For some of these applications, sufficiently low latency may only be achieved by skipping the video compression/decompression stages, for which hyper-high bit rates will be instrumental. Besides exchanging the copious captured data with ground stations in real time, these UAVs will have to be controlled with nearly unlimited flight range. In what follows, we provide a concrete industrial UAV application that is likely to require strong wireless cellular empowerment, and then detail how the 3GPP is targeting support for this and other upcoming use cases.


\subsection{3D Aerial Highways and the Delivery of Tomorrow}
\label{subsec:3Dhighways}

The retail industry---already undergoing a transformation from face-to-face stores to online shopping---is expected to go fully autonomous with upcoming 3D aerial highways. The survival and prosperity of online businesses and retailers, overwhelmed by an exponentially increasing number of online orders, relies on efficiently addressing the last-mile delivery bottleneck. To this end, the number of cargo-UAVs is expected to skyrocket, replacing delivery trucks and leading to less congested roads and more densely crowded skies \cite{CheJaaYan2020}.


Such vision is already embodied by projects of
e-commerce giants that aim at safely delivering packages of up to five pounds in under 30 minutes via small UAVs equipped with sense-and-avoid technology \cite{AmazonPrimeAir}. To this end, courier companies are to work with regulators to design an air traffic management system that will recognize who is flying, what, and where. Once provided with appropriate and enforced regulatory support, UAV carriers are expected to enhance current shipment services to millions of customers while increasing the overall safety and efficiency of the transportation system.

Ensuring safe operations of cargo UAVs entails continuously collecting a massive amount of data on speed and location, battery levels, and environmental conditions. Such information can be used to trigger control commands to prevent failures, aerial traffic congestion, and collisions. The corresponding communication key performance indicators include end-to-end latencies on the order of tens of milliseconds and a minimal tolerable lack of connectivity along the itinerary. Unlike present-day deployments, where UAV users are yet to become commonplace, 5G NR networks might have to meet the above requirements for a large number of connected UAVs.



\subsection{NR Use Cases and Requirements}
\label{subsec:NRusecases}

In response to these and other new industrial UAV applications, the traditionally ground-focused wireless cellular ecosystem is increasingly concerned with accommodating flying users. Table~\ref{tab:5G_use_cases} summarizes the most demanding UAV use cases identified by the 3GPP that could be enabled by 5G cellular support. These can be broadly classified into (\emph{i}) C2 links, and (\emph{ii}) payload data links. Table~\ref{tab:5G_use_cases} also quantifies the respective associated requirements in terms of bit rate, latency, reliability, UAV speed and altitude to be supported, as well as positioning accuracy.
In the following, we provide concrete examples of scenarios where a need for each of these link types might arise.

\subsubsection{Command and Control Links}
Owners of commercial UAVs will start using them to deliver mail-ordered goods to their customers' doorsteps. A controller can be employed to communicate with the UAV, where both the UAV and the controller are 5G-compliant devices. Once the owners receive clearance from the UAV traffic management (UTM) to fly their UAV, this comes along with a series of \emph{waypoints} that the UAV can follow while cruising from the warehouse to the recipient.
At times, a UAV owner might also be requested to take \emph{direct steering control} of the UAV. This might occur before climbing to cruise altitude, for dropping a package at destination, or to observe an accident that has occurred along the way. During direct steering control, the manoeuvring requires uplink (UL) video feedback from the UAV. The specifications for such feedback depend on whether the controller is in visual line-of-sight (VLoS), or beyond VLoS (BVLoS).
Autonomous flight is also possible via the UTM, which then provides predefined trajectories in the form of four-dimensional polygons while the UAV feeds back periodic position reports for tracking purposes \cite{AbdMar2020}.

\subsubsection{Payload Data Links}

UAVs are expected to enable users to experience virtual reality by means of panoramic \emph{8K video live broadcast}, e.g., through a $360^{\circ}$ spherical view camera on board the UAV that captures and uploads 8K video in real time to a cloud server. Such video stream will be received live by users through remote VR glasses. Besides high bit rates and low latency, this use case demands accurate positioning to avoid damage to life or property in densely populated regions. In building-intensive areas, where traditional satellite methods may lack accuracy, an excessive positioning offset might set off an anti-collision system, thus slowing down the UAV. As a countermeasure, UAVs are envisioned to employ \emph{4$\times$4K AI surveillance}, sending four-way 4K full-angle camera data to an AI controller that follows up by issuing timely control instructions. Other applications with similar requirements identified by the 3GPP are laser mapping, high-definition (HD) patrol, and remote UAV controller through HD video.


\subsubsection{UAVs as Radio Access Nodes}
In scenarios such as disaster monitoring, border surveillance, and emergency assistance, UAVs carrying a BS---denoted UxNB---can be quickly deployed.
The same might hold for coverage of hot-spot events, where broadband demand experiences a sudden surge. Although the exact specifications for this use case have not yet been defined by 3GPP, one can expect a combination of stringent C2 as well as payload data requirements. Indeed, these UxNBs will be autonomously controlled and dispatched as the network sees fit, and their required backhaul bit rates could be up to several Gbps depending on the traffic load that each UAV radio access node has to handle.


\subsection{NR Features in Support of UAVs}
\label{subsec:NRfeatures}

While not explicitly focused on cellular-connected UAVs, 3GPP Rel.~15 implements a number of features that go in the direction of satisfying the foregoing UAV connectivity requirements. Among these features, the most relevant
are:

\subsubsection{mMIMO and mmWave} The integration of a large number of antennas within sub-\SI{6}{GHz} BSs---a.k.a. mMIMO---and the wideband transmissions above \SI{24.25}{GHz}---a.k.a. mmWave---are widely deemed as key features of NR. Due to their relevance and depth, the significant impact of these features on cellular-connected UAVs is thoroughly treated in Sec.\ \ref{sec:massiveMIMO} and Sec.\ \ref{sec:mmWave}, respectively. 

\subsubsection{LTE/NR Dual Connectivity} 
LTE and NR are bound to coexist in the same geographical areas. Indeed, the first release of NR focused on non-standalone deployments, where at least one secondary NR BS complements the user plane of LTE through dual connectivity. In networks where both terrestrial and aerial devices coexist, LTE/NR dual connectivity provides the following benefits:
\begin{itemize}
    \item \emph{C2 traffic redundancy.} Thanks to information redundancy over LTE and NR links, dual-connected UAVs can enjoy more reliable communication than their single-connected counterparts \cite{8746579}. For instance, a more reliable sub-\SI{6}{GHz} NR link can be used to compensate for the errors that occur on the less reliable LTE interface. Conversely, an LTE link operating at lower frequencies could compensate for the link failures that typically occur when NR operates at mmWave frequencies, e.g., due to sudden blockages.
    \item \emph{UAV broadband UL traffic offloading.} Most LTE network resources serving GUEs are generally dedicated to the downlink (DL), given the higher amount of data traffic flowing in such direction. For example, typical LTE time-division duplexing (TDD) networks implement subframe configurations with a larger number of OFDM symbols for DL than for UL. A standalone LTE network would likely have to change this configuration to accommodate the large UL traffic requirements of UAVs. As a result, the incumbent LTE-connected GUEs would experience an immediate performance degradation, further exacerbated by the additional UAV-generated interference. This disruption could be alleviated when UAVs are dual-connected, since the UAV broadband UL transmissions could be offloaded to NR BSs.
    \item \emph{Aerial UL power control.} LTE/NR dual-connectivity is particularly appealing for UAVs because they are immune to one of the typical issues faced by GUEs: insufficient UL transmit power. In practice, this power limitation occurs because the LTE UL must always comply to power specified by its corresponding power control, with the NR UL only availing of the remaining power budget. This issue is unlikely to affect UAVs, which tend to use a much lower UL power than their GUE counterparts thanks to the generally satisfactory LoS propagation conditions they experience with their serving cell.
\end{itemize}

\subsubsection{Ultra-lean Design} One of the NR design criteria was to reduce the number of always-on transmissions, which generate interference and occupy precious time/frequency resources regardless of the network load. A prime example of this philosophy is the removal of the LTE cell-specific reference signals for channel estimation, transmitted by all LTE BSs in every DL subframe and unavoidably causing intercell interference. This simplified design is a stepping stone for the implementation of network-wide coordination techniques for efficient aerial communications. An illustrative example is the one where only a subset of NR BSs transmit towards UAVs in a dedicated part of the spectrum, while the remaining stay quiet; another instance is the cell-free architecture described in Sec. \ref{sec:cellfree}.

\begin{takeaway}
This decade is likely to see a proliferation of network-connected UAVs, whose stringent control and payload data requirements are to be satisfied through NR enhancements such as mMIMO, mmWave, and dual connectivity.
\end{takeaway}

\section{5G NR Massive MIMO for\\Enhanced UAV Cellular Support}
\label{sec:massiveMIMO}

In this section, we discuss how sub-\SI{6}{GHz} mMIMO, introduced to benefit the entire NR user population, can be particularly instrumental to tackle the otherwise severe initial access, cell selection, and interference challenges experienced by UAVs. 

\subsection{Tackling the Initial Access and Cell Selection Challenges} \label{subsec:massivMIMO1}

\begin{figure}
\centering
\includegraphics[width=\figwidth]{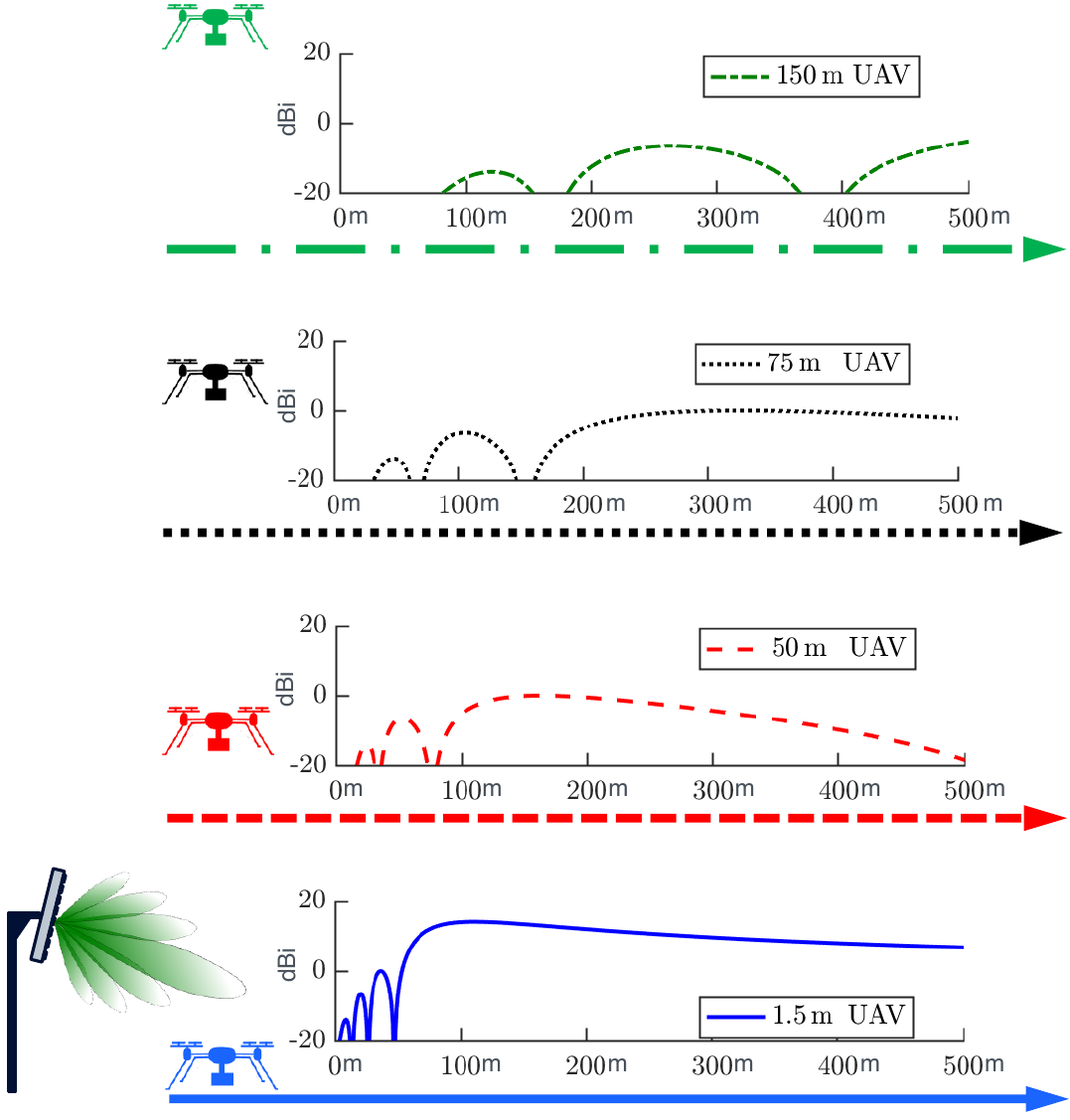}
\caption{Antenna gain [dBi] vs. 2D BS-to-UAV distance [m] for a BS deployed at \SI{25}{m} and UAVs of various heights, aligned to the BS’s horizontal bearing.}
\label{fig:mMIMO_Fig1}
\end{figure}

Prior to the reception and transmission of payload data, cellular-connected UAVs must access the network. For this purpose, cellular BSs regularly transmit synchronization signal blocks (SSBs) that facilitate their discovery. Up to LTE Advanced Pro networks, the radiation of SSB signals was determined exclusively by the BS antenna pattern \cite{azari2018reshaping, van2016lte, GerGarGal2018}. For the typical cellular network with downtilted BSs, this means that devices flying higher than BSs could only perceive SSB signals through antenna sidelobes.
This phenomenon is illustrated in Fig.~\ref{fig:mMIMO_Fig1}, which represents the antenna gain perceived by UAVs flying at (from top to bottom) $\{$\SI{150}{}, \SI{75}{}, \SI{50}{}, \SI{1.5}{}$\}$\,\SI{}{m} and moving away from a BS with a height of \SI{25}{m}. The BS is equipped with an $8\times 1$ vertical array downtilted by $12^{\circ}$ \cite{3GPP38901}. Fig.~\ref{fig:mMIMO_Fig1} shows how, while cellular devices near the ground generally experience reasonable and smooth antenna gain variations as they move away from the BS, this is not the case for UAVs, whose antenna gains are dramatically reduced and become highly irregular as the altitude increases and only BS antenna sidelobes can be perceived.

\begin{figure}[!t]
\centering
\includegraphics[width=\figwidth, trim = 0.8cm 0cm 1.6cm 0, clip]{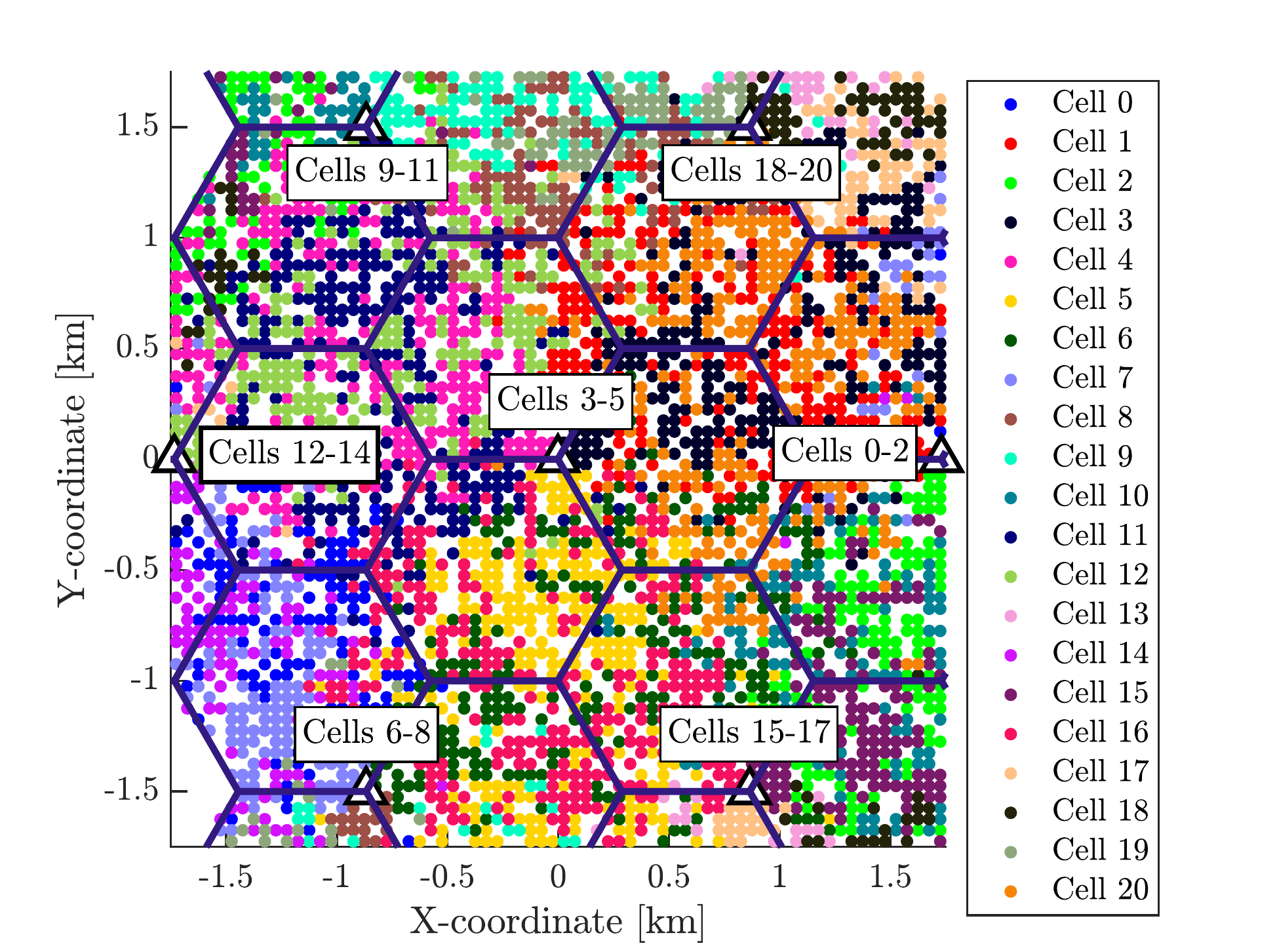}
\caption{Serving cell selection map based on RSRP for a UAV flying at \SI{50}{m}. White spots correspond to grid points where no UAVs were present.}
\label{fig:mMIMO_Fig2}
\end{figure}

Not only does the foregoing behavior pose a great challenge when UAVs attempt to initially access the network, but also when they fly at reasonable speeds and need to (re-)select their serving cell. To illustrate this, Fig.~\ref{fig:mMIMO_Fig2} represents a typical Urban Macro (UMa) network comprised of three-sector\footnote{In cellular networks, a deployment site is traditionally divided into three sectors, each covered by a different BS and considered as a different cell.} BSs with an intersite distance (ISD) of \SI{1.5}{km} and indicates, in grids of \SI{50}{m}\,$\times$\,\SI{50}{m}, the best serving cell, i.e., the one providing the strongest reference signal received power (RSRP), for UAVs flying at an altitude of \SI{50}{m} \cite{CheLinKha2020}. The map in Fig.~\ref{fig:mMIMO_Fig2} demonstrates that UAVs do not tend to associate to their physically closest cells, which is typically the case for GUEs.
Moreover, the figure shows that UAVs will execute a large number of handovers due to the changes in the best-serving cell they experience when moving from a given grid point to an adjacent one. 


\begin{figure}[!t]
\centering
\includegraphics[width=\figwidth]{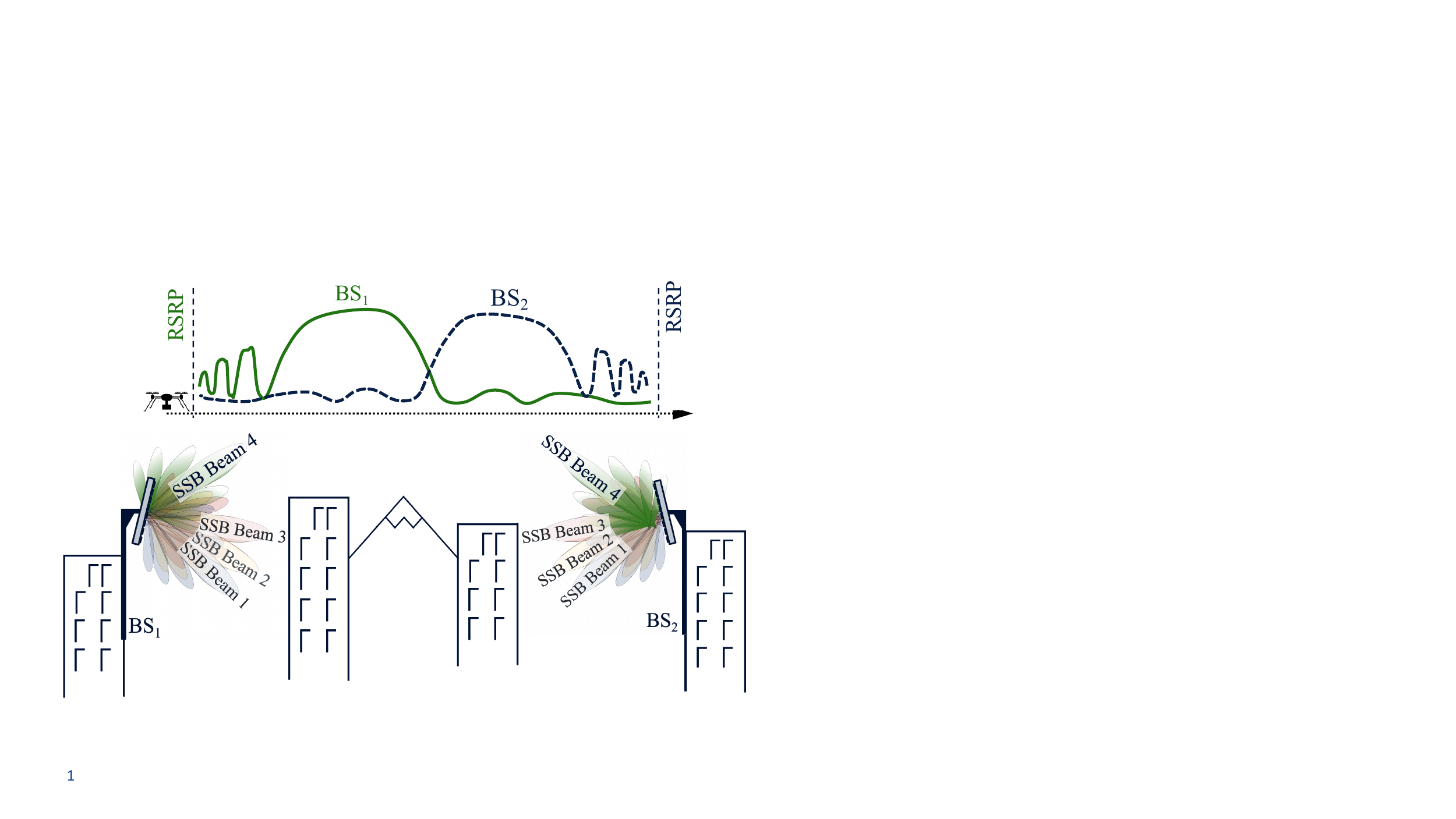}
\caption{Illustration of a SSB beam design optimized for serving UAVs.}
\label{fig:mMIMO_SSBOpt}
\end{figure}

In contrast to LTE Advanced Pro and preceding networks, the SSB signals originated by NR BSs can be transmitted towards multiple directions through time-multiplexed beam sweeping, denoted as an SS burst set \cite{dahlman20205g}. This capability is illustrated in Fig.~\ref{fig:mMIMO_SSBOpt}, where it can be observed that BSs operating between \SI{3}{GHz} and \SI{6}{GHz} can transmit up to four different SSB beams. 

The possibility of focusing beamformed SSBs towards the sky 
opens up a new way of mitigating the sidelobe-related issues previously described. Indeed, mobile network operators aiming at reliably serving UAVs may opt to perform a network-wide design of the beamformed SSBs to uniformly cover the sky, similarly to what is done for the ground. This concept is also exemplified in Fig. \ref{fig:mMIMO_SSBOpt}, where the UAV flying above the BSs now perceives a smooth RSRP transition that facilitates the initial cell selection and subsequent handovers.

\begin{takeaway}
The beamformed SSBs introduced in NR can greatly facilitate initial cell selection and subsequent handovers for UAV users, overcoming the issues associated with sidelobe-based UAV-BS association.
\end{takeaway}

\subsection{Tackling the Interference Challenge} \label{subsec:massivMIMO2}

\begin{figure}
\centering
\includegraphics[width=\figwidth, trim = 0.8cm 0 1.5cm 0, clip]{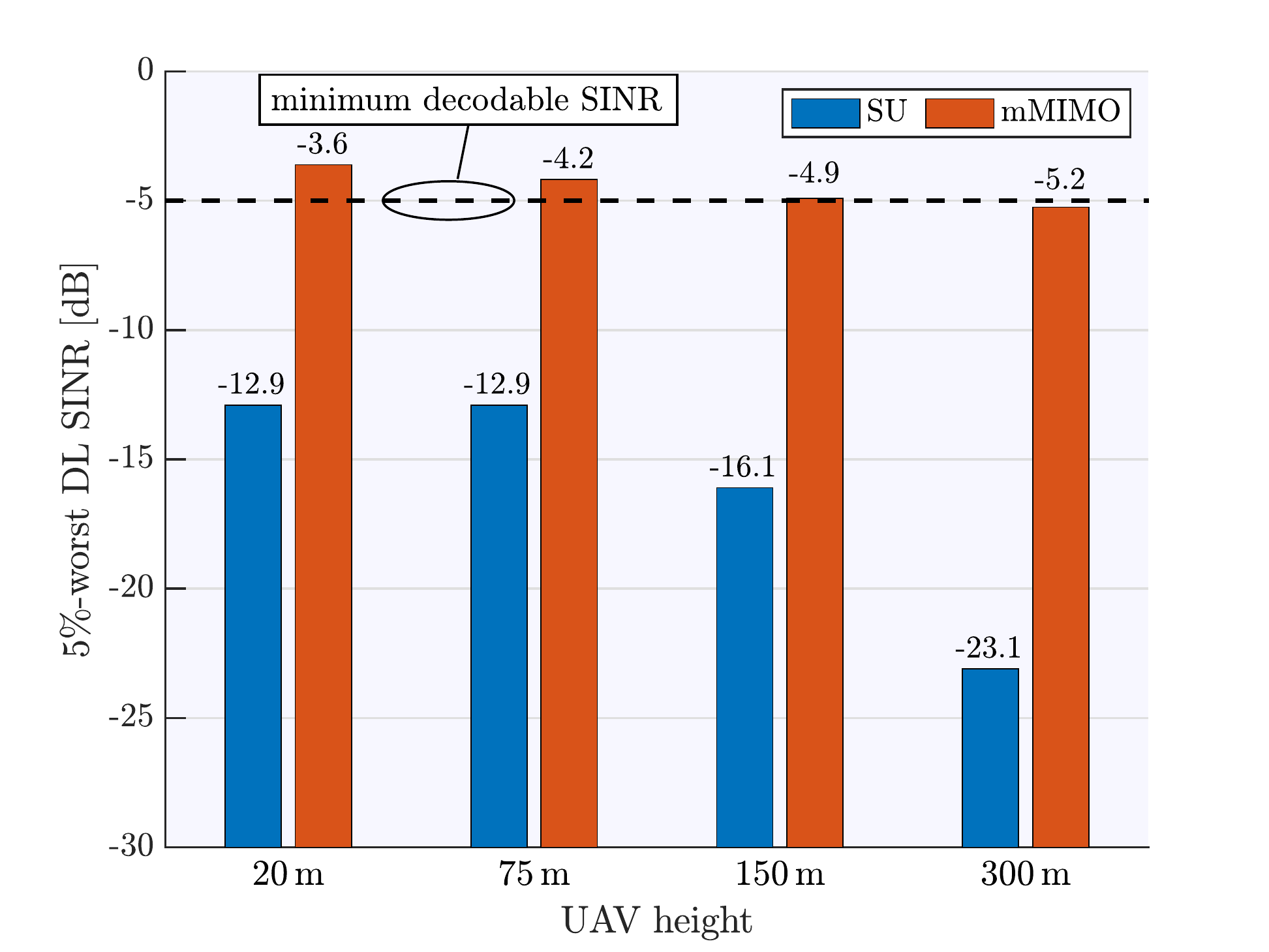}
\caption{5\%-worst DL SINRs per time/frequency block experienced by UAVs flying at different heights in single-user and mMIMO networks with perfect CSI.}
\label{fig:mMIMO_Fig3}
\end{figure}

Once UAVs connect to the network, the latter needs to guarantee a pervasive and sufficient signal quality to enable reliable C2 and data transmissions. The NR mMIMO capabilities are critical to achieve this objective \cite{MuzRafFak2020}. This observation is supported by the results shown in Fig.~\ref{fig:mMIMO_Fig3}, where the 5\%-worst DL signal-to-interference-plus-noise ratios (SINRs) per time/frequency physical resource block (PRB) of 7 OFDM symbols and 12 subcarriers experienced by single-antenna UAVs flying at different heights are shown for two typical fully loaded Urban Macro networks with an ISD of \SI{500}{m} \cite{GerGarGal2018}:
\begin{itemize}
    \item A single-user (SU) network comprised of BSs downtilted by $12^{\circ}$ and equipped with one RF antenna port connected to 8 vertically stacked cross-polarized antenna elements.
    \item A mMIMO network with BSs also downtilted by $12^{\circ}$, but equipped with 128 RF chains connected to an 8$\times$8 planar array of cross-polarized antenna elements. These BSs are capable both of focusing their transmissions (i.e., beamforming) and of spatially multiplexing up to eight devices\footnote{Multiplexing a larger number of devices might yield a higher cell spectral efficiency. However, it might prevent achieving a minimum guaranteed rate for all devices, which is the primary goal for supporting the UAV C2 link.} through---for the time being---zero-forcing precoding with perfect channel state information (CSI). 
\end{itemize}
In both networks, 14 GUEs and one UAV are deployed per cellular sector, as per 3GPP Case 3 \cite{3GPP36777}. The results of Fig.~\ref{fig:mMIMO_Fig3} demonstrate how NR mMIMO capabilities are essential to reliably serve the worst-performing UAVs in loaded cellular networks. This is mostly because UAVs receiving DL transmissions experience LoS propagation conditions with an increasing number of interfering BSs as their altitude increases \cite{AzaRosPol2019}. Cellular BSs with NR mMIMO capabilities can simultaneously increase the useful signal power (thanks to the boost provided by their beamforming gains) and reduce the intercell interference (thanks to the higher directionality of their transmissions).

\begin{figure}
\centering
\includegraphics[width=\figwidth, trim = 0.8cm 0 1.5cm 0, clip]{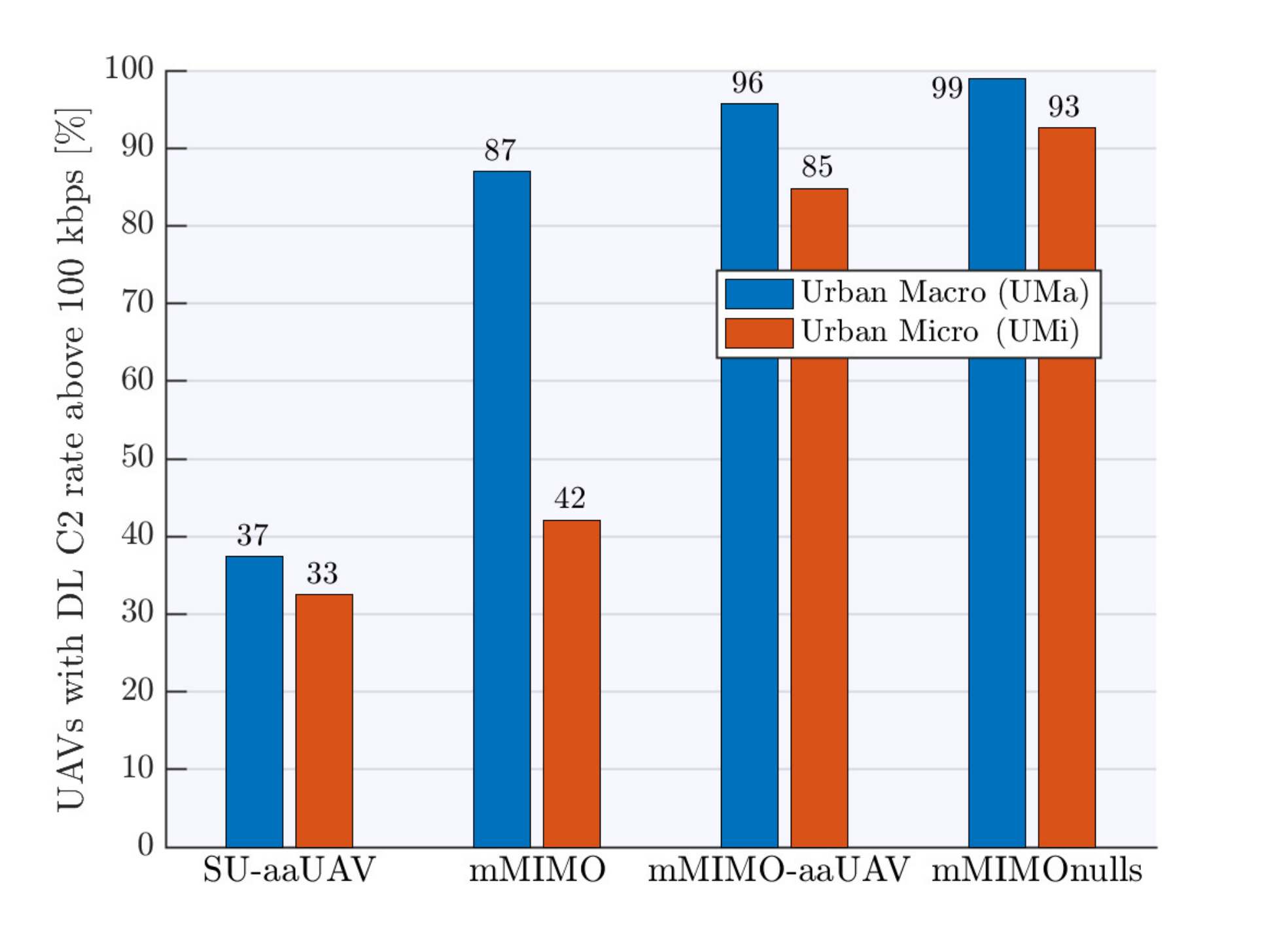}
\caption{Percentage of UAVs with DL C2 channel rates larger than \SI{100}{ kbps} when flying at \SI{150}{m} above UMa and UMi cells.}
\label{fig:mMIMO_Fig5}
\end{figure}

The CSI needed by mMIMO BSs to beamform and spatially separate UEs in TDD networks is acquired by observing UL sounding reference signals (SRSs) and leveraging UL-DL channel reciprocity \cite{dahlman20205g}. In practice, these estimates are imperfect since the finite number of SRSs have to be reused across multiple cells and they interfere with each other, a phenomenon commonly known as pilot contamination \cite{6375940,GalCamLop2018}. Fig.~\ref{fig:mMIMO_Fig5} considers a realistic CSI acquisition situation where eight SRSs are reutilized every three sectors and illustrates the percentage of UAVs flying at \SI{150}{m} that achieve DL C2 rates higher than \SI{100}{kbps} (the minimum requirement originally defined by 3GPP \cite{3GPP36777}). The figure considers both UMa and Urban Micro (UMi) deployments. In the latter, BSs are deployed at a height of \SI{10}{m} with an ISD of \SI{200}{m}. The performance is shown for UAVs and networks with the following enhanced capabilities \cite{GarGerLop2019}:
\begin{itemize}
    \item \emph{aaUAV.} UAVs with 2$\times$2 antenna arrays and a single RF chain capable of beamforming their UL/DL signals towards/from their serving BSs.
    \item \emph{mMIMOnulls.} mMIMO BSs capable of placing up to 16 radiation nulls towards the most interfering UAVs both during the UL SRS stage---therefore partially suppressing the harmful UAV-generated interference---and the data transmission stage---therefore mitigating the harmful intercell interference \cite{YanGerQue17}. mMIMO BSs can perform such null steering relying only on statistical CSI, i.e., by leveraging the strong directionality of the LoS channels towards the most harmful UAVs (in the UL) or the most interfered ones (in the DL) \cite{GarGerLop2019}.
\end{itemize}
The following key observations can be made based on the results of Fig.~\ref{fig:mMIMO_Fig5}:
\begin{itemize}
    \item While equipping UAVs with multiple antennas does not suffice to guarantee a satisfactory DL performance in SU networks---only 33\% of the UAVs in UMi SU networks experience rates larger than \SI{100}{kbps}---, substantial benefits can be found in networks with NR mMIMO BSs, where the number of UAVs reaching \SI{100}{kbps} in UMi networks doubles from 42\% to 85\%.
    \item The best-performing solution is the one where mMIMO BSs implement null steering capabilities, with almost all UAVs reaching the desired \SI{100}{kbps} in fully loaded UMa networks. This solution seems particularly attractive for mobile operators, since it does not rely on the features implemented by UAV manufacturers. 
    \item For all four systems considered, the performance of UAVs served in UMi networks is worse than that attained by UAVs flying through UMa networks. This can be explained as follows: the higher the network density, the more BSs will generate/receive interference to/from the LoS UAVs served by neighboring BSs, resulting in performance degradation \cite{8647634}.
\end{itemize}
For completeness, the interested reader can find in Table~III the DL C2 performance for UAVs flying over UMa deployments at other heights.

\begin{table}[t!]
    \vspace{1mm}
    \caption{Percentage of UAVs with DL C2 channel rates larger than \SI{100}{kbps} as a function of their height, for an UMa deployment.}
    \label{tab:mMIMO_Tab1}
    \centering
    \begin{tabular}{l|c|c|c|c|}
    \cline{2-5}
    {}
    & \cellcolor{BackgroundLightBlue} \shortstack{SU-aaUAV} 
    & \cellcolor{BackgroundLightBlue} \shortstack{mMIMO} 
    & \cellcolor{BackgroundLightBlue} \shortstack{\!\!\!mMIMO-aaUAV\!\!\!} 
    & \cellcolor{BackgroundLightBlue} \shortstack{\!\!mMIMOnulls\!\!} \\ \hline
    \rowcolor{BackgroundGray}
    \multicolumn{1}{|c|}{\SI{15}{m}}
    & 87\% & 97\% & 97\% & 99\% \\ \hline
    \rowcolor{BackgroundLightBlue}
    \multicolumn{1}{|c|}{\SI{75}{m}}
    & 79\% & 96\% & 98\% & 99\% \\ \hline
    \rowcolor{BackgroundGray}
    \multicolumn{1}{|c|}{\SI{150}{m}}
    & 37\% & 87\% & 96\% & 99\% \\ \hline
    \rowcolor{BackgroundLightBlue}
    \multicolumn{1}{|c|}{\SI{300}{m}}
    & 3\% & 33\% & 83\% & 98\% \\ \hline
    \end{tabular}
\noindent
\end{table}

\begin{figure}
\centering
\includegraphics[width=\figwidth, trim = 0.8cm 0 1.5cm 0, clip]{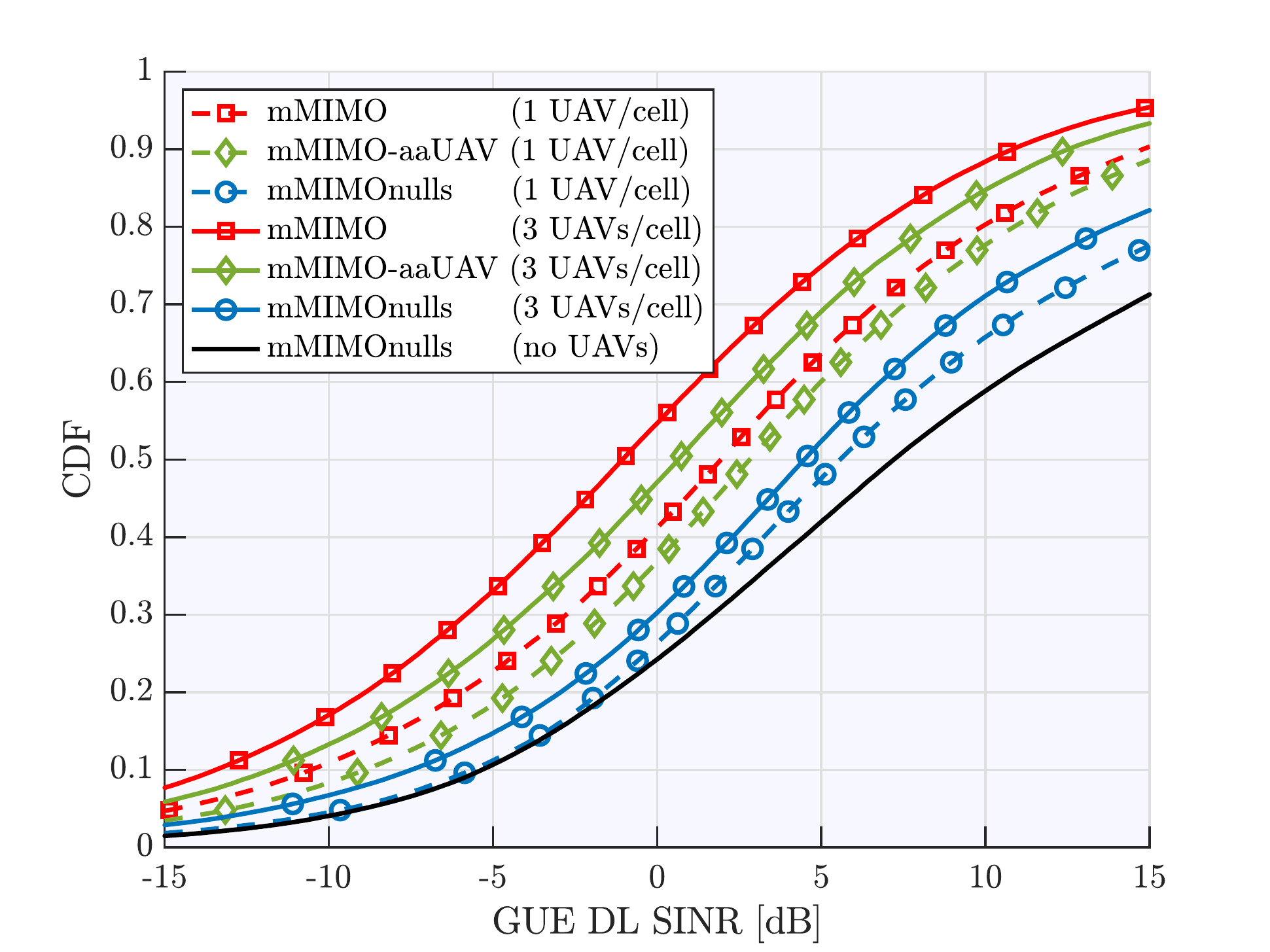}
\caption{CDF of the GUE DL SINR per time/frequency block under the presence of 1 and 3 UAVs per UMa cell flying at \SI{150}{m}.}
\label{fig:mMIMO_Fig6}
\end{figure}

\begin{takeaway}
At sub-\SI{6}{GHz}, the higher the network density, the more BSs that generate/receive interference to/from neighboring LoS UAVs. Enhancing NR mMIMO with UAV-aware intercell interference suppression capabilities can nearly double the UL rates of coexisting aerial and ground devices. Such approach is more effective than beamforming at the UAV side.
\end{takeaway}

While Figs.~\ref{fig:mMIMO_Fig1}--\ref{fig:mMIMO_Fig5} focus on the performance of the cellular-connected UAVs, one should not forget that these networks will have to satisfactorily continue to serve GUEs. Fig.~\ref{fig:mMIMO_Fig6} presents the cumulative distribution function (CDF) of the DL SINRs experienced by GUEs in the presence of 1 and 3 UAVs per cell at \SI{150}{m}. 
The GUE DL performance is seen to degrade as more UAVs are served. Somewhat surprisingly, the main reason for this can be found in the UL SRS transmission that precedes the mMIMO DL data transmission; the SRSs of high-altitude UAVs generate strong interference towards a large number of BSs that are simultaneously receiving SRSs from GUEs. In simple words, the larger the number of UAVs, the worse the DL channel estimates and the resultant DL beamforming and spatial multiplexing accuracy of the mMIMO BSs. Regardless of this phenomenon, equipping mMIMO BSs with null steering capabilities remains the best-performing solution.


\begin{figure}
\centering
\includegraphics[width=\figwidth, trim = 1cm 0 1.5cm 0, clip]{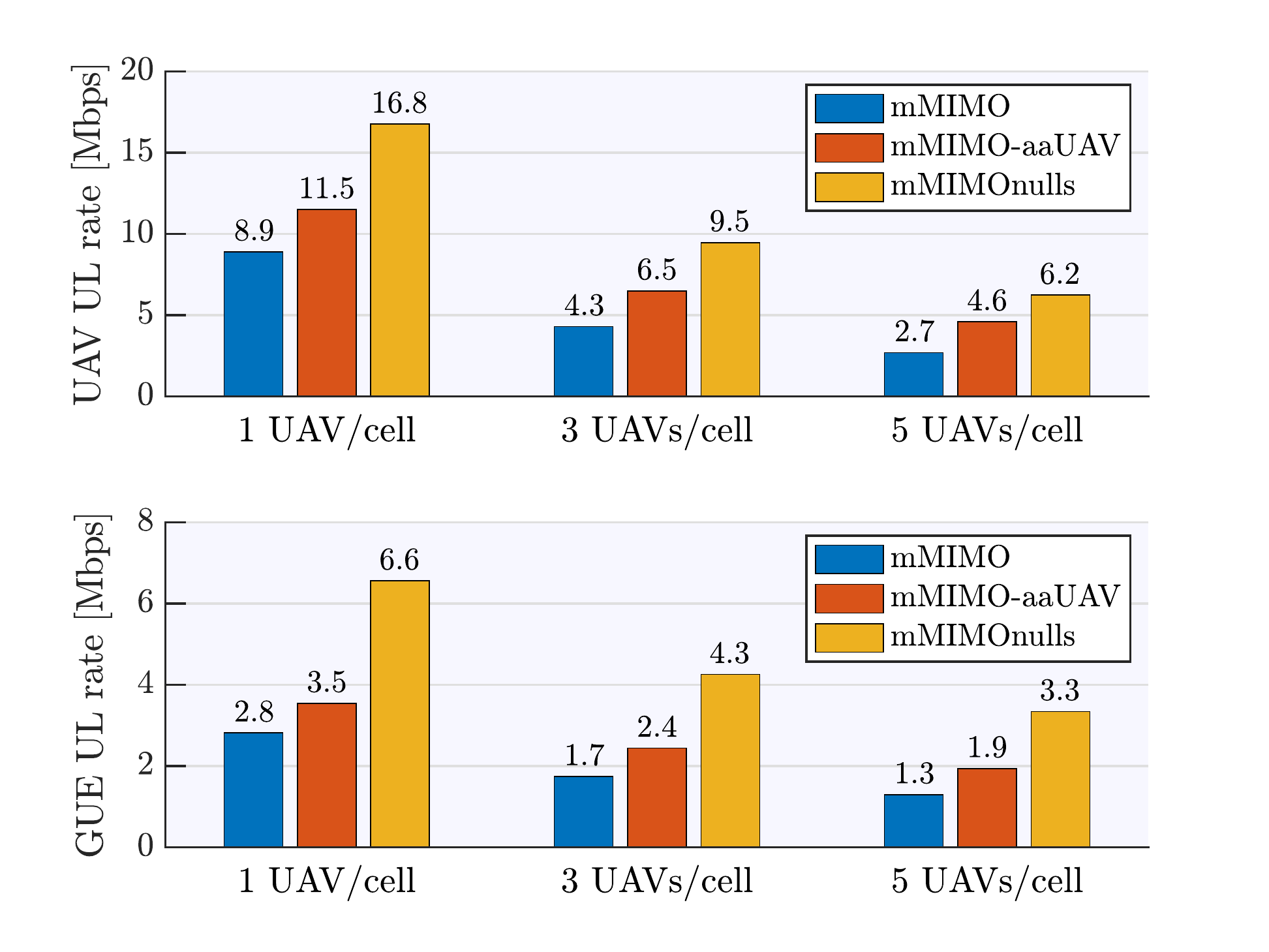}
\caption{Mean UL rates for UAVs (top) and GUE (bottom) for UMa cellular networks with 1, 3, or 5 UAVs per cell.}
\label{fig:mMIMO_Fig7}
\end{figure}

Finally, Fig.~\ref{fig:mMIMO_Fig7} looks directly into the UL performance of both UAVs and GUEs by showing the mean UL bit rates for UMa networks with 1, 3, and 5 UAVs per cell at altitudes uniformly distributed between \SI{1.5}{m} and \SI{300}{m}. The results of Fig.~\ref{fig:mMIMO_Fig7} exhibit similar trends to those of Fig.~\ref{fig:mMIMO_Fig6}, highlighting again the importance of enhancing baseline NR mMIMO networks with UAV-aware intercell interference suppression capabilities to substantially improve---nearly doubling in most cases---the UL rates of the coexisting aerial and GUEs.

\begin{takeaway}
When employing mMIMO, the performance of GUEs is affected by the presence of UAVs both in UL (due to interference) and in DL (due to the SRS pilot contamination incurred in the UL).
\end{takeaway}
\section{5G NR mmWave for UAV Capacity Boost} 
\label{sec:mmWave}

A relentless trend in the evolution of wireless communications is the hunger for bandwidth, and fresh bandwidth is only to be found at ever higher frequencies. The mmWave spectrum represents the current opportunity, and a formidable one at that, to unlock large swaths of bandwidth and enable the most ambitious 5G aerial use cases listed in Table~\ref{tab:5G_use_cases}. Moreover, in order to cope with the prohibitive isotropic propagation loss at such short wavelengths, mmWave links are highly directional and hence less prone to interference, which reduces the myriad of challenges detailed in Sec.\ \ref{subsec:massivMIMO2} and facilitates the coexistence of terrestrial and aerial devices. In this section, we quantify the achievable performance of mmWave-connected UAVs in both urban and suburban/rural scenarios. We focus on the UL, since this is the power-limited direction and the one typically demanding the highest capacity. 

\subsection{Urban mmWave Coverage} \label{subsec:mmWaveUrban}

Aerial coverage depends in an intricate manner on the angular and power distributions of signal paths along with the antenna patterns at the UAV and BS. To capture these distributions accurately in an urban environment, and given the lack of a calibrated mmWave aerial channel model, we conducted ray tracing simulations by means of the Wireless Insite tool by Remcom \cite{Remcom, KanMezLoz2021}.\footnote{As discussed in Section~\ref{sec:AI}, machine learning techniques can be employed to recreate such model, avoiding the computational burden of ray tracing simulations.}

\begin{figure}[!t]
\centering
\includegraphics[width=\figwidth]{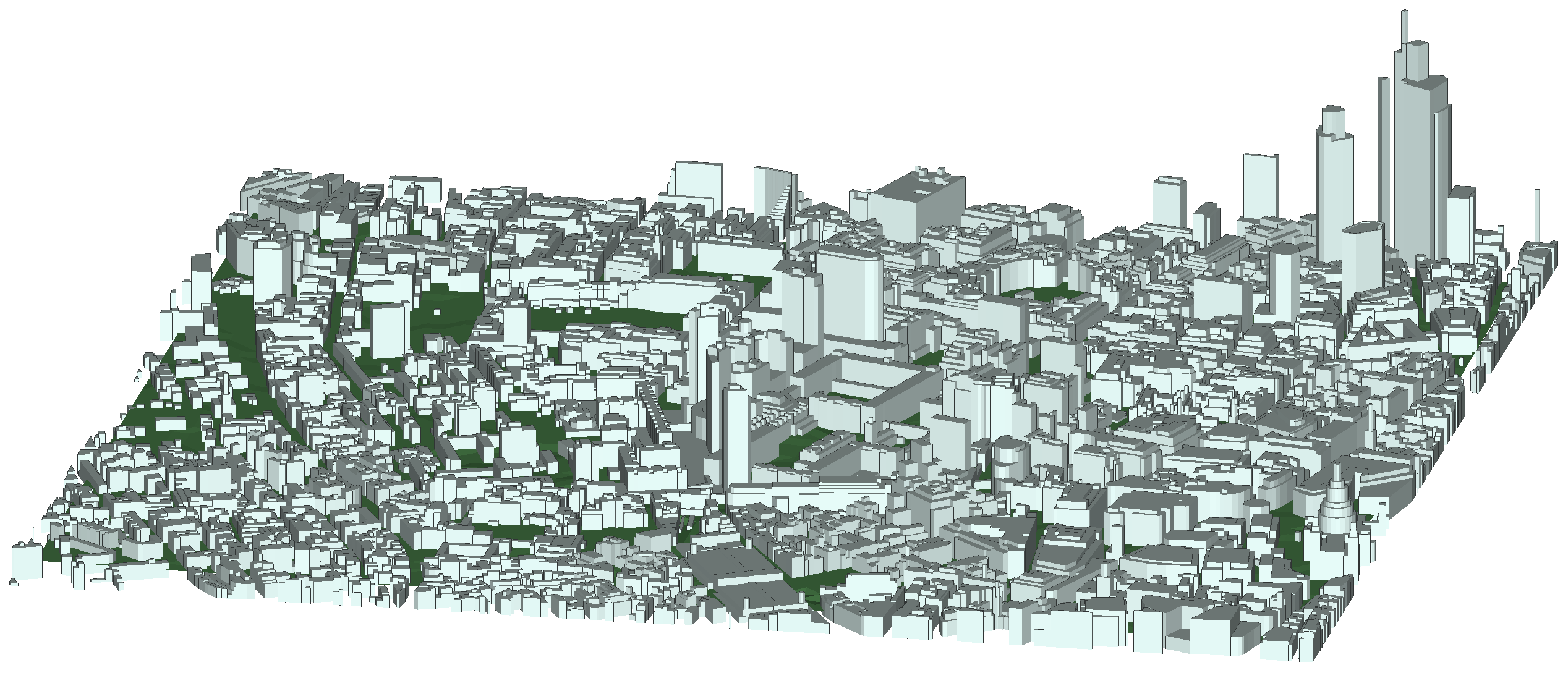}
\caption{3D map of a section of London employed, together with Remcom Wireless InSite ray tracer, to generate the propagation paths of $25800$ mmWave air-to-ground links.}
\label{fig:mmWave_london}
\end{figure}

\begin{figure}[!t]
\centering
\includegraphics[width=\figwidth]{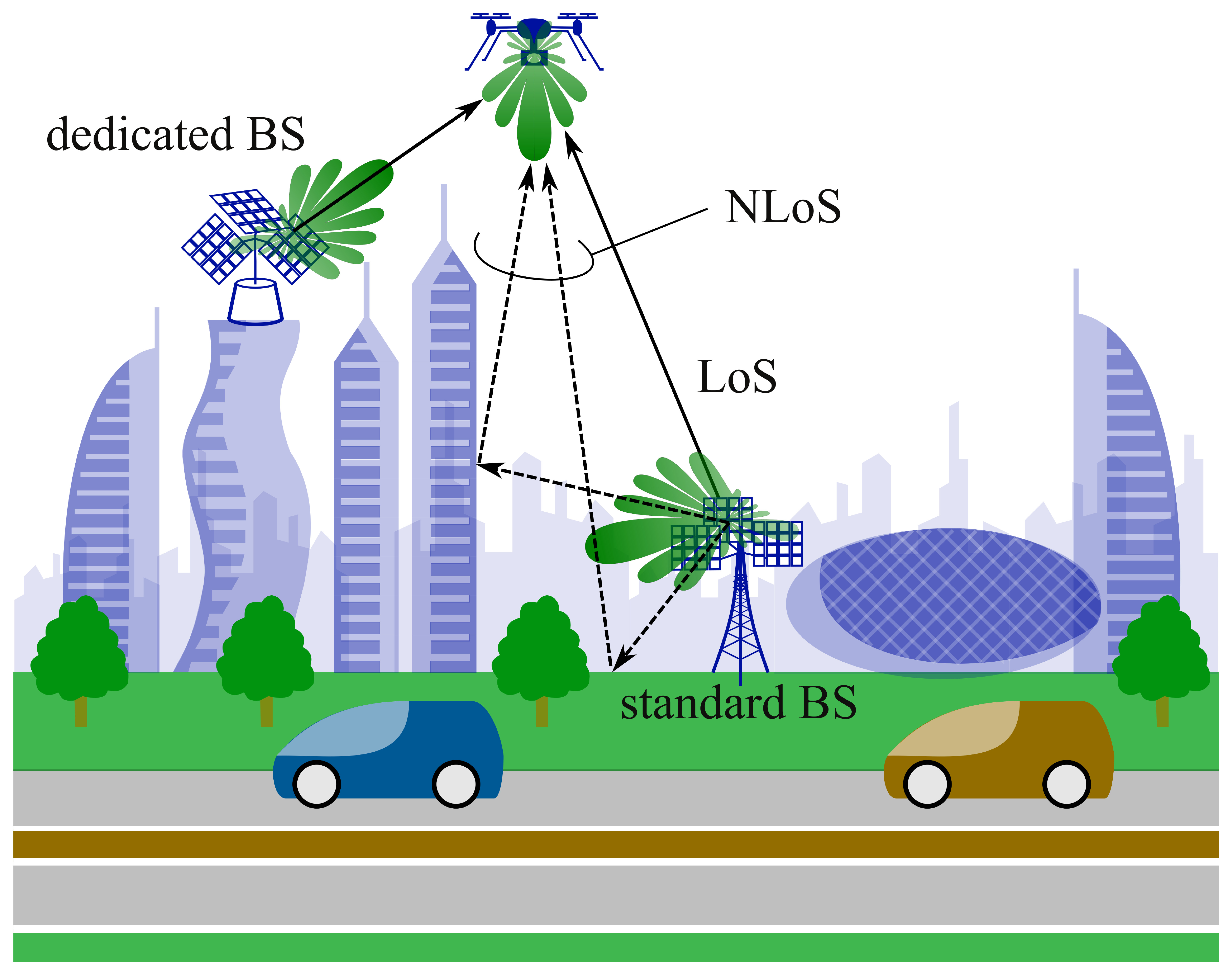}
\caption{Urban deployment featuring standard and dedicated BSs.}
\label{fig:mmWave_deployment}
\end{figure}

In particular, our study spanned a \SI{1}{km} $\times$ \SI{1}{km} 3D section of London, shown in Fig.~\ref{fig:mmWave_london}. As illustrated in Fig.~\ref{fig:mmWave_deployment}, such area features two complementary networks operating at $28$~$\mathrm{GHz}$ with a bandwidth of $400$~$\mathrm{MHz}$. The first network was comprised of street-level \emph{standard} NR mmWave BSs at heights between $2$ and $5$~$\mathrm{m}$, deployed uniformly at random with average ISD of $\ISDs$, each spawning three sectors with a $12^{\circ}$ downtilt. A second network of \emph{dedicated} mmWave BSs was deployed only on rooftops, also uniformly at random but with average ISD of $\ISDd$, and with a $45^{\circ}$ uptilt angle.  
The transmitting UAVs were distributed uniformly at specific heights, with their antennas pointed downwards. Each sector features an $8 \times 8$ uniform rectangular array (URA) while each UAV is equipped with a single $4 \times 4$ URA. In both cases, the antenna element radiation pattern is set according to \cite{3GPP38901}. The Python-based system-level simulator employed is freely available \cite{sj-github}.

\subsubsection{Coverage with Standard NR mmWave BSs}

\begin{figure}
\centering
\includegraphics[width=\figwidth, trim = 0.8cm 0 1.5cm 0, clip]{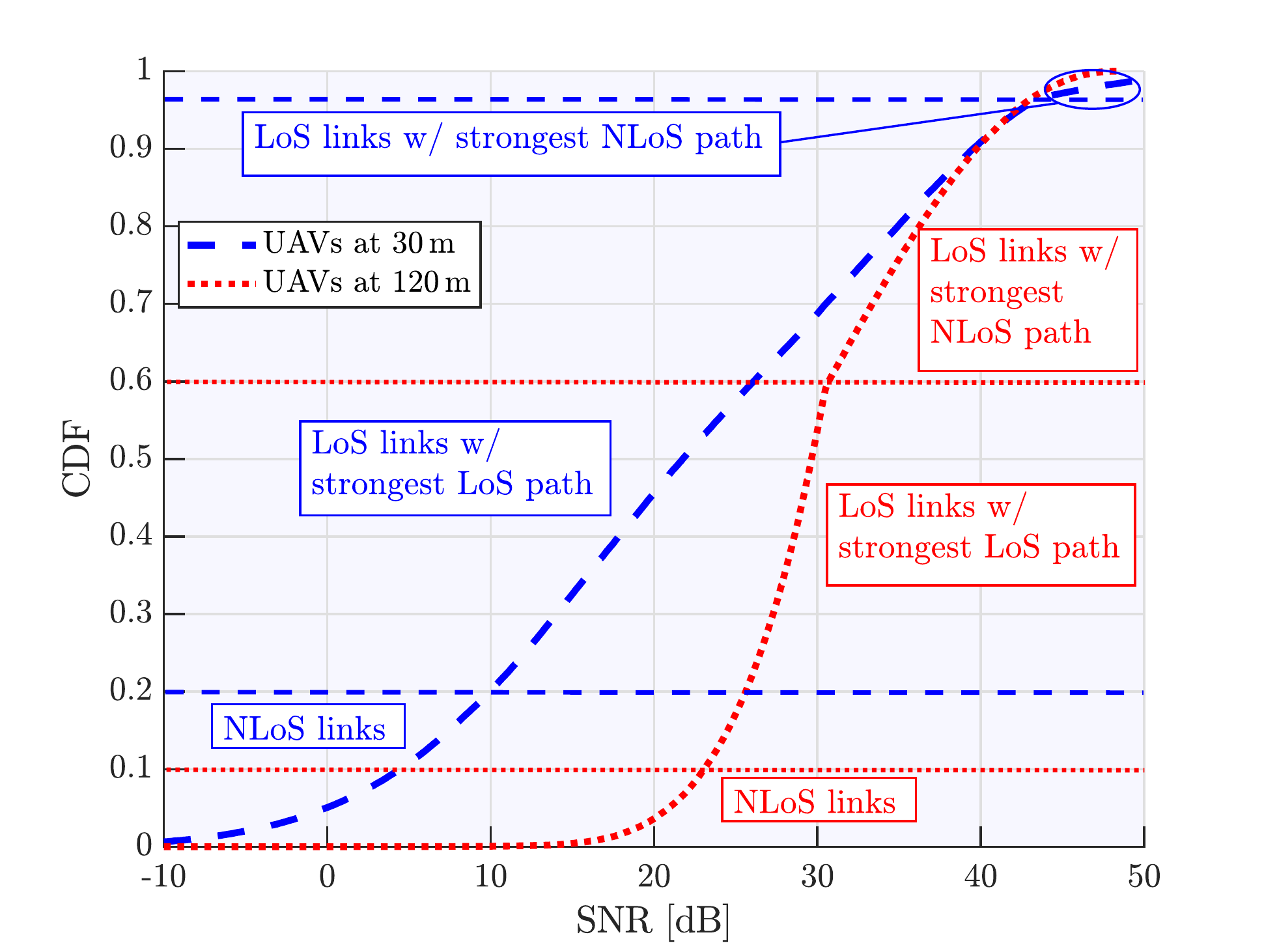}
\caption{SNR distribution for UAVs served by standard mmWave cells in an urban environment. Horizontal lines show the breakdown into three regions: NLoS-dominated (bottom), LoS-dominated (middle), and LoS-dominated with strong NLoS paths (top).}
\label{fig:mmWave_urban_1}
\end{figure}

Let us begin by considering a deployment reliant exclusively on standard BSs, to assess the extent to which such a deployment can provide satisfactory aerial coverage. With the signal-to-noise ratio (SNR) taken as a proxy for coverage, Fig.~\ref{fig:mmWave_urban_1} presents the distribution of the SNR with $\ISDs=$ \SI{200}{m}, and UAV altitudes of 30 and \SI{120}{m}. At least $98\%$ of UAVs at \SI{30}{m} have an SNR$>$ \SI{-5}{dB}, testifying to an acceptable coverage at low altitudes. Moreover, the coverage improves with altitude. By \SI{120}{m}, almost all UAVs achieve in excess of \SI{15}{dB}. This result may seem surprising given the downtilted directional nature of the antennas at the BSs, with a front-to-back ratio of \SI{30}{dB} \cite{3GPP38901}.

To understand how UAVs still enjoy satisfactory coverage, Fig.~\ref{fig:mmWave_urban_1} identifies three SNR regions with distinct dominant link behaviors. 
\begin{itemize}
\item The lower tail corresponds predominantly to non-LoS (NLoS) links.
\item The middle section is dominated by LoS links whose LoS component is in turn dominant.
\item The upper tail is mostly constituted by LoS links, but with a NLoS link being the strongest
link after accounting for the antenna gains.
\end{itemize}
Thus, in the lower tail, which is what determines the coverage, the connectivity rests mostly on NLoS communication, meaning through a conjunction of BS antenna sidelobes and reflected paths. Both these ingredients turn out to play a substantial role, revealing that the BS sometimes leverages NLoS paths that happen to be stronger than the LoS, especially when the UAV is horizontally close as illustrated in Fig.~\ref{fig:mmWave_deployment}.

As the UAV altitude increases, the LoS probability grows, but NLoS paths continue to play an important SNR-enhancing role. Altogether, a standard 5G mmWave deployment suffices for UAV coverage when $\ISDs=\SI{200}{m}$. Then, as $\ISDs$ extends beyond this value, coverage decays for low-altitude UAVs, and a progressively higher minimum altitude is required for service. Although not shown for brevity, our study reveals that for $\ISDs=\SI{400}{m}$, for instance, roughly $10\%$ of UAVs at \SI{30}{m} turn out to experience negative-dB SNRs, indicating that coverage is no longer guaranteed at this altitude.

\begin{takeaway}
At typical UMi densities, NR mmWave networks can provide satisfactory coverage to aerial links---in spite of employing highly directional and downtilted antennas---thanks to a favorable combination of antenna sidelobes and strong reflections.
\end{takeaway}

\subsubsection{Coverage Enhancement with Dedicated mmWave BSs}

\begin{figure}
\centering
\includegraphics[width=\figwidth, trim = 0.8cm 0 1.5cm 0, clip]{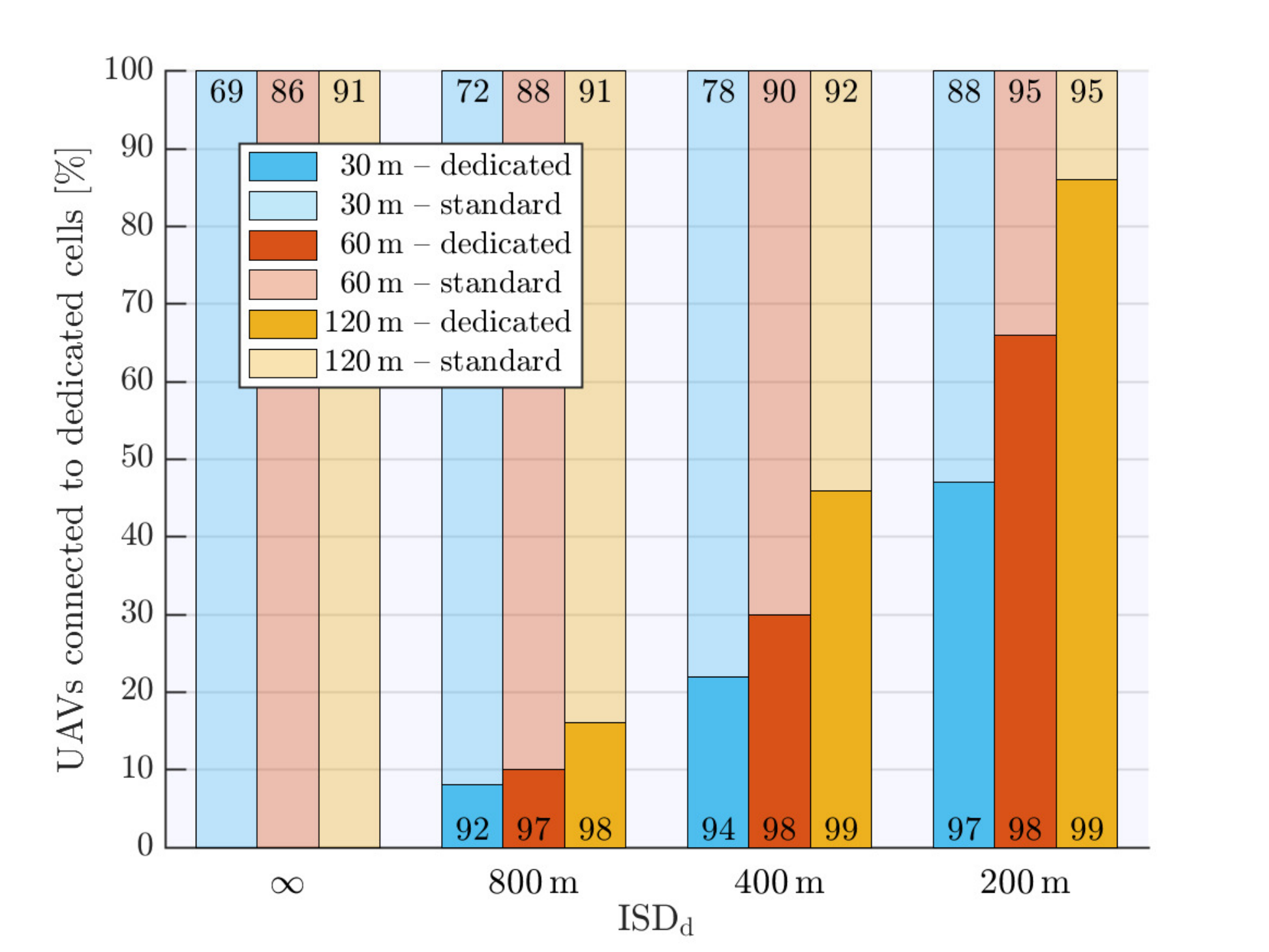}
\caption{Fraction of UAVs connected to standard (clear bars) and dedicated (solid bars) mmWave cells in an urban environment. Three UAV heights are considered and, for each case, the percentage of LoS links is indicated by the numbers at the top (for standard BSs) and at the bottom (for dedicated BSs). $\ISDs=$\SI{200}{m} is fixed for standard cells, whereas $\ISDd$ varies to capture scenarios with different dedicated cells densities.} 
\label{fig:mmWave_urban_2}
\end{figure}

Having characterized the coverage for a standard deployment, let us now quantify the impact of incorporating dedicated BSs with $\ISDd \geq \ISDs$. Fig.~\ref{fig:mmWave_urban_2} provides the following information:
\begin{itemize}
    \item The fraction of UAVs that choose to connect to dedicated BSs (solid bars) as the density of the latter increases, i.e., as $\ISDd$ decreases, rather than connect to standard BSs with $\ISDs= $\SI{200}{m} (clear bars). This fraction is presented for various UAV altitudes, indicating that it is the UAVs at higher altitudes that exhibit increased preference for the dedicated BSs. This is toned down at lower altitudes, but even then dedicated BSs are generally favored when their density equals that of the standard ones. As one would expect, the relevance of dedicated BSs dwindles as they become sparser.
    \item The percentage of links that are in LoS (numbers at the top and at the bottom for standard and dedicated BSs, respectively), providing another perspective on the effects of deploying dedicated BSs. The addition of dedicated BSs drastically brings such percentage close to 100\% at all the considered UAV altitudes and, with $\ISDd=\ISDs=$\SI{200}{m}, a vast majority of UAVs enjoy LoS connectivity to their serving BSs.
\end{itemize}

\begin{figure}
\centering
\includegraphics[width=\figwidth, trim = 1cm 0 1.2cm 0, clip]{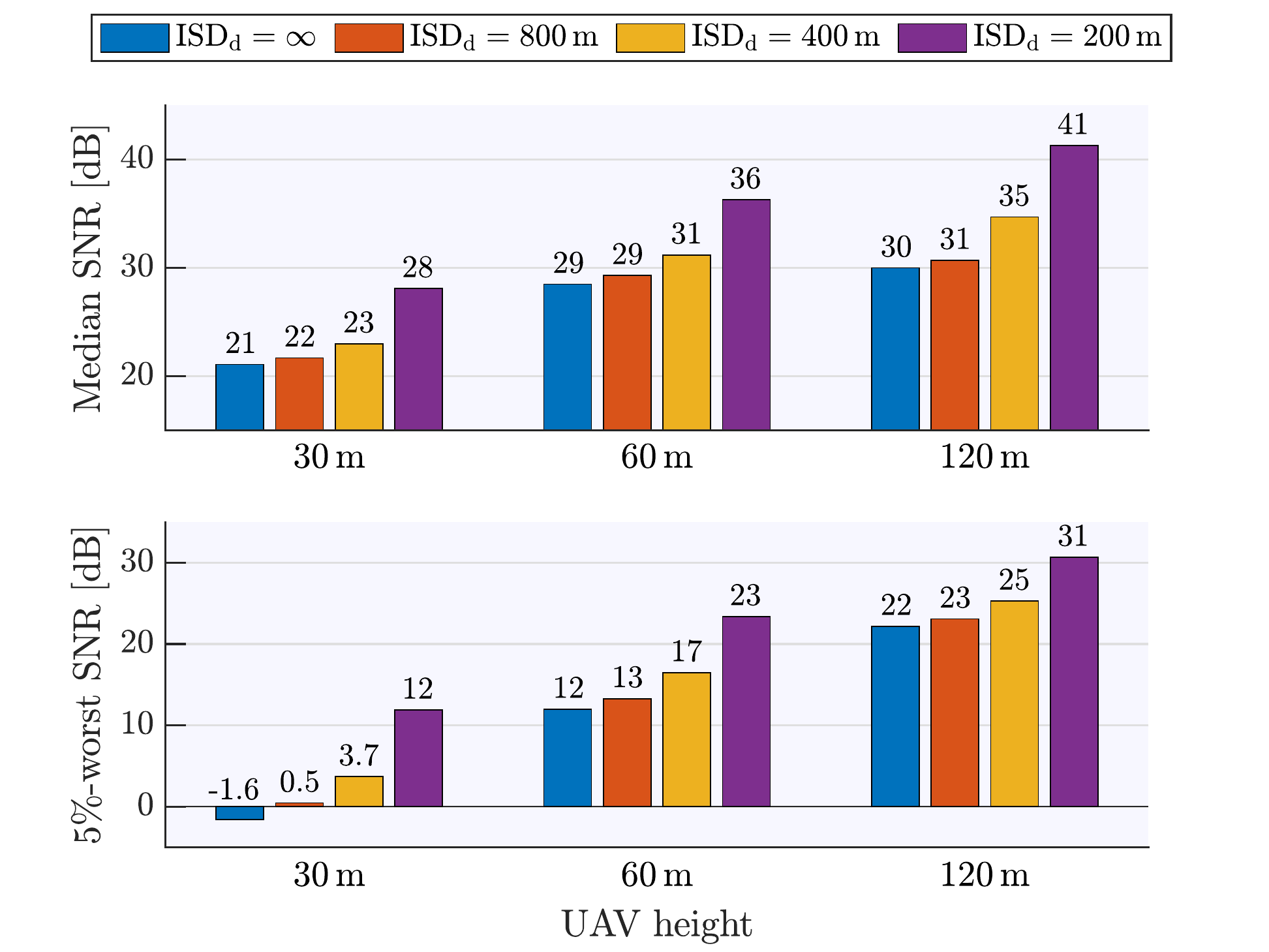}
\caption{Median and $5\%$-worst SNR for UAVs served by both standard and dedicated mmWave cells in an urban environment. $\ISDs=$\SI{200}{m} is fixed for standard cells whereas $\ISDd$ varies for dedicated cells of various deployment densities.}
\label{fig:mmWave_urban_3}
\end{figure}

The ensuing SNR distributions, with each UAV connecting to its preferred BS, either standard or dedicated, are presented in Fig.~\ref{fig:mmWave_urban_3} for UAV altitudes of 30, 60 and \SI{120}{m}. The SNR improvement with dedicated BSs is pronounced, especially for the $5\%$-worst SNR, provided those dedicated BSS are as dense as their standard counterparts. With sparser dedicated BSs, the improvement weakens, becoming anecdotal for $\ISDd = 4 \times \ISDs$.

\begin{takeaway}
As their density declines, the coverage from standard mmWave BSs becomes progressively less robust. For low-density deployments, dedicated BSs mounted on rooftops and uptilted could substantially enhance coverage.
\end{takeaway}

\subsection{Rural/Suburban mmWave Coverage} \label{subsec:mmWaveRural}

In contrast to the above, we now consider a UAV employed for a public safety application, taking place in a suburban or rural environment. Specifically, we employed actual drone flight traces from the Austin Fire Department, obtained using inertial measurement unit sensors and including position, velocity and orientation of a UAV that aims at wildfire prevention or behavior monitoring. The UAV motion traces were fed into an end-to-end 5G NR network simulator that updated the optimal beam pair every \SI{5}{ms}, consistently with the 3GPP beam management guidelines \cite{giordani2019tutorial}, accounted for the Doppler effect due to the UAV motion, and periodically generated a UDP payload of \SI{1500}{bytes}, for a total traffic source rate ranging from \SI{10}{} to \SI{1000}{Mbps} \cite{xia2019uav, mmw-ns3-nyu, Mez_COMST}. 

\begin{figure}
\centering
\includegraphics[width=\figwidth, trim = 0.9cm 0 0.9cm 0, clip]{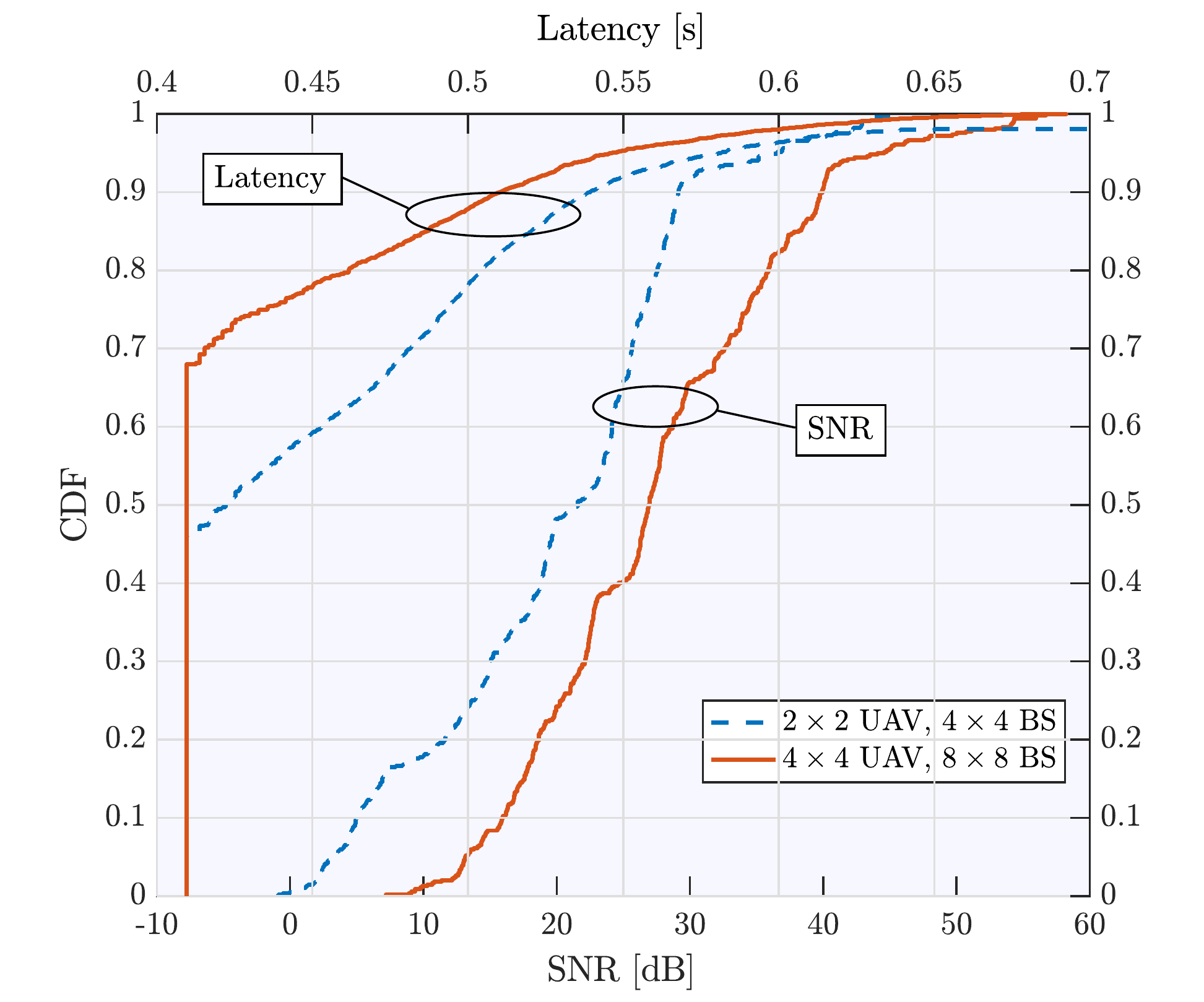}
\caption{CDF of the SNR and latency experienced by a UAV in a suburban/rural environment when served by mmWave links with two different antenna array configurations.}
\label{fig:mmWave_suburban_2}
\end{figure}

Fig.~\ref{fig:mmWave_suburban_2} captures the statistical distribution of the SNR and latency when operating at \SI{28}{GHz} with a bandwidth of \SI{1}{GHz}. The transmitting UAV is equipped with $2\times2$ and $4\times4$ antenna arrays whereas the receiving BS incorporates $4\times4$ and $8\times8$ arrays. As expected, a higher number of antennas can help sustain a better connection, especially when the UAV and the BS are further away, with direct benefits in both SNR and latency. 

\begin{figure}
\centering
\includegraphics[width=\figwidth, trim = 0.6cm 0 1.4cm 0, clip]{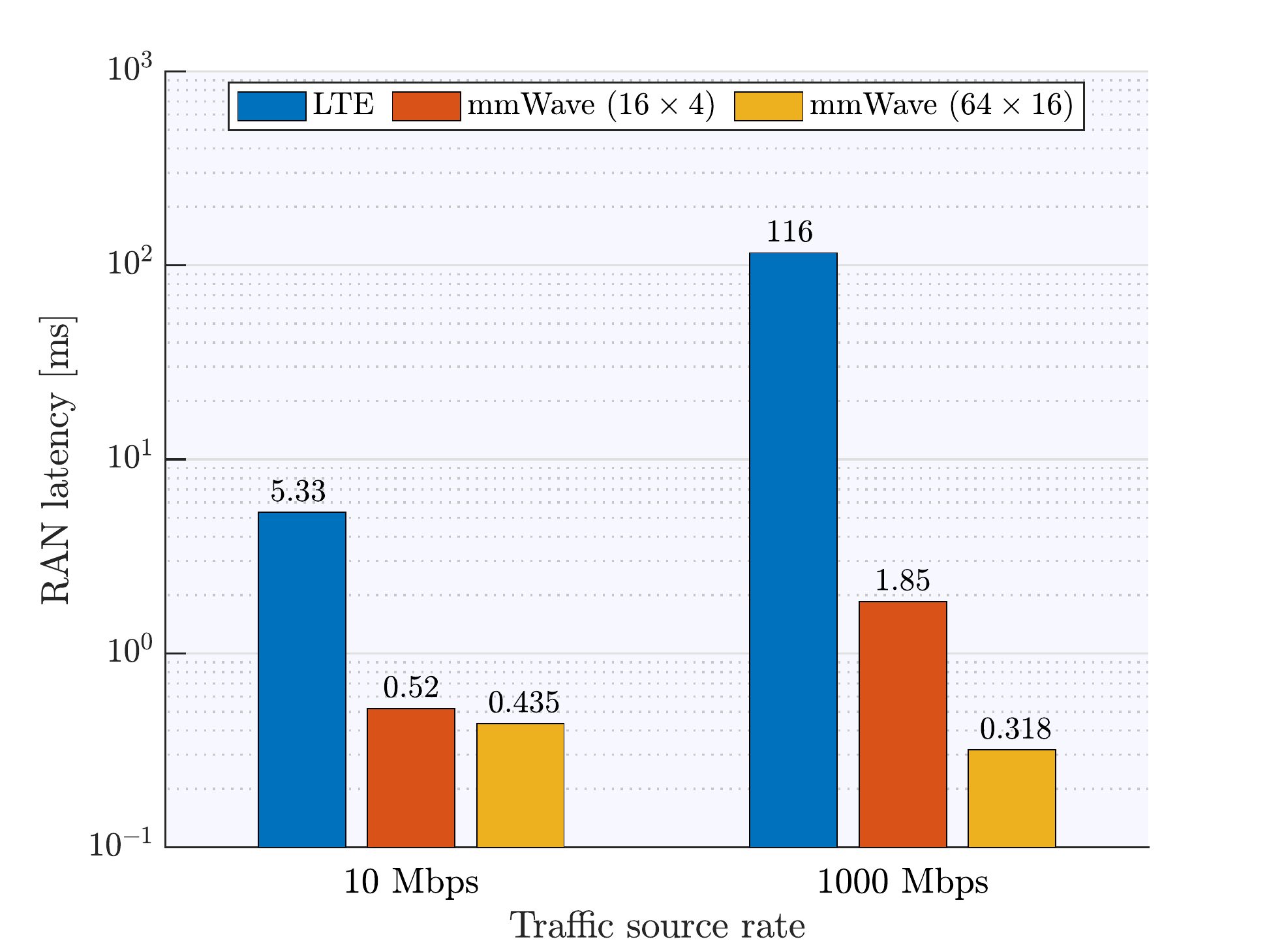}
\caption{Mean RAN latency experienced by UAV users served by LTE and mmWave links (two array configurations) under traffic loads of \SI{10}{} and \SI{1000}{Mbps}.}
\label{fig:mmWave_suburban_3}
\end{figure}

Fig.~\ref{fig:mmWave_suburban_3} further compares the mmWave setup to a baseline LTE system operating at \SI{2.1}{GHz} in terms of radio access network (RAN) latency. While incurring a reduced pathloss, the limited LTE bandwidth results in a much lower throughput. Moreover, the frame design in LTE does not support the sub-ms latency that can be achieved with NR mmWave~\cite{ford2017latency}. Consequently, LTE can yield mean latencies below \SI{10}{ms} when the traffic source rate is low. In contrast, the vast bandwidth availability in the mmWave spectrum enables Gbps connectivity with millisecond-latency even under a high traffic source rate \cite{dutta2017frame}.\footnote{Fig.~\ref{fig:mmWave_suburban_3} shows a slightly decreasing latency versus traffic load for a mmWave ($64 \times 16$) setup. This is simply owed to a numerical error due to the very small values of latency experienced under this setup.} 


\section{UAV-to-UAV Cellular Communications}
\label{sec:U2U}

Cellular device-to-device (D2D) communication---\emph{sidelink} in 3GPP terminology---, whereby devices communicate directly with each other bypassing ground BSs, were proposed for GUEs to facilitate the implementation of proximity services such as content sharing, public safety, and multi-hop transmission \cite{lin2014spectrum, asadi2014survey,george2015analytical}. This is because D2D communication can provide higher bit rates and lower latencies than those attained by the typical BS-centric communications when the relevant devices are in close proximity. 
Moreover, the establishment of a direct communication link might be essential when cellular coverage is not available. Other advantages of D2D communication include lowering the energy consumption, alleviating congestion, offloading cellular traffic, and possibly extending the range of communication via multiple D2D hops \cite{chun2017stochastic,tehrani2014device,liu2014device,feng2014device,lin2014overview}.
The 3GPP has been developing standards for D2D operations since LTE. In NR Rel.~16, sidelink solutions were specified for public safety and advanced automotive industry services, allowing vehicle-to-vehicle and vehicle-to-roadside-unit communication.
Such operations are being extended to cover new commercial use cases and requirements in NR Rel.~17 \cite{5GAmericas2021}. The availability and benefits of D2D communication motivate the use of such technology for UAVs, as discussed in the sequel.


\subsection{Applications and Distinct Features of Aerial D2D}
\label{subsec:U2Uapp}

Important use cases exist where a reliable and direct UAV-to-UAV (U2U) communication link can be of great value:
\subsubsection{UAV Swarms} A single UAV has limited processing capability, payload capacity for carrying various sensing devices, and range of communication. Accordingly, to meet specific goals such as environmental sensing, a group of UAVs may work together as a swarm. U2U communication is essential to deploy such UAV swarms.
Indeed, high-throughput U2U links are needed for the distribution of tasks in swarms where one or multiple responsible UAVs control the rest \cite{tahir2019swarms}.
\subsubsection{Autonomous Operation} To enable autonomous operation of independent UAVs performing different tasks, U2U communication links can be utilized to guarantee the low latencies required for collision avoidance and an adequate self-control of the aerial traffic.
\subsubsection{Aerial Relaying} U2U communication also enables aerial relaying for various purposes. For example, the coverage of ground BSs can be extended through multi-UAV hops to serve users in remote or disaster areas. Furthermore, U2U communication can be utilized to guarantee high throughput in areas without a reliable cellular service in the sky, e.g., through a hovering UAV head well connected to the ground.

In all above use cases, D2D communication in the sky exhibits distinct features w.r.t. its ground counterpart. Four different link types can be found in a U2U communication system reutilizing the ground cellular spectrum: 1) UAV-to-UAV, 2) UAV-to-BS, 3) GUE-to-BS, and 4) GUE-to-UAV. Each of these links experience different propagation conditions. For instance, UAV-to-UAV links are mostly in LoS conditions, subject to differences in the UAVs altitudes and building heights. The more favorable propagation conditions for direct UAV communication facilitates the reception of stronger signals, theoretically making U2U networking less dependent on the ground infrastructure for data exchange purposes due to their higher range. However, LoS propagation  may also be detrimental owing to the stronger interference to/from other UAVs and ground communications. This suggests that the design of D2D cellular communication must be revisited for aerial devices.

\subsection{U2U Communications: Key Design Parameters}
\label{subsec:U2Udesign}

\subsubsection{Power Control}

Power control is a key mechanism for efficient cellular D2D communications. In this context, a common approach known as fractional power control consists in utilizing a power proportional to the large-scale loss between the transmitter and receiver \cite{whitehead1993signal,UbeVilRos2008,BarGalGar2018GC}. Fractional power control intends to extend the battery life while providing a sufficient received signal quality and maintaining a constrained interference level. Under this scheme, a simplified expression of the transmission power per PRB, $\mathrm{P_{\{\text{UAV, GUE}\}}}$ [\SI{}{W}], is given by
\begin{equation}
\mathrm{P_{\{\text{UAV, GUE}\}}} = \min\{\mathrm{P_{max}/n_{\{\text{UAV, GUE}\}}},\mathrm{P_{ref}} \cdot \xi^{\epsilon_{\{\text{UAV, GUE}\}}} \},
\end{equation}
where $\mathrm{n_{\{\text{UAV, GUE}\}}}$ is the number of PRBs utilized for the communication by UAVs or GUEs, $\mathrm{P_{max}}$ is the maximum transmission power, $\epsilon_{\mathrm{\{\text{UAV, GUE}\}}} \in [0,1]$ is a factor controlling what fraction of the overall large-scale loss $\xi$ is compensated, and $\mathrm{P_{ref}}$ is a device-specific baseline power parameter.

\subsubsection{Spectrum Sharing Strategy}

Two main classes of spectrum sharing strategies can be considered for U2U communication, namely \textit{underlay} and \textit{overlay} \cite{AzaGerGar2020GC}.

\begin{itemize}
    \item \emph{Underlay spectrum sharing}. U2U links reuse a portion $\eta_{\mathrm{u}}$ of the GUE spectrum, which spans the entire available bandwidth \cite{AzaGerGar2019PIMRC}. Accordingly, the U2U links generate co-channel interference to the GUEs and vice versa. Intuitively, since a larger $\eta_{\mathrm{u}} \in [0, 1]$ results in further interference, both among UAVs and between UAVs and GUEs, $\eta_{\mathrm{u}} = \mathrm{n_{\text{UAV}}}/\mathrm{n_{\text{GUE}}}$ captures the aggressiveness of the underlay approach. In this setup, the U2U links may employ frequency hopping and randomly choose their $\mathrm{n_{\text{UAV}}}$ PRBs from the available $\mathrm{n_{\text{GUE}}}$ PRBs to reduce the UAV-to-UAV interference \cite{AzaGerGar2020}. 
    \item \emph{Overlay spectrum sharing}. The available spectrum is divided in a non-overlapping fashion between ground and U2U communications. As a result, GUEs are not interfered by UAVs at the expense of having access to a shrunk spectrum. Similarly, UAVs are only interfered by other co-channel UAVs. 
\end{itemize}

\subsection{Performance Analysis of U2U Communications}
\label{subsec:U2Uperformance}

We now illustrate the impact of the key design parameters as well as the spectrum sharing strategy on the performance of coexisting UL GUE and U2U communications. We consider an UMa cellular network operating at \SI{2}{GHz}, occupying \SI{10}{MHz} of bandwidth, and with a density of \SI{5}{BSs/km}$^2$ (corresponding to an average ISD of \SI{480}{m}). 
BSs schedule one GUE per PRB and U2U communication is held within UAV pairs with a density of \SI{1}{UAV/km}$^2$.
Unless otherwise specified, the receiving UAV is randomly and independently located around its associated transmitter following a truncated Rayleigh distribution with mean distance $\bar{R}_{\mathrm{UAV}} =$\SI{100}{m} \cite{AzaGerGar2020}.


\subsubsection{Impact of the UAV Altitude}
\begin{figure}
\centering
\includegraphics[width=\figwidth, trim = 0.9cm 0 1.5cm 0, clip]{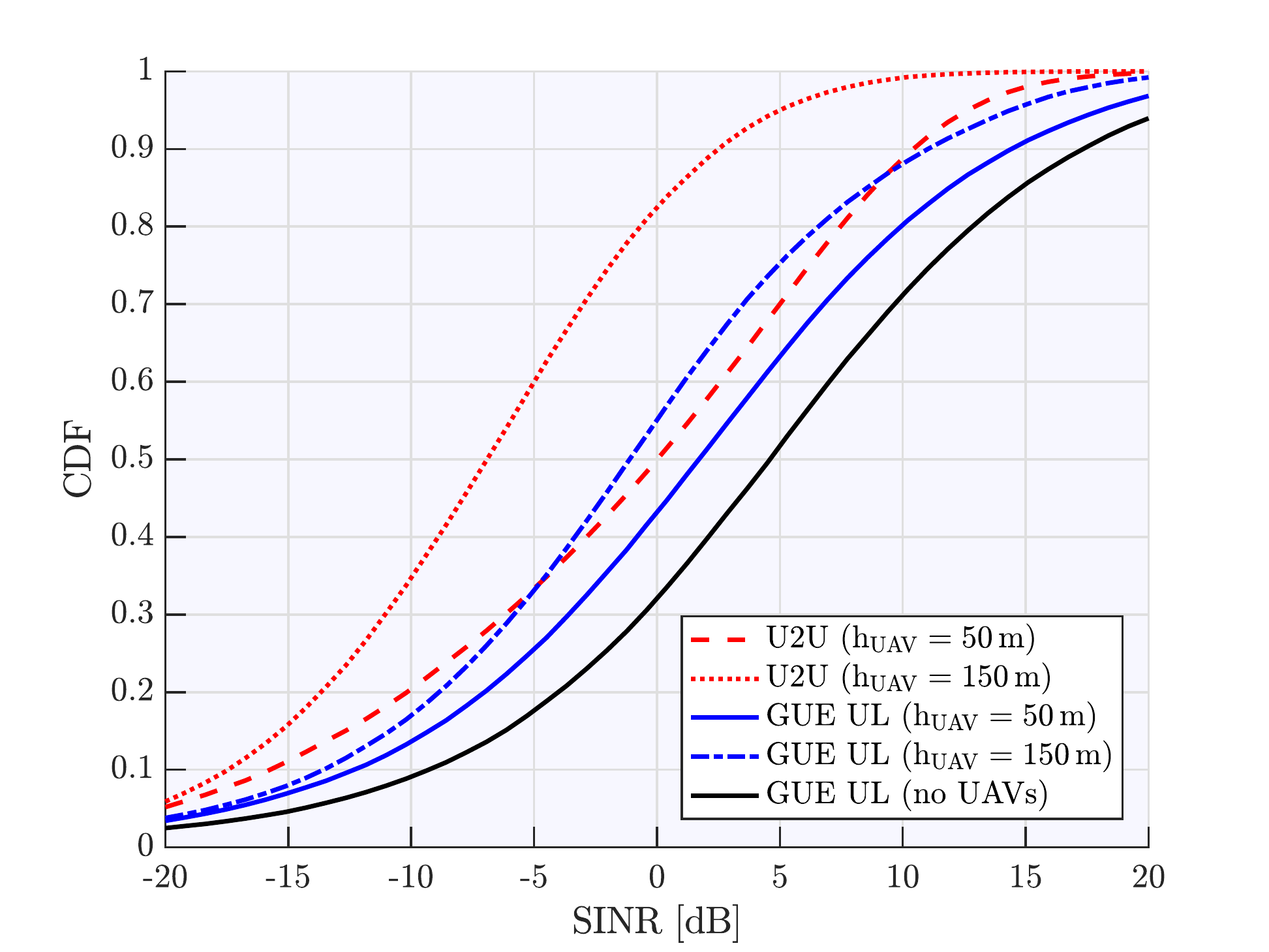}
\caption{CDF of the SINR experienced by: the GUE UL without U2U underlayed communications (black); the GUE UL with U2U underlayed communications and UAVs at different heights $\mathrm{h_{\mathrm{UAV}}}$ (blue); and U2U links underlayed with GUEs (red). $\mathrm{P_{max}}=$\SI{24}{dBm}, $\epsilon_{\mathrm{UAV}} = \epsilon_{\mathrm{GUE}} = 0.6$, and $\mathrm{P_{ref}} =$\SI{-58}{dBm}.}
\label{fig:U2U_Fig1}
\end{figure}

The impact of the UAV altitudes on the SINR performance of GUEs and UAVs is investigated in Fig. \ref{fig:U2U_Fig1}, which considers the most aggressive U2U underlayed approach where UAVs can access the whole UL GUE spectrum ($\eta_{\mathrm{u}}=1$). The results confirm that the UL performance of GUEs degrade in the presence of U2U links due to the co-channel interference generated by aerial devices. In spite of being limited due to the low powers employed by UAVs, such degradation becomes more pronounced when the UAVs fly at higher altitudes due to the increase in the number of ground BSs perceiving UAV-generated LoS interference. The performance of U2U communication links follows the same trend, since more interfering UAV-to-UAV and GUE-to-UAV links become LoS when the U2U pairs fly higher.

\begin{figure}
\centering
\includegraphics[width=\figwidth, trim = 0.9cm 0 1.5cm 0, clip]{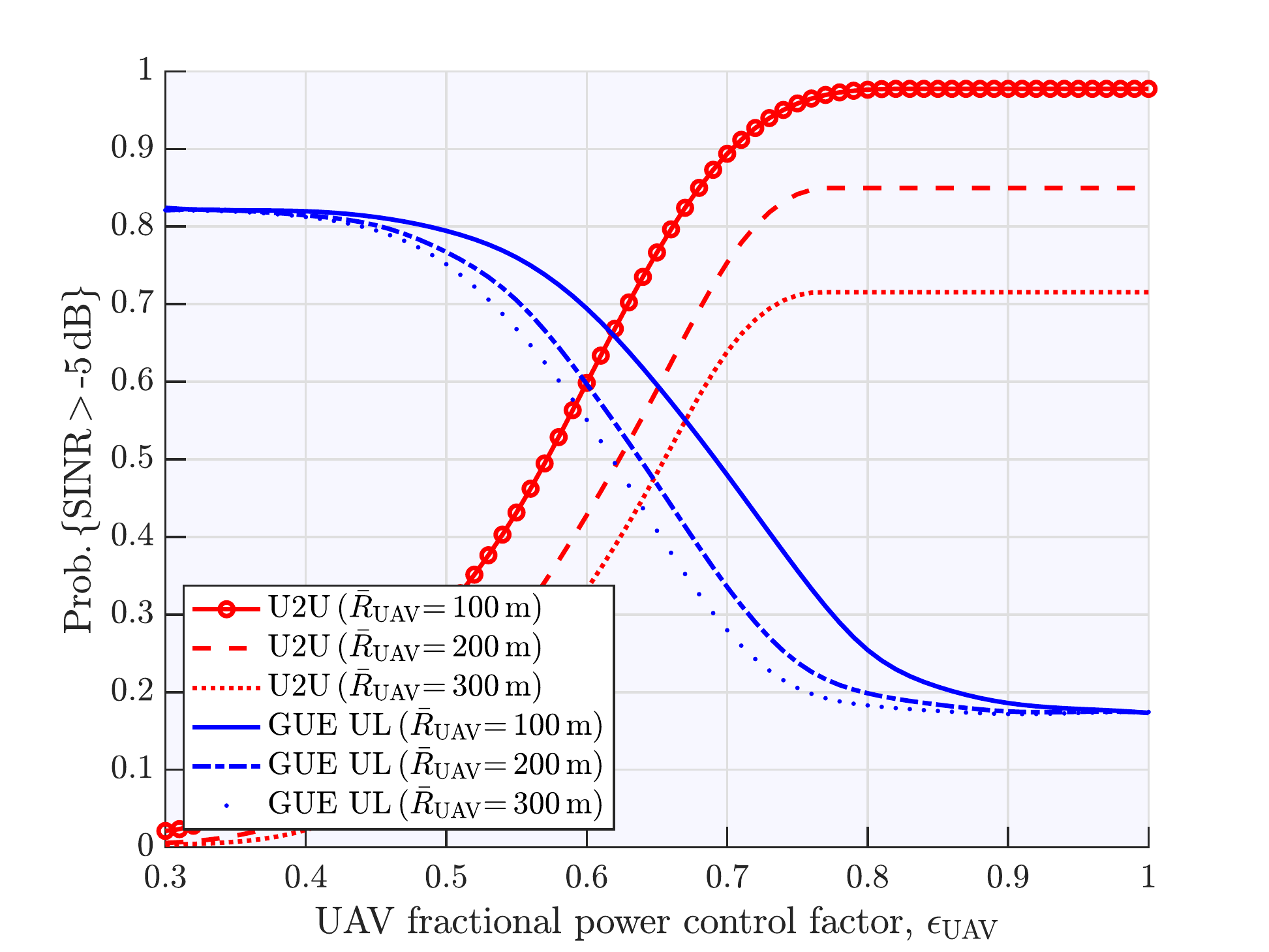}
\caption{U2U and GUE underlay coverage probabilities for different UAV power control factors and three values of the mean U2U link distance $\bar{R}_{\mathrm{UAV}}$.}
\label{fig:U2U_Fig2}
\end{figure}

\subsubsection{Impact of UAV Power Control}
Fig. \ref{fig:U2U_Fig2} captures the impact of the UAV power control factor $\epsilon_{\text{UAV}}$ for different mean U2U link distances $\bar{R}_{\mathrm{UAV}}$. The following broad-ranging observations can be made from the results of this figure, where the UAV altitude is \SI{100}{m}:
\begin{itemize}
    \item \emph{GUE UL performance}. Larger UAV power control factors lead to a lower GUE UL performance, a direct consequence of the higher interfering UAV powers. Such performance reduction is negligible when $\epsilon_{\text{UAV}}$ adopts extremely low values (where the UAV-generated interference is negligible compared to the ground interference) or extremely high values (where the majority of UAVs transmit at maximum power already). Fig. \ref{fig:U2U_Fig2} also shows that increasing the U2U communication link distances leads to a reduced GUE performance, since UAVs need to employ larger transmission powers to compensate for the increased U2U path loss.
    \item \emph{U2U performance}. Increasing $\epsilon_{\text{UAV}}$ improves the U2U communication performance on account of the growth in useful U2U signal power. Although this is at the expense of an increase in the interference among UAV pairs, it is worthwhile because the interference perceived by UAVs is mostly produced by the GUEs transmitting at high powers. Fig. \ref{fig:U2U_Fig2} also shows that the longer communication range between UAVs deteriorates the overall U2U performance.
\end{itemize}

\begin{figure}
\centering
\includegraphics[width=\figwidth, trim = 0.8cm 0 1.5cm 0, clip]{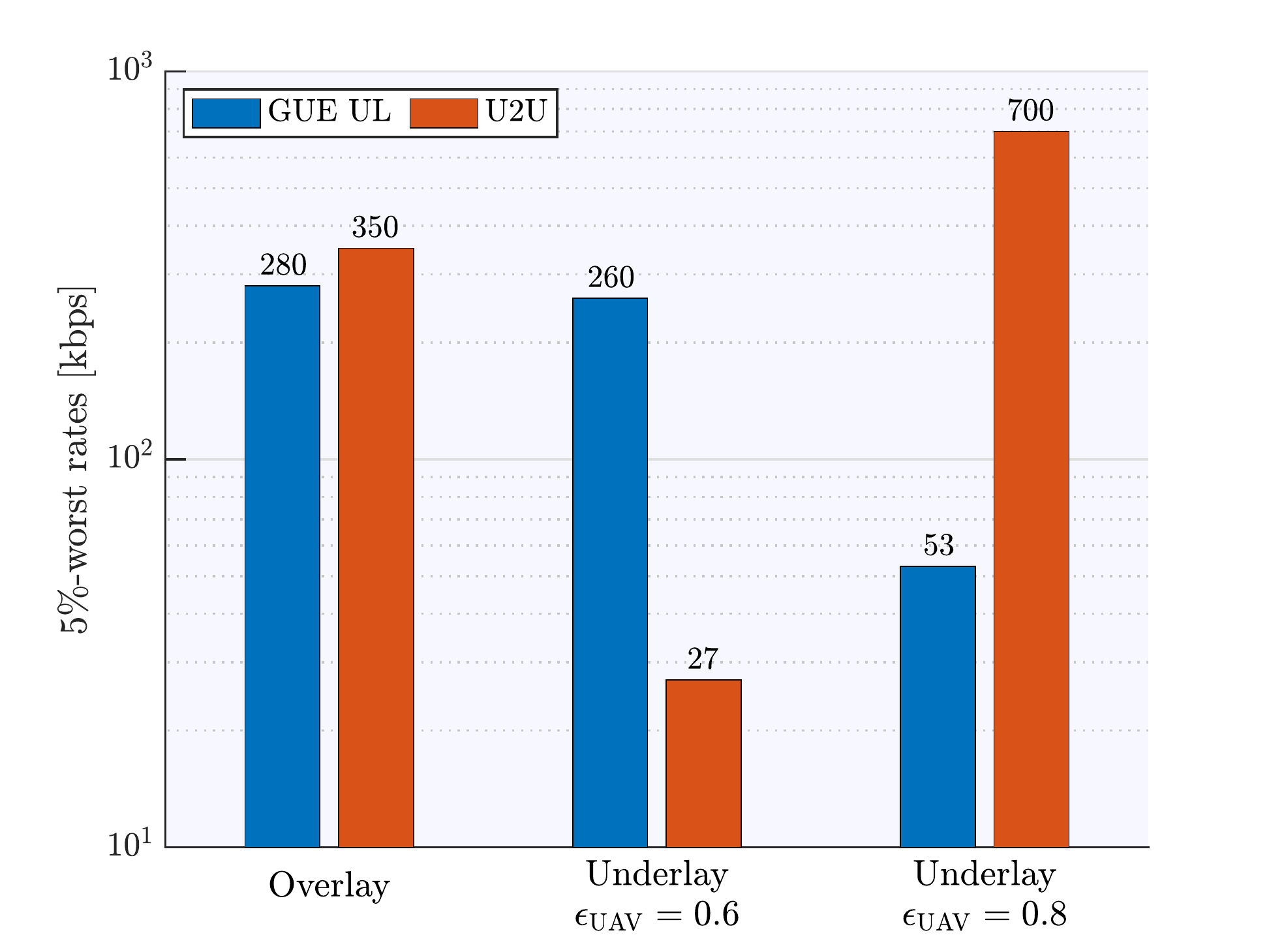}
\caption{5\%-worst rates for GUE UL and U2U links in overlay, where UAVs are restricted to access \SI{1}{MHz}, and underlay, where UAVs can access the entire \SI{10}{MHz}.}
\label{fig:U2U_Fig3}
\end{figure}

\subsubsection{Impact of the Spectrum Sharing Strategy}
Fig. \ref{fig:U2U_Fig3} compares the rate performance 
of the 5\%-worst GUE UL and U2U links in an overlay setup where UAVs are restricted to access \SI{1}{MHz} and in an underlay setup like the one in Fig. \ref{fig:U2U_Fig2} (considering two values of $\epsilon_{\mathrm{UAV}}$). We note that the impact of the UAV power control factor is negligible in the overlay, since both the signal power and the overall interference proportionally increase with $\epsilon_{\text{UAV}}$. Fig.~\ref{fig:U2U_Fig3} also reveals that the best guaranteed GUE UL performance is provided by the overlay setup, where the 5\%-worst U2U bit rates reach \SI{350}{kbps}. A similar GUE UL performance can be achieved in the underlay when UAVs utilize a reduced transmission power ($\epsilon_{\text{UAV}}=0.6$), although this comes at the expense of a 13-fold decrease in the 5\%-worst U2U rates. The U2U performance dramatically grows to \SI{700}{kbps} in the underlay scenario with $\epsilon_{\text{UAV}}=0.8$, where GUEs suffer due to the increased aerial interference. Overall, the results of Fig.~\ref{fig:U2U_Fig3} convey that the overlay spectrum sharing approach provides the best performance trade-off between direct aerial communication and legacy GUE communication.

\begin{takeaway}
Underlaying U2U links with the GUE UL incurs limited interference, which however becomes more pronounced when UAVs fly higher and most cross-tier links turn to LoS. UAV-specific fractional power control can trade off the performance of U2U links and GUE UL, whereby higher control factors favor the former at the expense of the latter.
\end{takeaway}

\subsection{mmWave U2U Communications}
\label{subsec:U2UmmWave}

\begin{figure}
\centering
\includegraphics[width=\figwidth, trim = 1.2cm 0 1.1cm 0, clip]{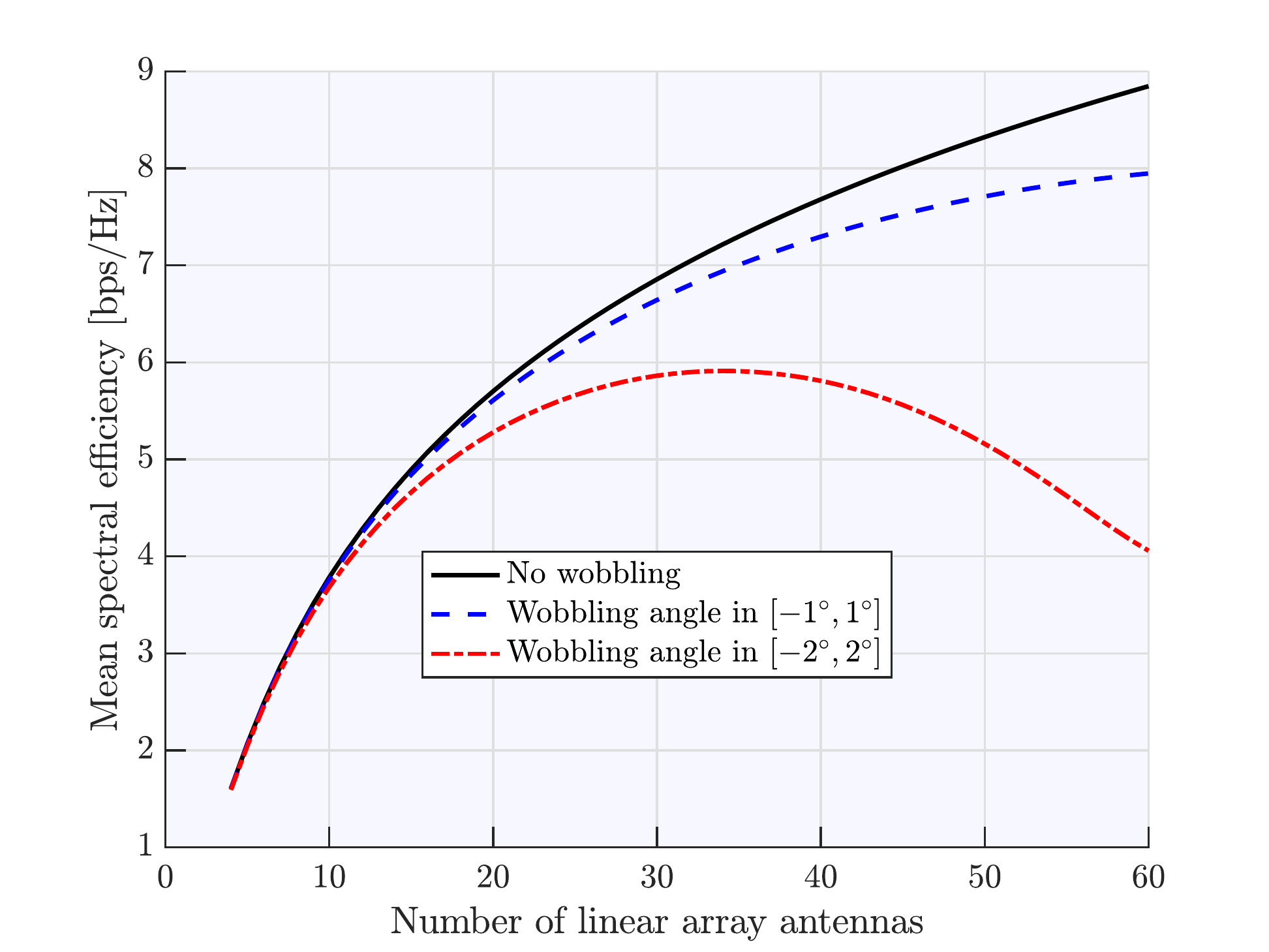}
\caption{Mean spectral efficiency versus number of antennas for a \SI{500}{m} U2U link at \SI{60}{GHz}, where both UAVs undergo wobbling around the local vertical angle.}
\label{fig:U2U_Fig4}
\end{figure}

As detailed in Sec.~\ref{sec:mmWave} for UAV-to-ground links, the large bandwidths available at mmWave frequencies could be exploited to boost the performance of U2U links too. Once again, highly directional antenna arrays should be employed to compensate for the higher propagation losses incurred at these frequencies. However, the fact that both end devices wobble and travel in U2U links will worsen the fluctuations and beam misalignments, resulting in link quality dips. This phenomenon is illustrated in Fig.~\ref{fig:U2U_Fig4}, which shows the mean spectral efficiency versus the number of antennas for a \SI{500}{m} U2U link at \SI{60}{GHz}. Both UAVs are equipped with a vertical uniform linear array and undergo wobbling. The wobbling is characterized through variations in the elevation angles of each UAV array, which follow a uniform random variable with maximum value $1^{\circ}$ or $2^{\circ}$. Fig.~\ref{fig:U2U_Fig4} shows that double UAV wobbling can have a significant impact on the optimal mmWave U2U antenna design. Depending on the extent of the wobbling and on the communication distance, a larger number of antennas could lead to a more frequent and prominent beam misalignment, and hence to a worse overall performance.

\section{UAV Cellular Communications Beyond 5G} 
\label{sec:6G}


Are you thrilled by the idea of mMIMO- and/or mmWave-connected UAVs soon delivering groceries to your doorstep? Brace yourself: in the next decade you might become the very passenger, taking the reliability requirements for the 

Autonomous cars-as-a-service and air taxis will redefine how we commute---and, in turn, where we live and work---with remarkable societal implications and emerging entrepreneurial opportunities. The 2020s might finally witness fully driverless cars hit the road, shifting mainstream production from Level~2---partial automation: humans monitoring all tasks and taking control when needed---to Level~5---full automation: vehicles going anywhere and anytime without human intervention \cite{Synopsis}. Sometimes in the 2030s, we might find ourselves above traffic in flying cars or levitating pods, all autonomous and empowered by widespread 6G connectivity \cite{Stuff}. Electric vertical takeoff and landing vehicles (eVTOLs), projected to be displacing some of the helicopters in operation today, may be the solution to reduce congestion in megalopolis and curb energy consumption \cite{Saracco}. Though hurdles like safety regulations \cite{stocker2017review}, noise concerns, and infrastructure needs could prolong projected launch dates, over 100 cities are already targeting urban air mobility solutions, with several companies planning to lift air taxis and a market of \$1.5 trillion by 2040 \cite{RolandBerger,Yahoo,Airbus,KittyHawk,Cartivator,Terrafugia,Aeromobil}.

Chinese drone company EHang recently had its two-seater passenger-grade autonomous UAV flying over a densely populated downtown area in Korea, in what the Mayor of Seoul defined as ``a dream of mankind for the future transportation” \cite{DroneLife}. Meanwhile, Uber Elevate seeks to build and scale Uber Air, a multimodal transportation product that seamlessly integrates time-saving first- and last-mile ground transportation, e.g., between city centers and airports, in a holistically sustainable way. Three target launch cities have been announced: Los Angeles, Dallas, and Melbourne, with interest spreading globally. Though initial operations will be limited to a handful of eVTOLs, this number might be scaling to possibly hundreds over a period of 5--10 years, as community acceptance and interest dictate \cite{UberAir}. In its quest, Uber is teaming up with AT\&T, at first to keep piloted aircrafts connected to 5G networks at low altitudes, to determine existing boundaries. The joint venture might then target 6G features to provide eVTOL safe and pilotless operations by 2030 \cite{ZDNet,SaeAlnAlo2020}.

While autonomous air taxis could eventually become one of the defining killer apps of the 6G era, they exhibit unprecedented C2 as well as connectivity requirements all around a 3D environment. Safety-wise, accurate cm-grade 3D localization and navigation will also be essential at heights ranging from tens to hundreds of meters. In this regard, 6G development towards even higher frequency ranges, wider bandwidths, and massive antenna arrays, will enable sensing solutions with very fine range, Doppler, and angular resolutions \cite{BouNolLie2020}. Improved positioning may also be achieved by using new reference nodes, such as the rooftop-mounted BSs (introduced in Sec.~\ref{subsec:mmWaveUrban}) and satellites or high-altitude platforms (HAPs), which may serve as aerial radio access nodes in 6G cellular networks as detailed in Sec.~\ref{sec:NTN} \cite{PanAlo2020}.

Besides supporting autonomous eVTOLs, future networks are meant to provide reliable, seamless, robust, and high-speed data connectivity for eVTOL passengers, enhancing their flight experience. Mobile operator Ooredoo recently connected a two-passenger, driverless air taxi to its 5G network. The test was conducted at The Pearl-Qatar, for about 20 minutes and at speeds of up to \SI{130}{km/h}, with passengers reportedly achieving rates up to \SI{2.6}{Gbps} \cite{5GObservatory}. While such experiment is indeed auspicious, many more air taxis with a plurality of flying passengers are expected to travel at hundreds of meters above ground, and next-generation mobile networks must be designed and planned appropriately.

In a quest for enabling the above bold ambitions---and many more non-UAV-related ones---the wireless community has already rolled up its sleeves in (re)search for candidate beyond-5G features. These enhancements will inevitably include more infrastructure, more bandwidth, and higher spectral efficiency via network intelligence. In the following sections, the spotlight is placed on the five main disruptive innovations being discussed in industrial fora and academia that are likely to be integrated in next-generation mobile systems: non-terrestrial networks, cell-free architectures, the use of AI, the deployment of reconfigurable intelligent surfaces, and communication at THz frequencies. Besides introducing these five paradigms, we focus specifically on how they are envisioned to benefit UAV communications and what hurdles stand on the way of this vision. 

\begin{takeaway}
Throughout the next decade, empowering air taxis, cargo-UAVs, and wirelessly backhauled flying BSs with sufficient data transfer capacity and link reliability will require one (or multiple) 6G paradigm shift(s).
\end{takeaway}


\section{Towards Resilient UAV Communication with Non-terrestrial Networks}
\label{sec:NTN}

\begin{figure}[!t]
\centering
\includegraphics[width=\figwidth]{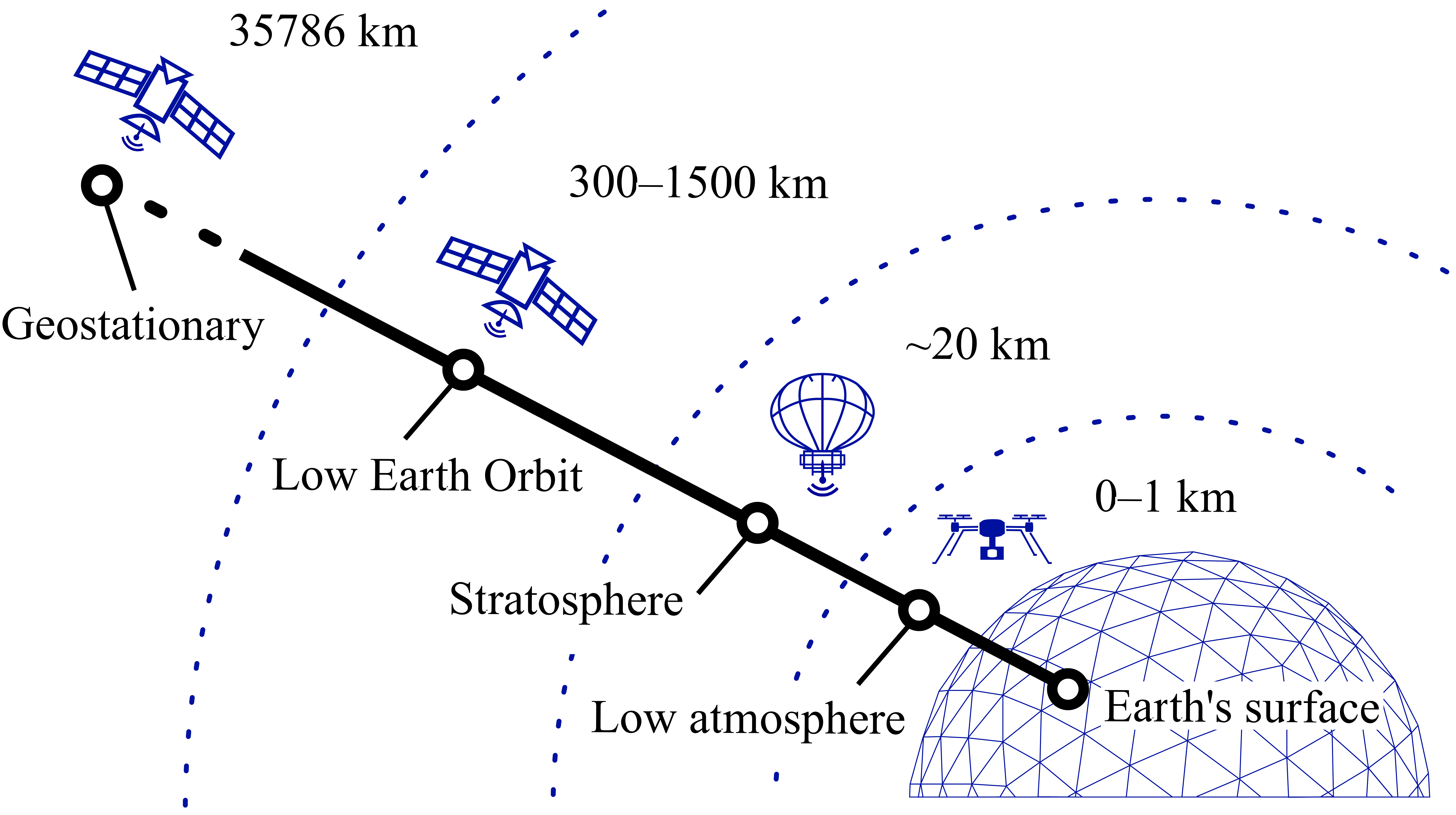}
\caption{Illustration of NTN platforms at different altitudes. From the top left: GEO, LEO, HAP, and UAV.}
\label{fig:NTN_Fig1}
\end{figure}

NTN is a termed coined within the 3GPP context to subsume all connections going through a flying component, i.e., UAV, HAP, or satellite, as illustrated in Fig.~\ref{fig:NTN_Fig1}. Satellite links have been around for over half a century, mostly in the geostationary equatorial orbit (GEO). Among other virtues, satellites are well positioned to reach where cables cannot, they can blanket vast areas with a LoS coverage umbrella, and their links are resilient to natural disasters on Earth \cite{noauthor_satellite_nodate,giordani2020non}. 
Lying in between us and space, HAPs are typically unmanned and operate in the stratosphere in quasi-stationary position. HAPs could well complement spaceborne platforms since they are cheaper to launch, can act as wireless relays, and achieve acceptable latencies owing to their lower altitudes \cite{EulLinTej2021,CaoYanAlz2018}.

The cellular standardization community has thus far focused on satellite communications for three main reasons: satellites have an established commercial base of manufacturers, operators, and consumers, both private and institutional; the entire sector has been redefining itself due to the recent interest shift from content broadcasting to multibeam data services; the introduction of more affordable insertions into the Low Earth orbit (LEO) is favoring the burgeoning of large non-geostationary constellations \cite{KodheliLMSSMDSC21}. The 3GPP effort has been mainly devoted to the following targets:
\begin{itemize}
    \item For Rel.~15, a selection of deployment scenarios, key system parameters, and channel models \cite{3GPP38811}.
    \item For Rel.~16, the definition of the architecture, higher-layer protocols, and physical layer aspects \cite{3GPP38821}.
    \item For Rel.~17, technical solutions---albeit LTE-focused---concerning timing relationships, UL synchronization, and hybrid automatic repeat request (HARQ) \cite{3GPP36763}.
    \item For Rel.~18, the agenda is currently being discussed.
\end{itemize}

\subsection{UAV-to-Satellite Communications}
\label{subsec:NTNUAVtosat}

The most promising use cases and technological enablers that combine UAVs and satellites can be identified as follows: 
\subsubsection{Integrated Terrestrial-NTN for UAV Users} 
\begin{figure}[!t]
\centering
\includegraphics[width=\figwidth]{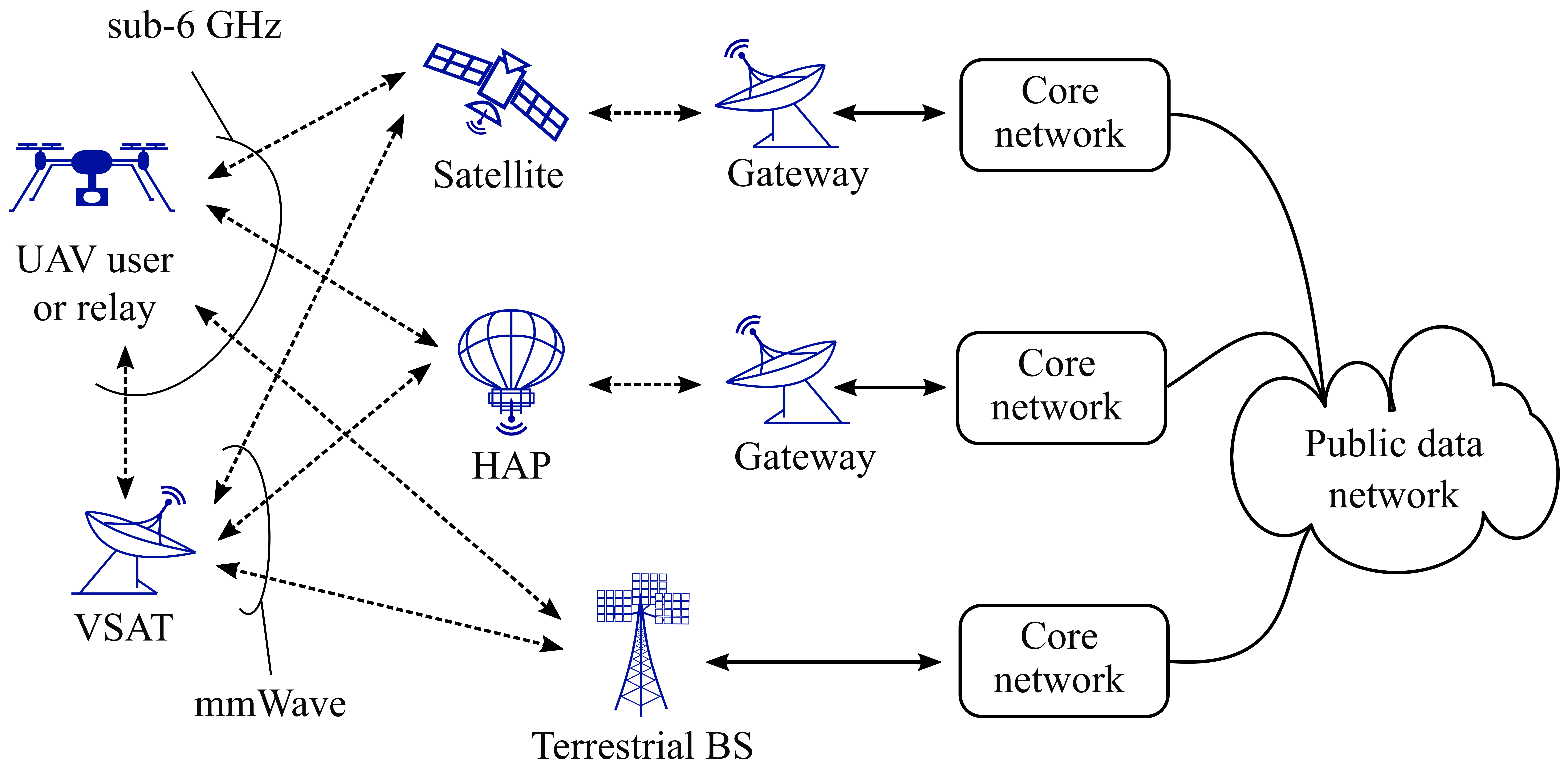}
\caption{Potential architecture for an integrated ground-air-space network supporting a UAV, with dashed and solid arrows respectively denoting wireless and wired links.}
\label{fig:NTN_Fig2}
\end{figure}

BVLoS control of UAVs entails zero outage in terms of connection availability, as connectivity gaps can jeopardize a UAV mission. The combination of terrestrial and NTN can potentially alleviate this shortcoming by providing blanket coverage over cells ranging from tens to hundreds of kms. Recently, such combination was used in a practical setup for transporting Covid-19 samples and test kits \cite{noauthor_covid_2021}. Unlike a hybrid terrestrial-NTN architecture, an integrated one avails of a common network management architecture \cite{DBLP:journals/tvt/GuidottiVCACMEA19}. The latter is required in the case of spectrum coexistence between terrestrial and NTN \cite{DBLP:journals/access/LagunasTSC20}.
Fig.~\ref{fig:NTN_Fig2} illustrates a potential architecture for an integrated ground-air-space network, including from the right gateway-to-air-or-space feeder links and service links. Depending on the carrier frequency, the latter can directly serve a UAV or be relayed through a very-small-aperture terminal (VSAT). 

\subsubsection{Backhauling to UAV Radio Access Nodes} Satellite broadband backhauling is one of the most rapidly developing areas mainly due to maritime and aeronautical demand. Backhauling to a UAV radio access node is a much more challenging endeavor, owing to the reduced terminal size for a data demand of roughly the same order. Due to the latency requirements, LEO satellites would be preferred for their proximity. Alternatively, GEO could be employed if integrated with the terrestrial network and with traffic classification \cite{8411107}. An instance of such configuration is content distribution for cache feeding \cite{VuPCMGR20}, where the traffic is latency-tolerant and mostly DL.
\subsubsection{Aggregators for Internet of Things (IoT) Services} IoT data collection from remote areas is a prime example where UAVs can shine as radio access nodes \cite{hong2020space}. However, there would still be a need for considerable UL capacity so that the data can be transferred to the main IoT servers for processing. NTN connectivity can facilitate this especially when the task is not latency-sensitive. Since IoT sensors typically have to be low-cost and low-form, direct satellite access does not seem a viable choice, except for LEO \cite{DBLP:journals/access/KodheliAMCZ19}. 

\subsection{Roadblocks to an Integrated Ground-Air-Space Network}
\label{subsec:NTNroadblocks}


Irrespective of the targeted use case, UAV-satellite communications face several challenges: 

\subsubsection{Hardware} Establishing a bidirectional connection with a satellite requires specialized antennas that might increase the power and weight requirements of the UAV. In this context, the main dilemma is choosing the most suitable satellite orbit. On the one hand, GEO requires a narrower field of view but also more power due to the larger propagation distance. LEO, on the other hand, incur a less severe path loss but their rapid movement requires a wider field of view, quick tracking capabilities, and considerable Doppler compensation. 
Furthermore, the NTN antenna would be very difficult to reuse for terrestrial connectivity, since it has to be positioned on the top of the UAV to enable LoS connection with the satellite. Despite the recent progress, low-cost production for massive civilian applications is still challenging.\footnote{In all above cases the assumption is that broadband NTN connectivity is used for both communication and control \cite{DBLP:journals/spm/Perez-NeiraVSMC19}. If the interest is limited to UAV control, lower-band narrow carriers could be used to simplify the antenna requirements.}

\subsubsection{PHY} In view of the satellite data service renaissance, a number of advanced technologies are actively being developed, such as multiuser precoding, active antennas, carrier aggregation, non-orthogonal multiple access, and beam hopping. Even though these are widely accepted techniques for spectrally efficient communication in terrestrial systems, their application to NTN, UAVs, or both, is often complicated due to the propagation environment and transponder peculiarities:
\begin{itemize}
    \item \emph{Multiuser precoding} appears to be the frontrunner since it was recently demonstrated over-the-air through a live GEO satellite \cite{9373415}. However, the applicability over LEO and for UAVs would bring about mobility-induced CSI variations, which also incur a significant latency in being reported back to the Gateway \cite{DBLP:journals/twc/ChristopoulosCO15}. 
    \item \emph{Active antennas}, in the sense of dynamic beamforming through electronically steerable arrays, are next in line, with commercial deployment only a couple of years away. Still, early deployments are expected to have limited capabilities in terms of beam granularity, update rate, and bandwidth.
    \item \emph{Satellite carrier aggregation}, a well established technology in 3GPP, seems promising when carriers from the same satellite are combined \cite{KibriaLMAC20}. However, aggregating carriers from different satellites---and even worse from different orbits---still appears to be an elusive target.
    \item \emph{Non-orthogonal multiple access} was shown to have some applicability in specific configurations, for example when users are at different altitudes, e.g., airplanes and UAVs. When it comes to constructing the superimposed waveform, airplanes could be seen as the strong users and UAVs as the weak ones \cite{DBLP:journals/tvt/WangLLPCO21}. The potential gains are however to be verified in practical settings.
    \item \emph{Beam hopping} is a satellite radio access alternative to frequency reuse, which allows allocating the entire spectrum for each beam by illuminating nonadjacent beams in a time-hopping pattern. Although this is gaining momentum with an experimental payload in the making, UAV applications might be sensitive to its intermittent access, especially if the C2 link is relayed over satellite \cite{DBLP:journals/access/LeiLYKCO20}.  
\end{itemize}

\subsubsection{Protocols} In satellite communications, the main protocol stack is based on digital video broadcasting - satellite (DVB-S). Even though the DVB consortium originally focused on broadcasting, it quickly became obvious that data services had to be accommodated too, and the relevant standards---e.g., DVB-S2X---have been adapted to support broadband connectivity. More recently, the connectivity convergence within 5G NR and beyond has attracted substantial interest from the satellite community, and there are currently work items focusing on enabling 3GPP standards over satellite. In this direction, there are two parallel integration approaches: 
\begin{itemize}
    \item \emph{At the architecture level}, where the underlying waveforms remain compatible with DVB-compliant equipment;
    \item \emph{At the waveform level}, where NR is adjusted to overcome the peculiarities of the ground-to-space channel.
\end{itemize}
The latter is more enticing since it would allow satellite equipment manufacturers to benefit from the 5G economies of scale. However, it is a more challenging one since the massive size of satellite cells poses a series of incompatibilities, e.g., in terms of HARQ timers, timing advances, and Doppler compensation \cite{DBLP:conf/asms-spsc/Al-HraishawiLC20}. While the priority of the 3GPP is addressing these issues for satellite-to-ground links, their technical solutions could pave the way for satellite-to-UAV cellular communication too.


\begin{takeaway}
Satellites could fill the inevitable ground coverage gaps that can currently jeopardize a UAV mission aiming for zero outage. Additionally, spaceborne cells could handle the most mobile UAVs across their large footprint, or backhaul aerial radio access nodes. However, important challenges need to be overcome for UAV-to-satellite communications in terms of hardware, PHY, and protocol design. 
\end{takeaway}
\section{Cell-free Architectures for UAVs} 
\label{sec:cellfree}


Looking towards 6G, there is the perception in many quarters that the cellular framework might be exhausted and should be transcended, at least for dense deployments.
Underpinning this perception is the recognition that, in the UL, intercell interference is merely a superposition of signals that were intended for other BSs, i.e., signals that happen to have been collected at the wrong place. If these signals could be properly classified and routed, they would in fact cease to be interference and become useful in the detection of the data they bear. A dual observation can be made about the DL, altogether motivating so-called cell-free networks where all BSs joint communicate with all users \cite{ngo2017cell}. Such networks have the potential of providing major benefits in terms of signal enhancement (both macro- and micro-diversity) and interference reduction \cite{venkatesan2007network,irmer2011coordinated}. 

For UAV communication, where LoS conditions generally exist to multiple BSs, with the correspondingly severe intercell interference, the appeal of a cell-free operation is reinforced even further \cite{DanGarGer2020,DanGarGer2020}. In fact, in terms of interference relief, such dense cell-free networks could be regarded as a competing alternative to macrocellular mMIMO; spatial degrees of freedom that are spatially distributed in the former versus concentrated in the latter.

To gauge the potential of cell-free architectures for UAV communication, we examine how the SINR distribution in a UMi environment ($\mathsf{ISD}=$\,\SI{200}{m}) changes as the network is rendered cell-free. Once more, the focus is on the UL, the most demanding link for UAV communication.
We begin with UAV-only service at \SI{2}{GHz}, detailed as:
\begin{itemize}
\item A \SI{20}{MHz} TDD channel.
\item Transmit power, noise figure, antenna gains, LoS probability, pathloss, and Ricean fading, all as per \cite{3GPP36777}.
\item BSs and UAVs randomly deployed, with the altitudes of the latter distributed between \SI{25}{} and \SI{150}{m}.
\item Fractional power control with $\epsilon_{\mathrm{UAV}}=0.7$ in all cases. (The cell-free extension of this power control technique is formulated in \cite{nikbakht2019uplink,nikbakht2020uplink}.)
\item SU per cell and single antenna per BS.
\end{itemize}
Channel estimation acquires heightened importance for cell-free operation because, as interference is removed, the residual estimation error becomes a limiting impairment. To provide a comprehensive view of the cell-free advantage, we thus quantify such advantage for the two extremes in terms of channel estimation: on one hand, perfect estimation; on the other hand, minimum mean-square error (MMSE) estimation based on an UL SRS occupying one symbol per time-frequency coherence interval \cite{lozano2008interplay}. Any operating point is bound to be somewhere in-between these extremes. 


Two cell-free variants are entertained, namely a first one where all the BSs jointly form matched-filter (MF) beams towards the users, maximizing the SNRs with no regard for the interference,
and a second one where the beams are designed under an MMSE criterion and the SINRs are maximized. The former is computationally less intense and may admit somewhat distributed implementations \cite{interdonato2019ubiquitous} while the latter, a decided shift towards the C-RAN paradigm, is where the full cell-free potential is unleashed \cite{pCell,bjornson2019making,attarifar2020subset}.

\begin{figure}[!t]
\centering
\includegraphics[width=\figwidth, trim = 0.8cm 0 1.3cm 0, clip]{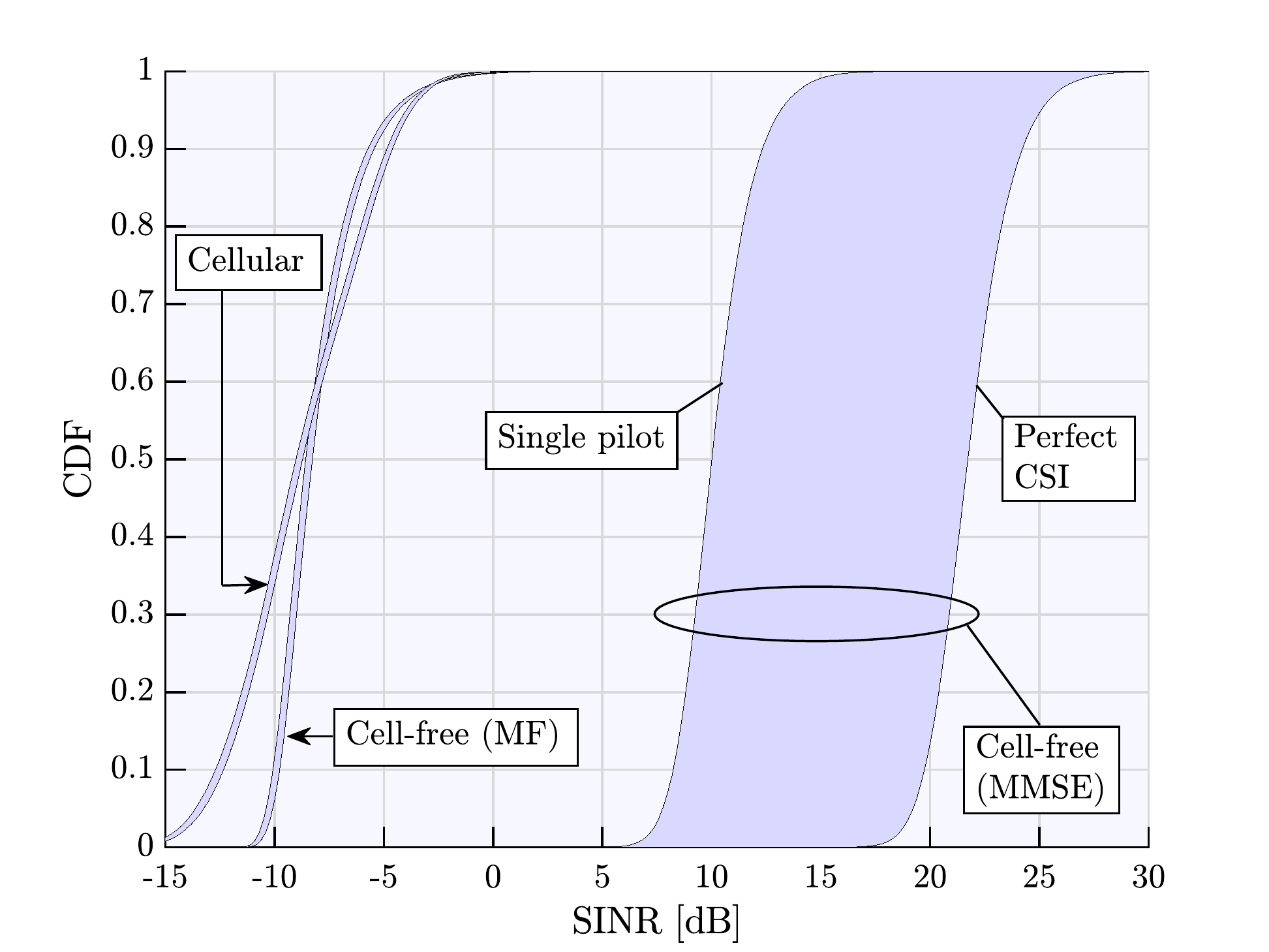}
\caption{SINR distribution in a UAV-only UMi cellular network alongside its cell-free counterparts. For each case, the shaded region spans the range between single-symbol SRS and perfect channel estimation.}
\label{fig:CF1}
\end{figure}

Presented in Fig. \ref{fig:CF1} is the CDF of the SINR, averaged over the small-scale fading, in cellular and cell-free networks having an equal number of BSs and users.
The cellular performance, whose lower tail corresponds to the values in Fig.~\ref{fig:mMIMO_Fig3}, is barely affected by the channel estimation precision: the estimation error is much weaker than interference-plus-noise, even with a single-symbol SRS per coherence interval.
The MF cell-free alternative, also not significantly impacted by channel estimation, increases the worst-case SINRs noticeably, but does not otherwise improve things. It is in its MMSE incarnation that cell-free operation dramatically improves the performance, with median SINR gains of \SI{20}{}--\SI{30}{dB} depending on the channel estimation accuracy, which takes center stage as anticipated.
As this accuracy goes hand in hand with the SRS overhead, to be discounted from the spectral efficiency, some intermediate operating point within the shaded range in the figure is desirable depending on the fading coherence. Regardless of its exact value, the benefit is hefty, and indeed even larger than for terrestrial service because of the increased LoS probability for UAVs.

\begin{figure}[!t]
\centering
\includegraphics[width=\figwidth, trim = 0.8cm 0 1.3cm 0, clip]{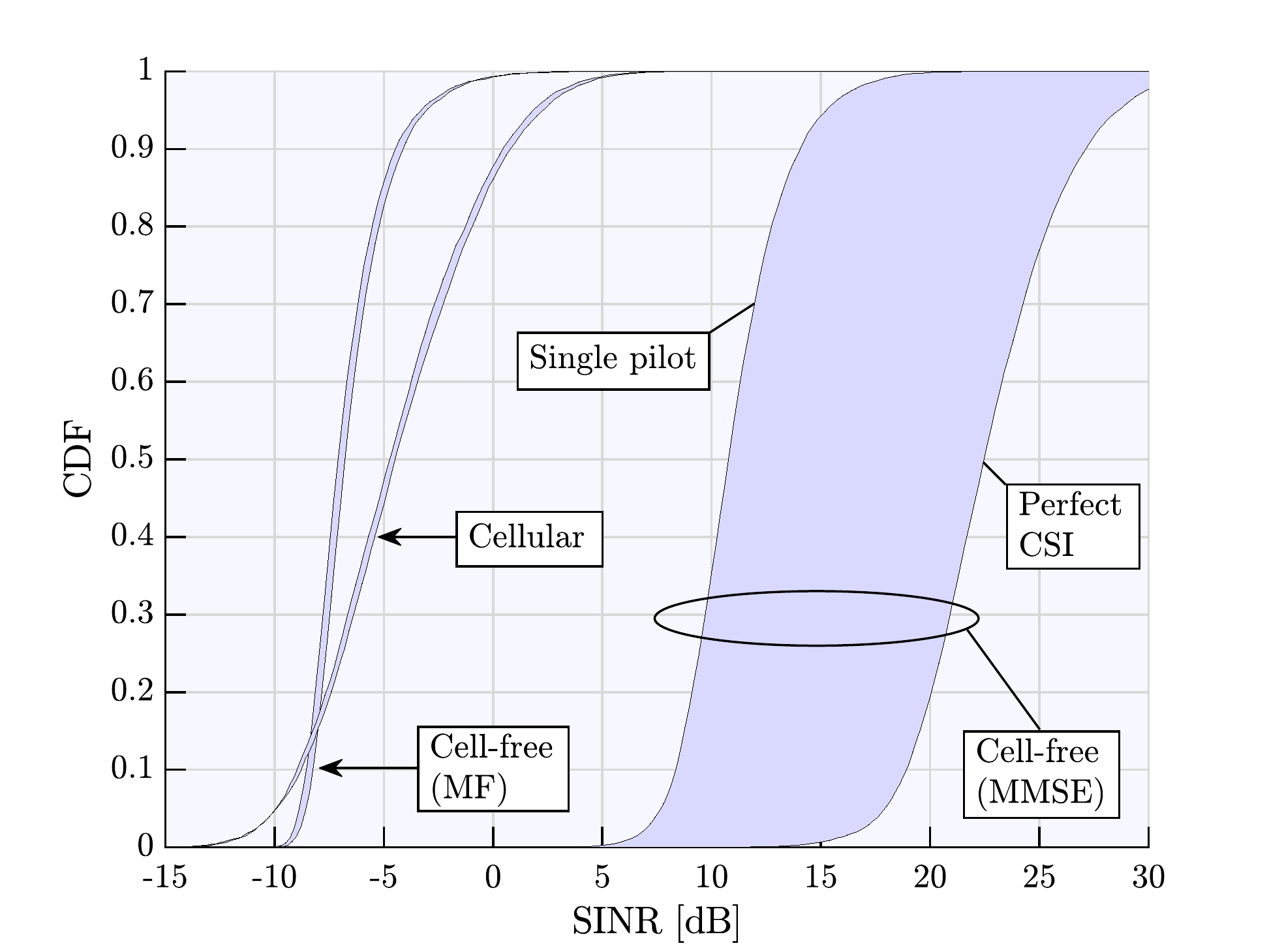}
\caption{SINR distribution for UAVs in a UMi cellular network having $7\%$ UAVs and $93\%$ GUEs, alongside its cell-free counterparts. For each case, the shaded region spans the range between single-symbol SRS and perfect channel estimation.}
\label{fig:CF2}
\end{figure}

Next, the exercise is repeated for a network where there are 14 GUEs for every UAV
\cite{3GPP36777}. The LoS probability and pathloss for GUEs are as per \cite{3GPP38901}. Shown in Fig.~\ref{fig:CF2} is the CDF of the SINR for the UAVs, which, on account of the reduced LoS probability for intercell interferers, shows a somewhat diminished degree of cellular outages. Nevertheless, the potential of the cell-free alternative in its MMSE form remains largely intact.

Research is ongoing on various aspects of cell-free operation that are still not fully resolved, including scalability \cite{bjornson2020scalable}, DL precoding \cite{attarifar2019modified,MMM20,interdonato2021enhanced}, or DL power allocation \cite{nayebi2017precoding,nikbakht2020unsupervised}, and new issues arise with the move to C-RAN structures, where the traditional shared resources of power and bandwidth are augmented by a new shared resource: computation \cite{valenti2014role}. All of this research is relevant to UAV service, and should be expanded to accommodate UAV specificity, chiefly the dependence on altitude and the strict battery charging schedules.

As anticipated in Sec. \ref{sec:5G}, UAVs may also act as flying BSs,
fashioning networks for disaster relief, emergency services, or special events \cite{XiaSemMez2020}. These networks would present a host of specific challenges, above all the wireless nature of the backhaul, the fact that the BSs would be battery-operated and themselves mobile, and the need to synchronize over the air.  
Also for such networks, which in a sense would be the dual of terrestrial networks serving UAVs, cell-free operation might be enticing \cite{zeng2016wireless}.

\begin{takeaway}
Cell-free architectures can provide macro-diversity and turn interference into useful signal, dramatically improving the worst-case performance scenarios and thus the overall reliability. For UAVs, undergoing LoS conditions to multiple BSs, the appeal of cell-free operations is reinforced even further.
\end{takeaway}

\section{Artificial Intelligence to Model and\\Enhance UAV Communications} 
\label{sec:AI}

Arguably, the pervasive application of AI frameworks across different parts of the cellular system design is widely regarded as one of the key constituents of the future 6G networks \cite{8869705, 9040264}. Below we present two proven and distinct use cases where UAV cellular communications greatly benefit from the application of AI frameworks.

\subsection{AI for Aerial Channel Modeling}
\label{subsec:AIchannel}

Despite their importance, accessible channel models---not based on computationally intensive ray tracing---for UAV communications that operate in the mmWave spectrum are not yet available. For instance, current 3GPP standard-defined aerial channel models are calibrated only for sub-\SI{6}{GHz} frequencies \cite{3GPP36777}. Attempts to create the first measurements-based mmWave aerial channel model  \cite{SemKanHaa2021,polesea2a,GarMohJai2020} focus exclusively on LoS links and
are still missing a statistical description of the multipath channel, which is essential to enable wireless coverage in a multitude of scenarios, as discussed in Sec.\ \ref{sec:mmWave}. Similarly, several other works such as \cite{shakhatreh2021modeling,KovMolSam2018,dabiri2019analytical,gapeyenko2018flexible} fall short of providing a 3D spatial channel model and are therefore not yet fully adequate to conduct a rigorous assessment of the performance that can be achieved by mmWave-connected UAVs.

Since mmWave systems rely on highly directional 
communication at both transmitter and receiver, channel models must provide
statistical descriptions of 
the full \emph{doubly directional} characteristics 
of the channel, which include the totality of path components as well as the angles of arrivals, angles of departure, gains and delays. Modern data-driven machine-learning methods become an attractive recourse to tackle this challenge. In particular, neural networks (NNs) have been used in \cite{stocker1993neural,chang1997environment,bai2018predicting,huang2018big,zhaodu2020} for indoor mmWave channel modeling, where the NN outputs parameters that represent a regression from the training dataset, similar to what is typically proposed with learning-based planning and prediction tools \cite{ostlin2010macrocell,dall2011channel,azpilicueta2014ray,kasparick2015kernel,ferreira2016improvement,romero2017learning,ma2017data,8647510}. 

\begin{figure}[!t]
    \centering
    \includegraphics[width=\figwidth, trim = 0.8cm 0 0cm 0, clip]{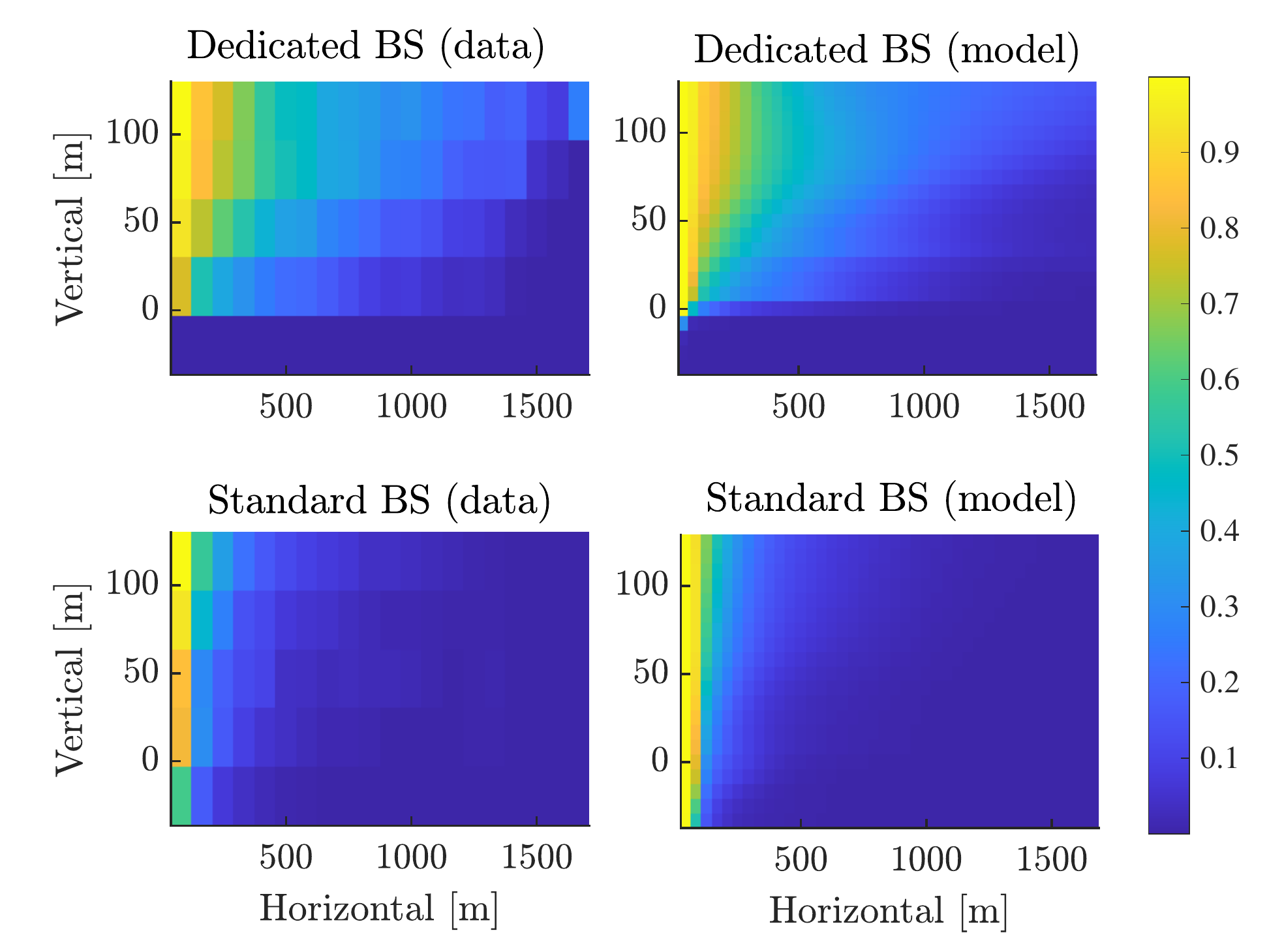}
    \caption{LoS link probabilities as a function of horizontal and vertical distance to the BS for standard and dedicated BS types (see Sec.\ \ref{subsec:mmWaveUrban}): empirical distribution of the test data (left) versus probability forecast by the trained link-state predictor (right).}
    \label{fig:AI_Fig1}
\end{figure}

\begin{figure}[!t]
    \centering
    \includegraphics[width=\figwidth, trim = 0.8cm 0 1cm 0, clip]{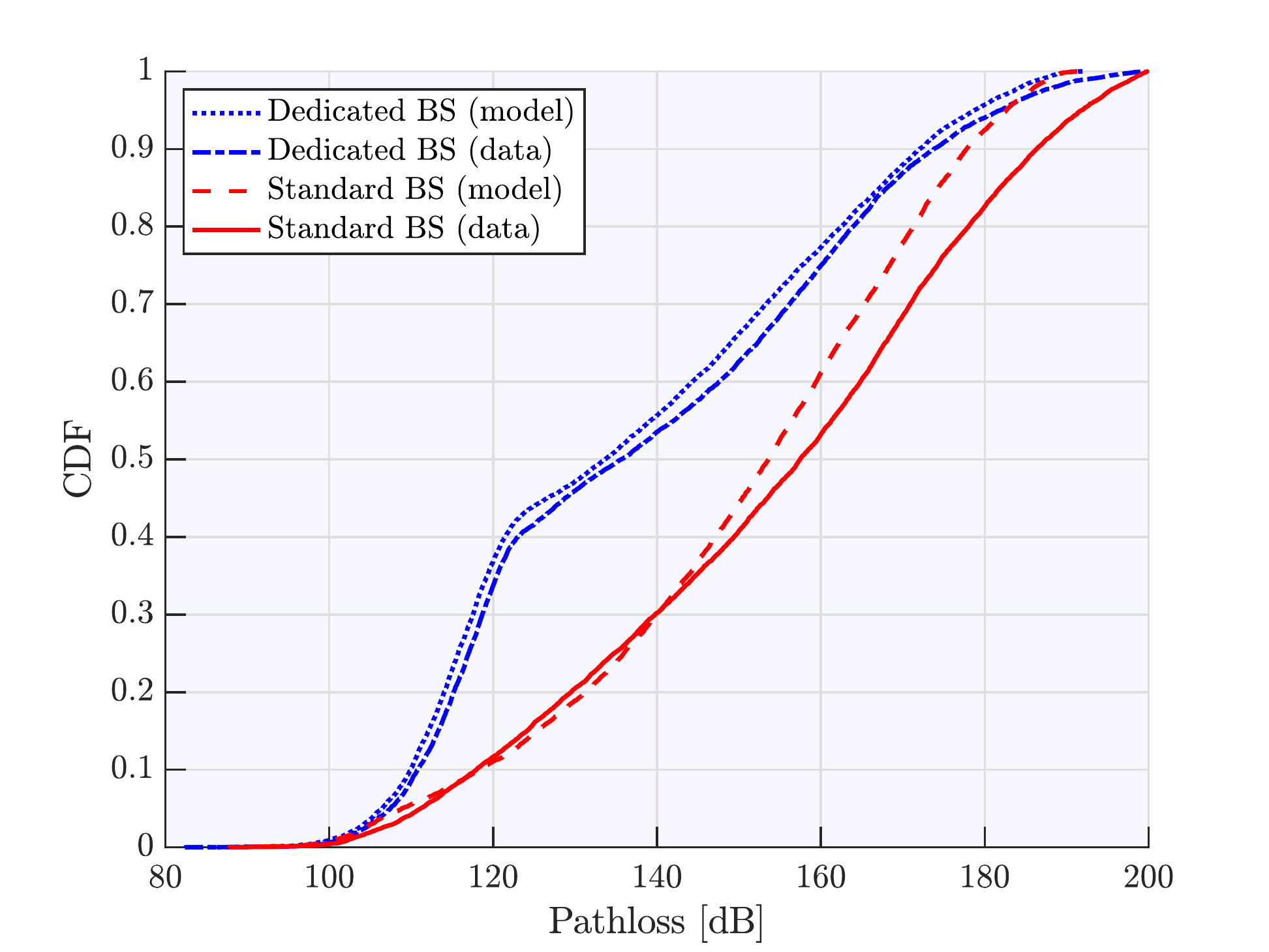}
    \caption{CDF of the path loss for the test data alongside the one generated by the trained model for the city of London.}
    \label{fig:AI_Fig2}
\end{figure}

To develop an accessible aerial channel model, we put forward a generative model
featuring a novel two-stage NN structure:
\begin{itemize}
    \item A first NN determines if the link is in a state of LoS, NLoS, or outage.
    \item A conditional variational autoencoder is employed to generate the path parameters  given that link state.
\end{itemize}
This work is extensively discussed in \cite{XiaRanMez2020,XiaRanMez2020a}, and 
a massive ray-tracing-based urban dataset suitable for training purposes as well as the ensuing
statistical generation of the 3D spatial channel model is publicly available \cite{mmw-github-nyu}.
The accuracy of the AI-based model is corroborated in Figs.~\ref{fig:AI_Fig1} and \ref{fig:AI_Fig2} for the section of London portrayed in Fig.~\ref{fig:mmWave_london}. Fig.~\ref{fig:AI_Fig1} compares the actual LoS probabilities in the test data, obtained via ray tracing, with the output of the NN-based link-state predictor. Similarly, Fig.~\ref{fig:AI_Fig2} contrasts the CDF of the path loss corresponding to the test data
with the CDF of the path loss generated by the trained model. 

\subsection{AI for UAV Mobility Management}
\label{subsec:AImobility}

As highlighted in Section~\ref{subsec:massivMIMO1}, cellular-connected UAVs require novel and robust solutions for mobility management support, leveraging their knowledge of the network-controlled UAV trajectory \cite{LinWirEul2019,StaKovKoz2018,FakBetHay2019,AmeSaaMar2020,Amer2020a}. While new radio resource control signaling has been introduced in LTE for this purpose \cite{MurLinMaa2018}, the intricacy and dynamic nature of the problem calls for AI and machine learning approaches \cite{azari2020mobile,hu2020reinforcement,azari2020machine,challita2019interference,galkin2020reqiba,liu2020artificial}. In this regard, reinforcement learning appears particularly suitable to optimize handover decisions and therefore reduce the number of unnecessary handovers \cite{CheLinKha2020,sutton2018reinforcement}. 

\begin{figure}[!t]
    \centering
    \includegraphics[width=\figwidth, trim = 0.8cm 0 1.5cm 0, clip]{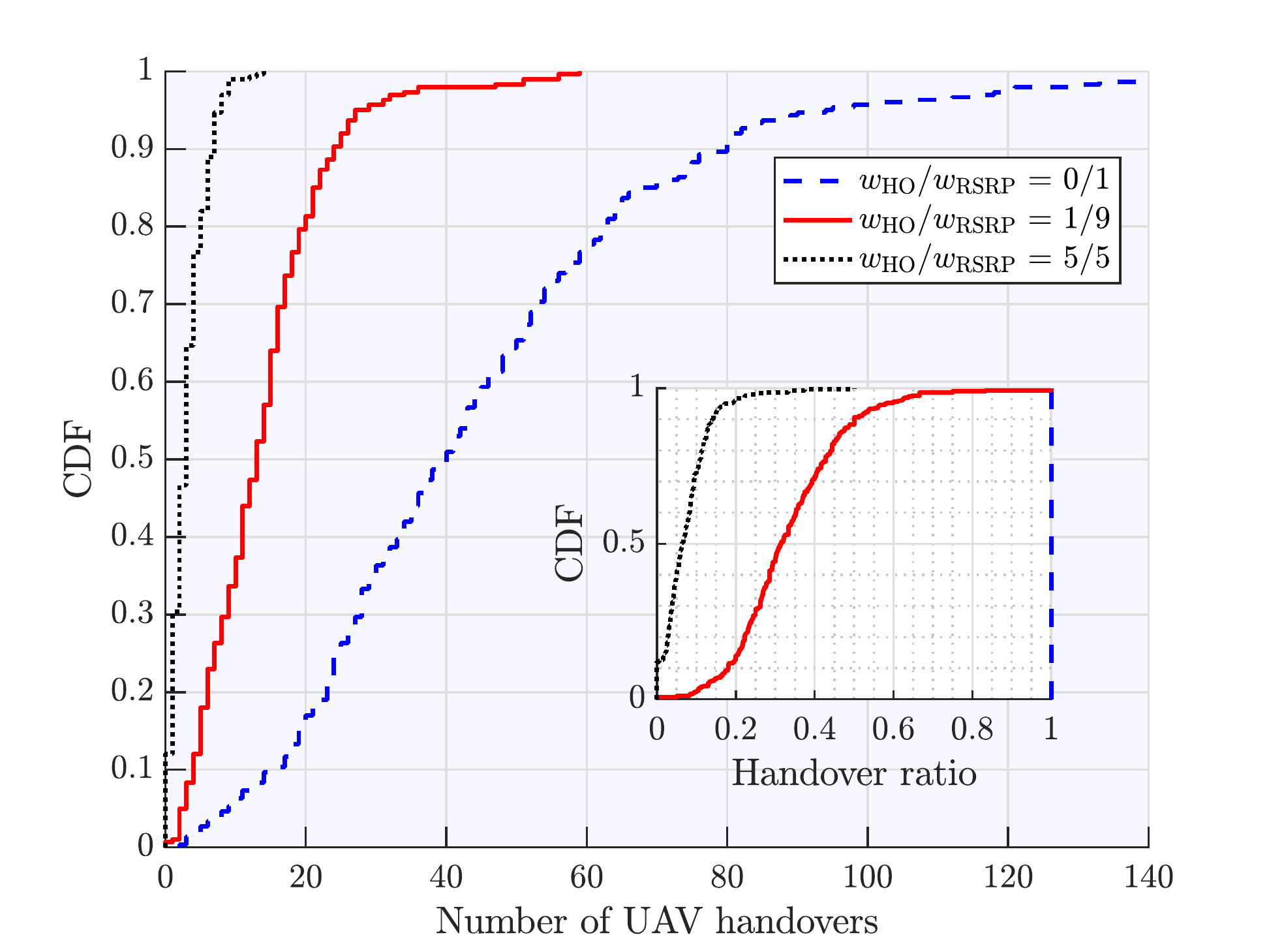}
    \caption{CDF of the number of handovers for a UAV, with the corresponding handover ratio shown in the inset, under a reinforcement learning approach for various weight combinations.}
      \label{fig:AI_Fig3}
\end{figure}

Fig.~\ref{fig:AI_Fig3} considers the same setup as in Fig.~\ref{fig:mMIMO_Fig2}, where this time a flexible Q-learning algorithm is employed with the purpose of reaching a trade-off between the number of handovers performed by UAVs (to be minimized) and their observed RSRP values (to be maximized). The Q-learning algorithm essentially decides when a UAV with a given position and travelling trajectory should execute a handover. The figure illustrates the CDF of the number of handovers for different combinations of weights $w_{\mathrm{HO}}$ and $w_{\mathrm{RSRP}}$, respectively assigned to the handover cost and to the RSRP value. Note that the case $w_{\mathrm{HO}}/w_{\mathrm{RSRP}}=0/1$ corresponds to a conventional greedy scheme, where the UAV always connects to the strongest cell. The inset also shows the corresponding handover ratio, defined as the ratio between the number of handover under each weight combination and the one under the greedy approach. Fig.~\ref{fig:AI_Fig3} shows how the number of handovers decreases as the ratio $w_{\mathrm{HO}}/w_{\mathrm{RSRP}}$ increases, and that by properly tuning the weights, such number can be reduced significantly. For instance, for $w_{\mathrm{HO}}/w_{\mathrm{RSRP}}=5/5$, where handover cost and RSRP are considered equally important, the highest 10\% handover ratio is reduced by 85\% with respect to the greedy baseline setup with $w_{\mathrm{HO}}/w_{\mathrm{RSRP}}=1/1$. Learning-based techniques thus generalize traditional mobility management schemes and, more broadly, open up new possibilities for optimal UAV-aware network design and operations.

\begin{figure}[!t]
    \centering
    \includegraphics[width=\figwidth, trim = 0.8cm 0 1.5cm 0, clip]{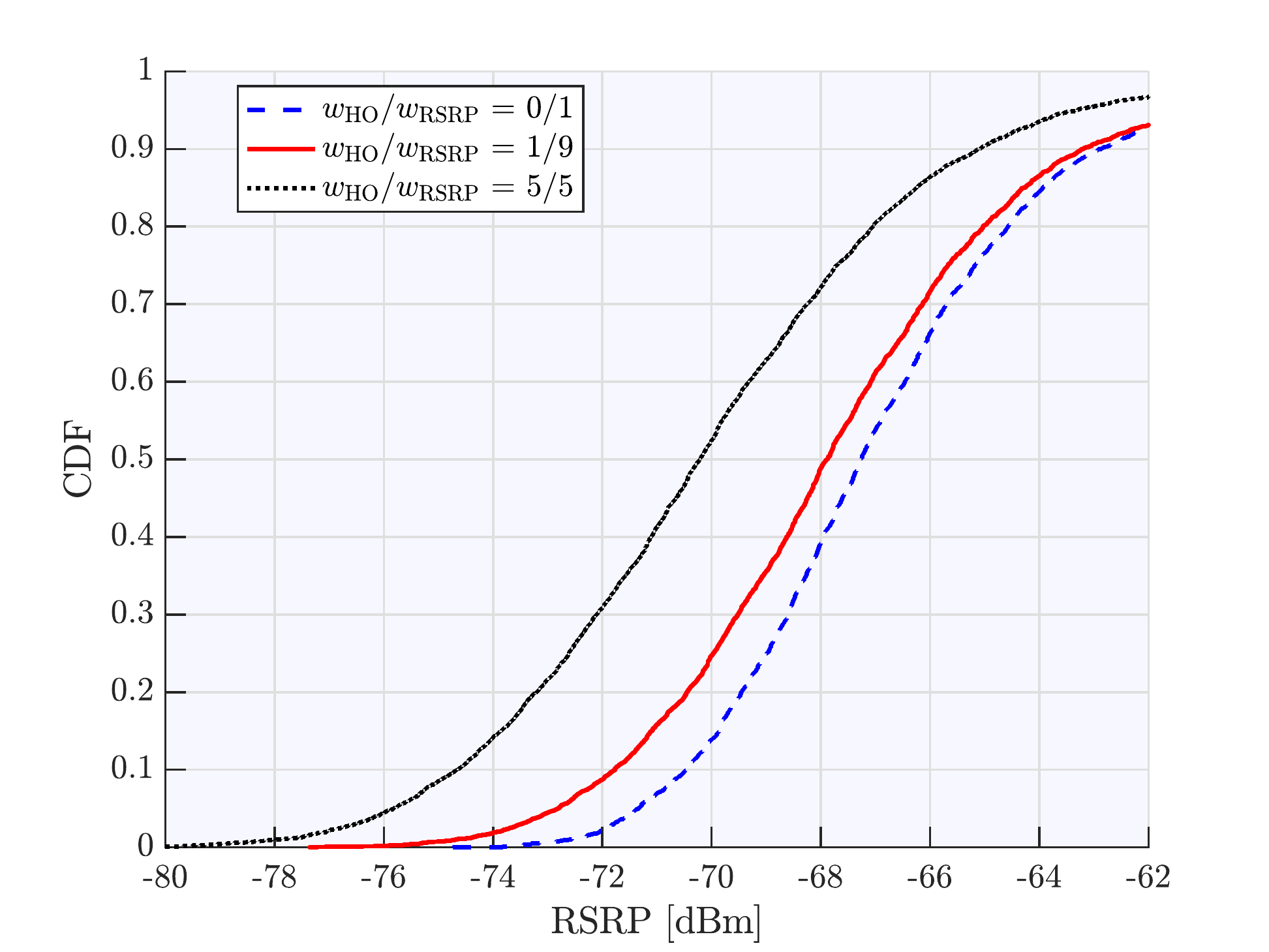}
    \caption{CDF of the RSRP observed by a UAV under a reinforcement learning approach for various weight combinations.}
    \label{fig:AI_Fig4}
\end{figure}

Fig.~\ref{fig:AI_Fig4} shows the CDF of the RSRP observed by a UAV for the same three weight combinations, illustrating how the RSRP drops as the ratio $w_{\mathrm{HO}}/w_{\mathrm{RSRP}}$ increases. For instance, we can observe how the 5\%-worst UAVs lose around \SI{4.5}{dB} when comparing $w_{\mathrm{HO}}/w_{\mathrm{RSRP}} = 5/5$ with the greedy baseline, whereas setting $w_{\mathrm{HO}}/w_{\mathrm{RSRP}}=1/9$ only leads to a \SI{1.8}{dB} deficit. In all three cases considered, the minimum RSRP is always above \SI{-80}{dBm}, corresponding to a \SI{33}{dB} SNR over a bandwidth of \SI{1}{MHz}, and typically sufficient to provide reliable connectivity. Depending on the specific scenario, the weights may be configured to operate at an acceptable RSRP level while minimizing the number of handovers and thus both the associated signaling overhead and failure probability.

\begin{takeaway}
Data-driven machine-learning methods are an attractive resource to fill the gaps in aerial channel modeling, enabling novel and rigorous performance assessments of UAV communications. Additionally, AI-based approaches can generalize traditional mobility management schemes and, more broadly, open up new possibilities for optimal UAV-aware network design and operations.
\end{takeaway}
\section{Assisting UAV Cellular Communications with Reconfigurable Intelligent Surfaces}  
\label{sec:RIS}

As mentioned, the autonomous UAV cargos and taxis of tomorrow will afford minimal lacks of connectivity along their flights.
Unfortunately, UAVs might happen to navigate complex and unpredictable propagation environments, with the air-to-ground links occasionally blocked by trees or high-rise buildings. Besides enhancing the cellular infrastructure---by densifying ground deployments (Sec.~\ref{sec:massiveMIMO},~\ref{sec:mmWave},~\ref{sec:cellfree}) or reaching for space (Sec.~\ref{sec:NTN})---an alternative means to boost reliability is the one of manipulating the wireless propagation environment itself to one's own advantage. Such smart manipulation can capitalize on an emerging transmission technology known as RISs \cite{DBLP:journals/ejwcn/RenzoDHZAYSAHGR19,DBLP:journals/access/BasarRRDAZ19,DBLP:journals/jsac/RenzoZDAYRT20,DBLP:journals/ojcs/RenzoNSDQLRPSZD20,DBLP:journals/wc/HuangHAZYZRD20,DBLP:journals/corr/abs-2007-02759,DBLP:journals/cm/WuZ20,DBLP:journals/corr/abs-2104-06265,2020arXiv200703435L}.


RISs realize programmable and reconfigurable wireless propagation environments through nearly passive and tunable signal transformations. An RIS is a planar structure engineered with properties that enable a dynamic control of the electromagnetic waves. An RIS is usually made of a large number of passive scattering elements whose response to electromagnetic waves can be adaptively configured through simple and low-cost electronic circuits such as PIN diodes or varactors. Conceptually, an RIS can be thought of being made of many sub-wavelength antenna dipoles (or patch antenna elements), which can be controlled by tunable lumped loads \cite{DBLP:journals/corr/abs-2009-02694}. By tuning the load, the scattered field can be optimized and, for instance, a plane wave that impinges upon an RIS from a given direction can be steered towards a direction of reflection, different from that of incidence and corresponding to the location of an intended user \cite{DBLP:journals/corr/abs-2011-14373,DBLP:journals/corr/abs-2102-07155}.

An RIS can realize sophisticated and exotic signal transformations, which include anomalous refractions, polarization conversion, collimation, and focusing \cite{DBLP:journals/jsac/RenzoZDAYRT20}. A remarkable instance of these capabilities is that of dual-function RISs, which can simultaneously reflect and refract signals adaptively to ensure full coverage on both sides of an RIS, depending on the spatial distribution of the users around \cite{DBLP:journals/corr/abs-2011-00765}.  Besides making complex wireless environments programmable, RISs can be used to design single-RF but multistream transmitters at a low complexity, power, and cost \cite{DBLP:journals/corr/abs-2009-00789}, and they can serve as an ingredient in reconfigurable backscatter communication systems \cite{DBLP:journals/corr/abs-2103-08427}.

\begin{figure}[!t]
    \centering
    \includegraphics[width=\figwidth]{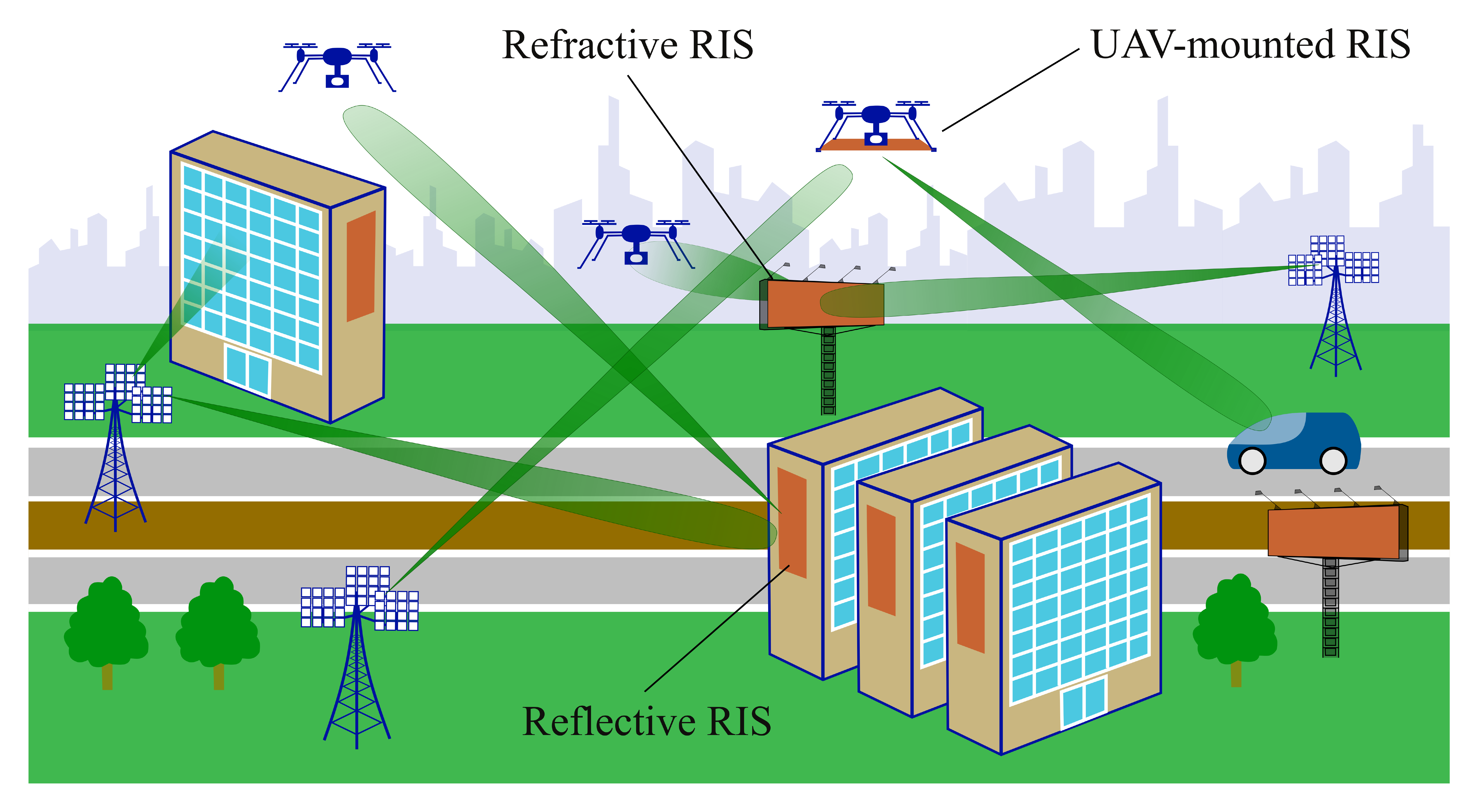}
    \caption{Examples of RIS-assisted UAV cellular communications.}
    \label{fig:RIS_Fig1}
\end{figure}

As exemplified in Fig.~\ref{fig:RIS_Fig1}, RISs could be employed in different ways to overcome some of the limitations of UAV communication networks, making them more efficient and reliable:
\begin{itemize}
    \item \emph{Smart reflections.} RISs could be deployed on the façades of high-rise buildings and controlled to create signal reflections that enhance the reliability of air-to-ground links when the latter are, e.g., blocked by other buildings.
    \item \emph{Smart refractions.} RISs could be installed on billboards and optimized to create signal refractions that enhance the link reliability when UAV users transit behind the billboard. In some cases, refractive RISs could also be placed on windows to assist outdoor-to-indoor communications, e.g., between a UAV BS and an in-building user.
    \item \emph{UAV-carried smart surfaces.} RISs could be mounted on UAVs too, acting as relays to provide a $360^{\circ}$ panoramic view and on-demand coverage through controllable ground-to-air-to-ground signal reflections.
\end{itemize}

Ongoing studies are focusing on whether and how to exploit the potential synergies between UAVs and RISs \cite{DBLP:journals/corr/abs-2012-04775,DBLP:journals/corr/abs-2103-07151,DBLP:journals/corr/abs-2010-09317}. A large body of research has been concerned with the performance evaluation of RIS-assisted UAV communications, accounting
for the UAV heights, the RIS features, and the potential presence of phase errors \cite{DBLP:journals/tvt/YangMZHR20,MaDinHas2020,DBLP:conf/icc/CaiWHNY20,DBLP:journals/tcom/ShafiqueTH21,DBLP:journals/iotj/RanjhaK21a,DBLP:PhaseError,DBLP:3DBeamFlattening,DBLP:journals/corr/abs-2101-09145}. However, a deployment optimization of RIS-assisted UAV cellular communications should be carried out on realistic large-scale networks
\cite{DBLP:journals/ejwcn/RenzoS19}. The marriage between UAVs and RISs also brings about the unique problem of jointly optimizing the UAV trajectories and the signal transformations 
applied by the RISs 
\cite{DBLP:journals/wcl/LiDYLR20,DBLP:conf/iccchina/LiL20,DBLP:journals/access/GeDZWY20,DBLP:journals/corr/abs-2007-14029,DBLP:THz,DBLP:journals/twc/WeiCSNYZS21}. In some cases, due to its algorithmic or modeling deficiencies, such optimal design has been approached with machine learning tools, achieving more computationally efficient implementations \cite{DBLP:journals/tcom/ZapponeRD19,DBLP:journals/corr/abs-1808-01672,DBLP:journals/corr/abs-2002-11040,DBLP:journals/corr/abs-2104-06758,DBLP:conf/globecom/ZhangSB19,DBLP:AgeofInformation,DBLP:DRL}.

\begin{takeaway} 
RISs might enhance the efficiency and reliability of future air-to-ground links through optimally controlled reflections and/or refractions. Nonetheless, major outstanding issues in this domain range from the electromagnetic-consistent modeling of RISs all the way to the performance evaluation and deployment optimization in practical large-scale networks featuring UAVs.
\end{takeaway}

\section{UAV Communication at THz Frequencies}
\label{sec:THz}

While 5G is seizing the mmWave band, the attention of researchers is shifting already to the THz range, broadly defined as \SI{100}{GHz}--\SI{10}{THz} \cite{piesiewicz2007short,elayan2019terahertz}. Because of reduced diffraction, THz communication is mostly circumvented to LoS channels, but that is compatible with UAV situations where extremely high bit rates may be required, say wireless fronthaul to UAVs acting as BSs \cite{THz120Gbps}. Furthermore, the THz band opens the door to a synergistic integration of communication with positioning and even imaging \cite{8732419}. 

The understanding of THz propagation is still incipient and comprehensive channel models are lacking for terrestrial applications, let alone for aerial settings \cite{TekEktKur2020}. On the one hand, 
behaviors that begin to arise at mmWave frequencies become prominent in the THz realm, including pronounced peaks of molecular absorption \cite{akyildiz2014terahertz}, angular sparsity, high omnidirectional pathloss, and wavefronts that are nonflat over the span of large arrays \cite{jiang2005spherical}.
Likewise, the extremely broad bandwidths bring to the fore phenomena that were hitherto muted, including spatial widening over large arrays and the ensuing beam squinting \cite{8443598}.
On the other hand, some of these same behaviors lead to new opportunities. Wavefront curvature over the arrays, for instance, opens up the possibility of MIMO communication in strict LoS conditions: even in the absence of multipath propagation, high-rank channels can then be created based only on the array apertures \cite{Driessen:99,bohagen2007design,torkildson2011indoor,DooCM2020}.
Moreover, because the attributes of such channels hinge on sheer geometry, they can be controlled through the design and disposition of the arrays themselves \cite{DoLeeLoz2020}.

\begin{figure}
\centering
\includegraphics[width=\figwidth, trim = 0.8cm 0 1.3cm 0, clip]{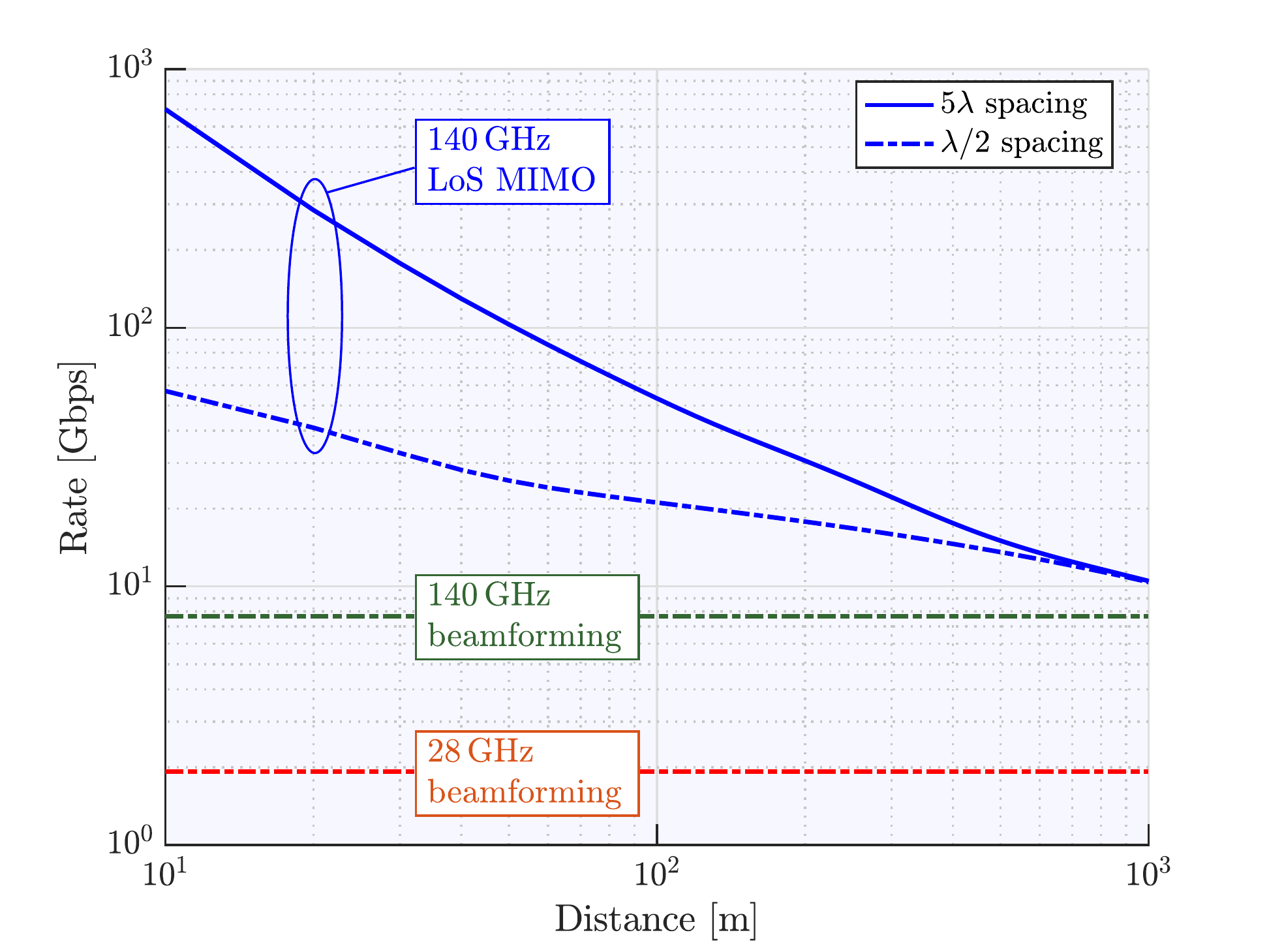}
\caption{Information-theoretic bit rate as a function of distance for \SI{28}{GHz} and \SI{140}{GHz} LoS transmissions. At \SI{28}{GHz}: 16 transmit and 64 receive antennas. At \SI{140}{GHz}: 256 transmit and 1296 receive antennas.}
\label{fig:THz1}
\end{figure}

The potential of THz communication can be gauged in Fig.~\ref{fig:THz1}, which presents the information-theoretic bit rates achievable as a function of the distance on a free-space channel. A comparison is drawn between a \SI{400}{MHz} transmission at \SI{28}{GHz} and a \SI{1.6}{GHz} transmission at \SI{140}{GHz}. The transmit power, noise figure, and antenna gains, are all as per \cite{3GPP38901}. At \SI{28}{GHz}, the transmitter features a 16-antenna planar array with half-wavelength spacing (\SI{1.6}{cm} $\times$ \SI{1.6}{cm}) and the receiver features a 64-antenna array, also planar and with half-wavelength spacing (\SI{3.75}{cm} $\times$ \SI{3.75}{cm}); beamforming is applied. At \SI{140}{GHz}, the noise figure is increased by \SI{3}{dB} and the same array areas and antenna spacings are used, now accommodating 256 antennas at the transmitter and 1296 antennas at the receiver; beamforming is again applied. The spectral efficiencies are capped at \SI{4.8}{bps/Hz} per antenna to prevent artifact values corresponding to excessively large constellations \cite{XiaSemMez2020}; this results in steady bit rates up to \SI{1000}{m}, with the \SI{140}{GHz} transmission exhibiting a four-fold advantage by virtue of the combined effects of more bandwidth and more antennas (within the same real estate).
An even more dramatic leap in performance is possible by resorting to LoS MIMO, even if the compact half-wavelength antenna spacings are retained, and let alone with these relaxed to $5\lambda$ so the respective arrays occupy \SI{16}{cm} $\times$ \SI{16}{cm} and \SI{37.5}{cm} $\times$ \SI{37.5}{cm}. Then, truly stupendous bit rates are theoretically feasible at short and intermediate distances and, even if only a fraction of these are realized, the upside is extraordinary.

Enter UAVs, with new challenges in the domain of channel characterization such as the effects of altitude and trajectories, the uncertainties associated with wind \cite{8974403}, the orientation and wobbling of the arrays \cite{6868215,9206092}, or possible signal blockages by the body of the UAV \cite{semkin2021lightweight}.
The power consumption aspect also becomes exacerbated, and additional limitations might be imposed because of size and weight considerations. Chief among the power consumption issues at THz frequencies stand the analog-to-digital
converters, whose power consumption grows linearly with the bandwidth and exponentially with the resolution \cite{lee2008analog}. Despite these hurdles, which call for multidisciplinary research efforts, UAV communication at THz frequencies is an exciting proposition with vast potential, if not as a primary means of communication, at least as an enhancing mode with fallback to lower frequencies.

\begin{takeaway}
UAV communication at THz frequencies has vast potential, putting in play truly enormous bandwidths and allowing for MIMO transmissions even in strict LoS conditions. Issues such as power consumption, attenuation, and sensitivity to vibrations suggest confining THz operations to short-to-intermediate distances and to a non-standalone mode.
\end{takeaway}

\section{Conclusion}

While the understanding of UAV cellular communications has been advancing over the last few years, many fundamental challenges remain to be addressed.
In this article, we blended academic and industrial views, embarking on a journey that took us from 5G to 6G UAV use cases, requirements, and enabling technologies. 

Through tangible results, we detailed how NR enhancements will greatly help satisfying the stringent control and payload data demands of network-connected UAVs throughout this decade. Among them, beamformed control signals facilitate UAV cell selection and handovers with respect to sidelobe-based association. mMIMO paired with UAV-aware nullsteering or UAV-based beamforming is essential to guarantee reliable cellular connectivity in both UMa and denser UMi scenarios. Somewhat surprisingly, NR mmWave networks too can provide satisfactory coverage of the sky, thanks to a favorable combination of antenna sidelobes and strong reflections. Albeit subject to regulations, such coverage could be further enhanced by rooftop-mounted uptilted mmWave cells. For UAV-to-UAV applications, underlaying these direct links with the ground uplink incurs limited mutual interference, which however becomes more pronounced when UAVs fly higher. UAV-specific power control can effectively trade off the UAV-to-UAV performance with its ground-link counterpart. 

Shifting gears to the next decade, empowering air taxis with sufficient data transfer capacity and a minimal lack of connectivity will require a 6G paradigm shift in 2030. NTNs could fill the inevitable ground coverage gaps that can currently jeopardize a UAV mission, also handling the more mobile UAVs across their large footprint. Cell-free architectures and reconfigurable smart wireless environments could turn interference into useful signal and boost UAV coverage, dramatically improving worst-case performance and thus overall reliability. AI can already enable aerial channel modeling and will also help with optimal UAV-aware network design and operations, e.g., for mobility management. Finally, THz frequencies could provide enormous bandwidths for UAVs, allowing MIMO even in strict LoS conditions. As tailoring these technologies to UAVs is nothing short of challenging, we also pointed out numerous open problems and made well-grounded suggestions on much-needed future work. We hope this article will foster new research and breakthroughs, bringing the wireless community one step closer to the fly-and-connect era.
\section*{Acknowledgments}

G.~Geraci and A.~Lozano were supported by ERC grant 694974, by MINECO's Project RTI2018-101040, by the Junior Leader Fellowship Program from ``la Caixa" Banking Foundation, and by the ICREA Academia program. 

M.~M.~Azari and S.~Chatzinotas were supported by the FNR 5G-SKY project under grant agreement number 13713801, and the SMC funding program through the Micro5G project.

M.~Mezzavilla and S.~Rangan were supported by NSF grants 1302336, 1564142,  1547332, and 1824434,  NIST, SRC, and the industrial affiliates of NYU WIRELESS. 

M.~Di~Renzo was supported in part by the European Commission through the H2020 ARIADNE project under grant agreement number 871464 and through the H2020 RISE-6G project under grant agreement number 101017011.

\bibliographystyle{IEEEtran}
\bibliography{bibl}

\section*{Biographies}
\balance
\small

\noindent
\textbf{Giovanni Geraci} (S'11 - M'14 - SM'19) is an Assistant Professor at Univ. Pompeu Fabra in Barcelona. He was previously a Research Scientist with Nokia Bell Labs and holds a Ph.D. from UNSW Sydney. He serves as an IEEE ComSoc Distinguished Lecturer and as an Editor for the IEEE Transactions on Wireless Communications and IEEE Communications Letters. He is co-inventor of a dozen patents, co-edited the book ``UAV Communications for 5G and Beyond'' by Wiley---IEEE Press, and has delivered a dozen IEEE ComSoc tutorials. Giovanni received ``la Caixa'' Junior Leader and ``Ram\'{o}n y Cajal'' Fellowships, the Nokia Bell Labs Ireland Certificate of Outstanding Achievement, the IEEE PIMRC'19 Best Paper Award, and the IEEE ComSoc EMEA Outstanding Young Researcher Award.

\vspace{0.2cm}
\noindent
\textbf{Adrian Garcia-Rodriguez} (S'13 - M'17) is a Senior Research Engineer at Huawei R\&D Paris, France, where he focuses on advanced cellular network optimization techniques and UAV communications. Previously, he was a Research Scientist in Nokia Bell Labs, Ireland, which he joined after obtaining his Ph.D from University College London. He is a co-inventor of 25+ filed patent families, a recipient of the Nokia Bell Labs Ireland Certificate of Outstanding Achievement, and a co-author of 40+ IEEE publications with 1k+ citations, including the Best Paper Award at PIMRC'19 for his work on ``UAV-to-UAV cellular communications''.

\vspace{0.2cm}
\noindent
\textbf{M. Mahdi Azari} (S'15 - M'19) obtained his PhD from KU Leuven (Belgium), and his MSc and BSc degrees from University of Tehran (Iran), in electrical and communication systems engineering. Currently, he is a research associate at SnT, University of Luxembourg. Prior to this, he was a PostDoc researcher at CTTC (Spain). He has (co-)authored various scientific articles and a book chapter on communication systems, all in recognized venues. For his work on ``Cellular UAV-to-UAV Communications'' he received the Best Paper Award at IEEE PIMRC'19. He is a silver medalist of Iran’s National Mathematical Olympiad and recipient of the INEF award. 

\vspace{0.2cm}
\noindent
\textbf{Angel Lozano} (S'90 - M'99 - SM'01 - F'14) is a Professor at Univ. Pompeu Fabra. He received his Ph.D. from Stanford University in 1998. In 1999, he joined Bell Labs (Lucent Technologies, now Nokia) in Holmdel, NJ, where he was with the Wireless Communications Department until 2008. He serves as Area Editor for the IEEE Transactions on Wireless Communications and as Editor for the IEEE Communication Technology News. He co-authored the textbook ``Foundations of MIMO Communication'' (Cambridge University Press, 2019), received the 2009 Stephen O. Rice Prize,	the 2016 Fred W. Ellersick prize, and the 2016 Communications Society \& Information Theory Society joint paper award. He holds an Advanced Grant from the ERC and was a 2017 Highly Cited Researcher.

\vspace{0.2cm}
\noindent
\textbf{Marco Mezzavilla} (S'10 - M'14 - SM'19) is a research faculty at the NYU Tandon School of Engineering. He received his Ph.D. (2013) in information engineering from the University of Padova. His research focuses on the design and validation of algorithms and communication protocols for next-generation wireless technologies.

\vspace{0.2cm}
\noindent
\textbf{Symeon Chatzinotas} (S'06 - M'09 - SM'13) is currently Full Professor and Head of the SIGCOM Research Group at SnT, University of Luxembourg. He is coordinating the research activities on communications and networking, acting as a PI for more than 20 projects.
He was the co-recipient of the 2014 IEEE Distinguished Contributions to Satellite Communications Award and Best Paper Awards at EURASIP JWCN, CROWNCOM, ICSSC. He has co-authored more than 450 technical papers.
He is currently in the editorial board of the IEEE Transactions on Communications, IEEE Open Journal of Vehicular Technology and the International Journal of Satellite Communications and Networking.

\vspace{0.2cm}
\noindent
\textbf{Yun Chen} received her M.Sc. degree in electrical and computer engineering from the University of Texas at Austin in 2020, and currently she is a Ph.D. student associated with Wireless Sensing and Communication Lab at North Carolina State University. Her interest lies in sensor fusion and machine learning (deep learning) aided vehicle-to-everything (V2X) communications based on 5G mmWave MIMO systems.

\vspace{0.2cm}
\noindent
\textbf{Sundeep Rangan} (S'94 – M'98 – SM'13 – F'16) is an ECE professor at NYU and associate director of NYU WIRELESS. He received his Ph.D. from the University of California, Berkeley. In 2000, he co-founded (with four others) Flarion Technologies, a spinoff of Bell Labs that developed the first cellular OFDM data system. It was acquired by Qualcomm in 2006, where he was a director of engineering prior to joining NYU in 2010.

\vspace{0.2cm}
\noindent
\textbf{Marco Di Renzo} (F'20) is a CNRS Research Director (CNRS Professor) with the Laboratory of Signals and Systems of Paris-Saclay University – CNRS and CentraleSupelec, Paris, France, and the Coordinator of the Communications and Networks Research Area of Paris-Saclay University’s Laboratory of Excellence DigiCosme. He serves as the Editor-in-Chief of IEEE Communications Letters and as the Founding Chair of the Special Interest Group on Reconfigurable Intelligent Surfaces of the Wireless Technical Committee of the IEEE Communications Society. Also, he is a Distinguished Speaker of the IEEE Vehicular Technology Society and a Member of the Emerging Technologies Standing Committee of the IEEE Communications Society. Dr. Di Renzo is a Fellow of the IET and a Highly Cited Researcher.

\end{document}